%% file: main.tex
\documentclass{svjour3}                 
 \smartqed  

 \usepackage{url}
\usepackage{hyperref} 
\hypersetup{ 
    colorlinks,%
    citecolor=black,%
    filecolor=black,%
    linkcolor=black,%
    urlcolor=black,%
pdftitle={Finding Desirable Objects under Group Categorical Preferences},
pdfauthor={Nikos Bikakis, Karim Benouaret, Dimitris Sacharidis},
pdfsubject={Ranking Preferable Objects for Group of Users} ,
pdfkeywords={{Preferable object} {Multiple users} {Group preferences} {personalized skylines} {group recommendations} {Collective dominance} {multi-users recommendations} {Collectively maximals}  {Rank aggregation} {Skyline queries} {ranking scheme} {group skyline} {Collective preferences} {hierarchical  attributes} {collective skyline} {group recommender}  {aggregate preferences} {pareto ranking}  {group recommendations} {BBS} {group maximal} {group ranking} {BNL} {multiple users ranking} {desirable objects} {objective aggregation} {group top-k} {preferable objects} {preferences ranking} {Categorical attributes} {scalable group recommendations} {preferable group objects}  {top-k} {preferences aggregation} {collective maximals} {recommender system} {multi-source skyline} {index group recommendations} {skyline aggregation} {group skyline} {ranking aggregation} {group suggestions} {R-tree} {ranking model} {top-k aggregation} {group dominance} {Collective ranking} {pareto aggregation} {hierarchical   data} {objective ranking} {multi-users ranking}	 {multi-users preferences aggregation} {efficient group recommendations} }
} 

 \usepackage{marvosym}
\usepackage{wrapfig}
\usepackage[caption=false]{subfig}
\usepackage{color}
\usepackage[usenames,dvipsnames,svgnames,table]{xcolor}
\usepackage{multirow}
\usepackage{rotating}
\usepackage{graphicx}
\usepackage[linesnumbered, vlined, ruled]{algorithm2e}
\usepackage{booktabs}
\usepackage{amssymb} 
\usepackage{amsfonts} 
\usepackage{amsmath} 
 \usepackage{soul}
\usepackage{etex}
\usepackage{extarrows}
\usepackage{relsize}
\usepackage{epsfig}
\usepackage{epstopdf}
\usepackage{xspace}
\usepackage{amsfonts}
\usepackage{enumerate}
\usepackage{tabularx}
 \usepackage{mathtools}

\newcommand{\eat}[1]{}
\newcommand{\alt}[2]{
#1 
}
\newcommand{\stitle}[1]{\vspace{0cm}\medskip\noindent\textbf{#1}}

\newcommand{\T}{\ensuremath\mathcal{T}}
\renewcommand{\O}{\ensuremath\mathcal{O}}
\newcommand{\R}{\ensuremath\mathcal{R}}
\newcommand{\U}{\ensuremath\mathcal{U}}  
\newcommand{\A}{\ensuremath\mathcal{A}}

\renewcommand{\H}{\ensuremath\mathcal{H}}

\newcommand{\mat}[1]{\ensuremath\langle #1 \rangle}

\DeclareMathOperator{\rank}{rank}

\renewcommand{\qed} {\hfill\Squarepipe}

\newcommand{\myFontH}[1]{{\fontfamily{phv}\selectfont #1}} 

\newenvironment{myproof}
{\medskip\setlength{\leftskip}{10pt}\setlength{\rightskip}{0pt}\par\noindent\ignorespaces 
   \textsc{Proof.}}
{\medskip\par}

\newcounter{lem}

\newcounter{proposcnt}
\newenvironment{myproposition}
{\refstepcounter{proposcnt}\medskip\setlength{\leftskip}{5pt}\setlength{\rightskip}{5pt}\par\noindent\ignorespaces 
   \textbf{Proposition~\theproposcnt.}}
{\par}

\newcounter{prob}
\newenvironment{myprob}[1][]
{\refstepcounter{prob}\par\setlength{\leftskip}{5pt}\setlength{\rightskip}{5pt}\medskip\noindent\ignorespaces
   \textbf{Problem~\theprob. [#1]}}
{\medskip\par}

\newcounter{propertycnt}
\newenvironment{myproperty} [1][]
{\refstepcounter{propertycnt}\par\setlength{\leftskip}{5pt}\setlength{\rightskip}{5pt}\medskip\noindent\ignorespaces
    \textbf{Property~\thepropertycnt. [#1]}}
{\par}

\newcommand{\specialcell}[2][c]{%
  \begin{tabular}[#1]{@{}c@{}}#2\end{tabular}}

\newcommand{\tline}  {\specialrule{0.8 pt}{0pt}{1pt}}		 
\newcommand{\bline}  {\specialrule{0.8 pt}{1pt}{0pt}}		
\newcommand{\dline}  {\specialrule{0.4 pt}{0pt}{2pt} \specialrule{0.4 pt}{0pt}{2pt}}		

\begin{document}


\title{Finding Desirable Objects under Group Categorical Preferences 
$^\star$ \thanks{$^\star$To appear in Knowledge and Information Systems Journal (KAIS), 2015}
}

\author{ Nikos Bikakis \and 
         Karim Benouaret   \and 
       Dimitris~Sacharidis}
 
\institute{
			Nikos Bikakis \at
			National Technical University of Athens, Greece \&
 			ATHENA Research Center, Greece \\
         	\email{bikakis@dblab.ntua.gr}           
            \and
        	Karim Benouaret \at
        	Inria Nancy, France    \\       
        	\email{karim.benouaret@inria.fr}           
            \and
			\mbox{Dimitris Sacharidis} \at
 			Technische Universit\"at Wien, Austria\\
        	\email{dimitris@ec.tuwien.ac.at}           
}

\maketitle

 -\vspace{-10mm}

 \input{abstract}

%


\input{intro}
\input{related}

\input{define}

\input{algo}

\input{rank}

\input{ext}

\input{exp}

\input{concl}

\bibliographystyle{abbrv}
\bibliography{biblio}

\end{document}

%% file: abstract.tex

\begin{abstract}
Considering a group of users, each specifying individual preferences over
categorical attributes, the problem of determining a set of objects that are
objectively preferable by all users is challenging on two levels. First, we
need to determine the preferable objects based on the categorical preferences
for each user, and second we need to reconcile possible conflicts among users'
preferences. A na\"ive solution would first assign degrees of match between
each user and each object, by taking into account all categorical attributes,
and then for each object combine these matching degrees across users to
compute the total score of an object. Such an approach, however, performs two
series of aggregation, among categorical attributes and then across users,
which completely obscure and blur individual preferences. Our solution,
instead of combining individual matching degrees, is to directly operate on
categorical attributes, and define an objective Pareto-based aggregation for
group preferences. Building on our interpretation, we tackle two distinct but
relevant problems: finding the Pareto-optimal objects, and objectively ranking
objects with respect to the group preferences. To increase the efficiency when
dealing with categorical attributes, we introduce an elegant transformation of
categorical attribute values into numerical values, which exhibits certain
nice properties and allows us to use well-known index structures to accelerate
the solutions to the two problems. In fact, experiments on real and synthetic
data show that our index-based techniques are an order of magnitude faster
than baseline approaches, scaling up to millions of objects and thousands of
users.

\keywords{
 Group recommendation \and   Rank aggregation \and Preferable objects \and Skyline queries \and Collective dominance \and Ranking scheme \and  Recommender systems
}

\end{abstract}

%% file: intro.tex

\section{Introduction}
\label{sec:intro}

Recommender systems have the general goal of proposing objects (e.g., movies,
restaurants, hotels) to a user based on her preferences. Several instances of
this generic problem have appeared over the past few years in the Information
Retrieval and Database communities; e.g.,   \cite{AT05,IBS08,BOHG13,SKP11}.
More recently, there is an increased interest in \emph{group recommender
systems}, which propose objects that are well-aligned with the preferences of
a set of users \cite{JS07,MAS11,CC12,BC11}. Our work deals with a class of
these systems, which we term \emph{Group Categorical Preferences} (GCP), and
has the following characteristics. (1) Objects are described by a set of
categorical attributes. (2) User preferences are defined on a subset of the
attributes. (3) There are multiple users with distinct, possibly conflicting,
preferences. The GCP formulation may appear in several scenarios; for instance,
colleagues arranging for a dinner at a restaurant, friends selecting a
vacation plan for a holiday break.

\begin{table}[h]
{{\caption{New York Restaurants}}}
\label{tab:rest}
\begin{tabular}[t]{cccccc}
\tline
& \multicolumn{5}{c}{\textbf{Attributes}}\\ \cmidrule{2-6}
 \textbf{{Restaurant}} & \textbf{{ Cuisine}} & {\textbf{Attire}} & {\textbf{Place} } &  \textbf{{Price}} &  \textbf{{Parking}}\\ \dline
$o_1$ & Eastern & Business casual & Clinton Hill  & \$\$\$ & Street \\
$o_2$ &French  & Formal  & Time Square & \$\$\$\$ & Valet\\
$o_3$ & Brazilian & Smart Casual & Madison Square &\$\$ & No \\
$o_4$ & Mexican & Street wear &Chinatown & \$ & No \\
\bline
\end{tabular} 
\end{table}
\vspace{-5mm}

\begin{table}[h]
\caption{User preferences}
\label{tab:users}
\begin{tabular}{cccccc}
\tline
& \multicolumn{5}{c}{\textbf{Preferences}}\\ \cmidrule{2-6}
\textbf{User} & \specialcell{\textbf{Cuisine}} & {\textbf{Attire}}& \textbf{Place}  & \textbf{Price} & \textbf{Parking}\\ \dline
$u_1$ & European & Casual & Brooklyn & \$\$\$ & Street\\
$u_2$ & French, Chinese &  -- & -- & -- & Valet\\
$u_3$ & Continental & -- & Time Square, Queens & -- & -- \\
\bline
\end{tabular}
 \end{table}

\begin{figure}[t]
\subfloat[Attire attribute]{ \includegraphics[scale=0.7]{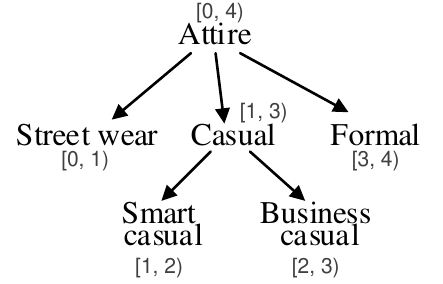} \label{fig:att}}
\subfloat[Cuisine attribute]{ \hspace{30pt} \includegraphics[scale=0.7]{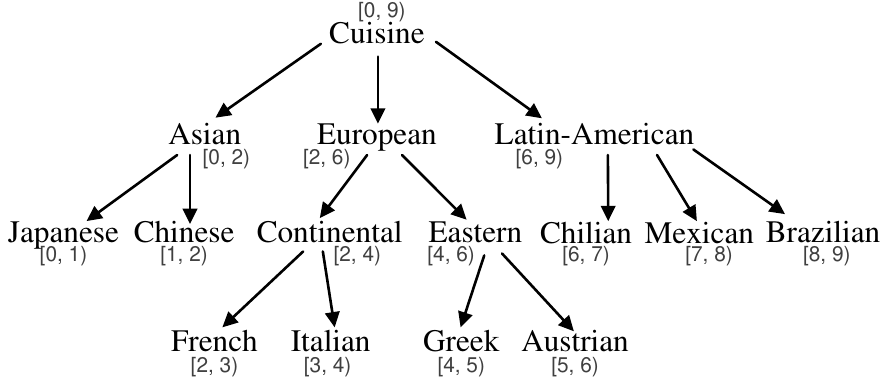} \label{fig:cus}} \\
\subfloat[Parking attribute]{ \includegraphics[scale=0.7]{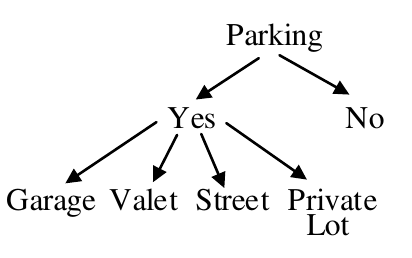} \label{fig:park} }
\subfloat[Place attribute (part of New York places)]{ \hspace{20pt}\includegraphics[scale=0.7]{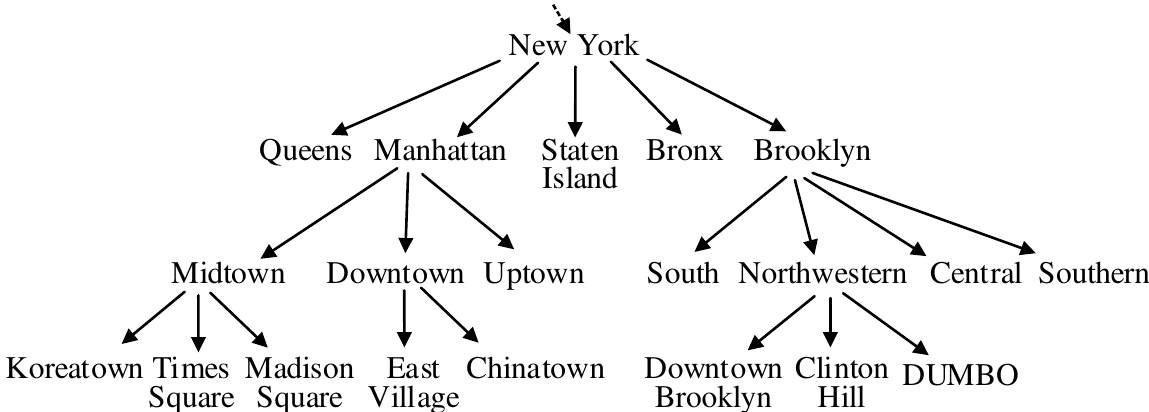}\label{fig:place}} \\
{\caption{Attribute hierarchies}}
\label{fig:hierarchies}
\end{figure}

To illustrate GCP, consider the following example. Assume that three friends
in New York are looking for a restaurant to arrange a dinner. Suppose that,
the three friends are going to use a Web site (e.g.,
Yelp\footnote{www.yelp.com}) in order to search and filter restaurants based
on their preferences. Note that in this setting, as well as in other Web-based
recommendation systems, categorical description are prevalent compared to
numerical attributes. Assume a list of available New York restaurants, shown
in \autoref{tab:rest}, where each is characterized by five categorical
attributes: \textit{Cuisine}, \textit{Attire}, \textit{Place}, \textit{Price}
and \textit{Parking}. In addition, \autoref{fig:hierarchies} depicts the
hierarchies for these attributes. Attire and Parking are three-level
hierarchies, Cuisine and Place are four-level hierarchies, and Price (not
shown in \autoref{fig:hierarchies}) is a two-levels hierarchy with four leaf
nodes ($\$, ..., \$\$\$\$$). Finally, \autoref{tab:users} shows the three
friends' preferences. For instance, $u_1$ prefers European cuisine, likes to
wear casual clothes, and prefers a moderately expensive ($\$\$\$$) restaurant in
the Brooklyn area offering also street parking. On the other hand, $u_2$
likes French and Chinese cuisine, and prefers restaurants offering valet
parking, without expressing any preference on attire, price and place.

Observe that if we look at a particular user, it is straightforward to
determine his ideal restaurant. For instance, $u_1$ clearly prefers $o_1$,
while $u_2$ clearly favors $o_2$. These conclusions per user can be reached
using the following reasoning. Each preference attribute value $u_j.A_k$ is
\emph{matched} with the corresponding object attribute value $o_i.A_k$ using a
matching function, e.g., the Jaccard coefficient, and a matching degree per
preference attribute is derived. Given these degrees, the next step is to
``compose'' them into an overall matching degree between a user $u_j$ and an
object $o_i$. Note that several techniques are proposed for ``composing''
matching degrees; e.g., \cite{MAS11,JM04,K02,SKP11,C03}. The simplest
option is to compute a linear combination, e.g., the sum, of the individual
degrees. 
Finally, alternative aggregations models (e.g., Least-Misery,
Most-pleasure, etc.) could also be considered.

Returning to our example, assume that the matching degrees of user $u_1$ are:
$\langle 1/2, 1/2, 1/6, 1, 1 \rangle$ for restaurant $o_1$, $\langle 1/4, 0,
0, 0, 0 \rangle$ for $o_2$, $\langle 0, 1/2, 0, 0, 0 \rangle$ for $o_3$, and
$\langle 0, 0, 0, 0, 0 \rangle$ for $o_4$ (these degrees correspond to
Jaccard coefficients computed as explained in \autoref{sec:define}). Note that
for almost any ``composition'' method employed (except those that only, or
strongly, consider the Attire attribute), $o_1$ is the most favorable
restaurant for user $u_1$. Using similar reasoning, restaurant $o_2$, is ideal
for both users $u_2$, $u_3$.

When all users are taken into consideration, as required by the GCP formulation,
several questions arise. Which is the best restaurant that satisfies the entire
group? And more importantly, \emph{what does it mean to be the best restaurant}? A
simple answer to the latter, would be the restaurant that has the highest
``composite'' degree of match to all users. Using a similar method as before,
one can define a collective matching degree that ``composes'' the overall
matching degrees for each user. This interpretation, however, enforces an
additional level of ``composition'', the first being across attributes, and
the second across users. These compositions obscure and blur the individual
preferences per attribute of each user.

To some extent, the problem at the first ``composition'' level can be
mitigated by requiring each user to manually define an importance weight among
his specified attribute preferences. On the other hand, it is not easy, if
possible at all, to assign weights to users, so that the assignment is fair.
There are two reasons for this. First, \emph{users may specify different sets
of preference attributes}, e.g., $u_1$ specifies all five attributes, while $u_2$
only Cuisine and Parking. Second, even when considering a particular preference attribute,
e.g., Cuisine, \emph{users may specify values at different levels of the
hierarchy}, e.g., $u_1$ specifies a European cuisine, while $u_2$ French cuisine, 
which is two levels beneath. Similarly, objects can also have attribute
values defined at different levels. Therefore, any ``composition'' is bound to
be \emph{unfair}, as it may favor users with specific preferences and objects
with detailed descriptions, and disfavor users with broader preferences and
objects with coarser descriptions. This is an inherent difficulty of the GCP
problem.

In this work, we introduce the \emph{double Pareto-based aggregation}, which
provides an objective and fair interpretation to the GCP formulation without
``compositing'' across preference attributes and users. Under this concept,
the matching between a user and an object forms a \emph{matching vector}. Each
coordinate of this vector corresponds to an attribute and takes the value of
the corresponding matching degree. The first Pareto-based aggregation is
defined over attributes and induces a partial order on these vectors.
Intuitively, \emph{for a particular user}, the first partial order objectively
establishes that an object is \emph{better}, i.e., more preferable, than
another, if it is better on all attributes. Then, the second Pareto-based
aggregation, defined across users, induces the second and final partial order on
objects. According to this order, an object is better than another, if it is more
preferable according to \emph{all users}.

Based on the previous interpretation of the GCP formulation, we seek to solve
two distinct problems. The first, which we term the \emph{Group-Maximal
Categorical Objects} (GMCO) problem, is finding the set of maximal, or Pareto-optimal,
objects according to the final partial order. Note that since this order is
only partial, i.e., two objects may not be comparable, there may exist
multiple objects that are maximal; recall, that an object is maximal if there
exists no other object that succeeds it in the order considered. In essence,
it is the fact that this order is partial that guarantees objectiveness. The
GMCO problem has been tackled in our previous work \cite{BBS14}.

The second problem, which we term the \emph{Group-Ranking Categorical Objects}
(GRCO) problem, consists of determining an objective ranking of objects.
Recall that the double Pareto-based aggregation, which is principal in
guaranteeing objectiveness, induces only a partial order on the objects. On
the other hand, ranking implies a total order among objects. Therefore, it is
impossible to rank objects without introducing additional ordering
relationships among objects, which however would sacrifice objectiveness. We
address this contradiction, by introducing an objective \emph{weak order} on
objects. Such an order allows objects to share the same tier, i.e., ranked at
the same position, but defines a total order among tiers, so that among two
tiers, it is always clear which is better.

The GMCO problem has at its core the problem of finding maximal elements
according to some partial order. Therefore, it is possible to adapt an
existing algorithm to solve the core problem, as we discuss in
\autoref{sec:basic_algo}. While there exists a plethora of main-memory
algorithms, e.g., \cite{Kung1975,Bentley1990}, and more recently of external memory
algorithms (termed skyline query processing methods), e.g.,
\cite{BKS01,CGGL03,PTFS05}, they all suffer from two performance limitations.
First, they need to compute the matching degrees and form the matching vectors
for all objects, before actually executing the algorithm. Second, it makes
little sense to apply index-based methods, which are known to be the most
efficient, e.g., the state-of-the-art method of \cite{PTFS05}. The reason is
that the entries of the index depend on the specific instance, and need to be
rebuilt from scratch when the user preferences change, even though the
description of objects persists.

To address these limitations, we introduce a novel index-based approach for
solving GMCO, which also applies to GRCO. The key idea is to index the set of
objects that, unlike the set of matching vectors, remains constant across
instances, and defer expensive computation of matching degrees. To achieve
this, we apply a simple transformation of the categorical attribute values to
intervals, so that each object translates to a rectangle in the Euclidean
space. Then, we can employ a space partitioning index, e.g., an R$^*$-Tree, to
hierarchically group the objects. We emphasize that this transformation and
index construction is a one-time process, whose cost is amortized across
instances, since the index requires no maintenance, as long as the collection
of objects persists. Based on the transformation and the hierarchical
grouping, it is possible to efficiently compute upper bounds for the matching
degrees for \emph{groups} of objects. Therefore, for GMCO, we introduce an
algorithm that uses these bounds to guide the search towards objects that are
more likely to belong to the answer set, avoid computing unnecessary matching
degrees.

For the GRCO problem, i.e., finding a (weak) order among objects, there has
been a plethora of works on the related topic of combining/fusing multiple
ranked lists, e.g., \cite{FS93,DKNS01,AM01,MA02,FV07,MAS11}. However, such
methods are not suitable for our GCP formulation. Instead, we take a different
approach. We first relax the unanimity in the second Pareto-based aggregation,
and require only a percentage $p\%$ of users to agree, resulting in the p-GMCO
problem. This introduces a \emph{pre-order} instead of a partial order, i.e.,
the induced relation lacks antisymmetry (an object may at the same time be
before and after another). Then, building on this notion, we define tiers
based on $p$ values, and rank objects according to the tier they belong, which
results in an objective weak order. To support the effectiveness of our
ranking scheme, we analyze its behaviour in the context of rank aggregation
and show that it posseses several desirable theoretical properties.

 \stitle{Contributions.}
The main contributions of this paper are summarized as follows.
\begin{itemize}

\item
We introduce and propose an objective and fair interpretation of  group categorical preference (GCP) recommender systems, based on double Pareto-based aggregation. 

\item
We introduce three problems in GCP systems, finding the group-maximal objects (GMCO), finding relaxed group-maximal objects ($p$-GMCO), and objectively ranking objects (GRCO).

\item 
We present a method for transforming the 
hierarchical domain of a categorical attribute into a numerical domain. 

\item We propose index-based algorithms for all problems, which employ 
a space partitioning index to hierarchically group objects.

\item
We theoretically study the behaviour of our ranking scheme and present a
number of theoretical properties satisfied by our approach.

\item
We present several extensions involving the following issues: 
multi-values attributes, non-tree hierarchies, subspace indexing, and objective attributes.

\item
We conduct a thorough experimental evaluation using both real and synthetic data.
 
\end{itemize}

\stitle{Outline.}
The remaining of this paper is organized as follows. \autoref{sec:define} contains the necessary definitions for the GCP formulation. Then, \autoref{sec:algo} discusses the
GMCO problem, \autoref{sec:pGMCO} the $p$-GMCO problem, and
\autoref{sec:rank} the GRCO problem. \autoref{sec:ext} discusses various
extensions. 
\autoref{sec:exp} contains a detailed experimental study. 
\autoref{sec:related} reviews related work, while
\autoref{sec:concl} concludes this paper.


%% file: related.tex

\section{Related Work}
\label{sec:related}

This section reviews work on recommender systems and algorithms for
Pareto aggregation.

 \vspace{-10pt}
\subsection{Recommender Systems}

There exist several techniques to specify \textit{preferences} on objects \cite{SKP11,LL87}. The
\textit{quantitative preferences}, e.g., \cite{AW00,HKP01,KI04}, assign a
numeric score to attribute values, signifying importance. For example, values
$a$, $b$, $c$ are assigned scores 0.9, 0.7, 0.1, respectively, which implies
that $a$ is more preferable than $b$, which in turn is more preferable than
$c$. There also exist \emph{qualitative preferences}, e.g., \cite{K02,C03},
which are relatively specified using binary relationships. For example, value
$a$ is preferred over $b$ and $c$, but $b$, $c$ are indifferent. This work
assumes the case of boolean quantitative preferences, where a single attribute
value is preferred, while others are indifferent.

The general goal of \textit{recommendation systems} \cite{AT05,BOHG13,YuHSD14,KannanIP14} 
 is to identify those objects that are most aligned to a
user's preferences. Typically, these systems provide a \emph{ranking} of the objects by
\emph{aggregating} user preferences. Particularly,
the work in \cite{AW00} defines generic functions that merge quantitative preferences. The works in
\cite{CBC+00,HKP01} deal with linear combinations of preference scores and propose index and view
based techniques for ranking tuples. For preferences in general,
\cite{K02,C03} introduce a framework for composing or accumulating interests.
Among the discussed methods is the Pareto composition, which is related to the
skyline computation, discussed below.

Recently, several methods for \textit{group recommendations} are proposed
\cite{JS07,MAS11,CC12,BC11}. These methods, recommend  items to a group of
users, trying to satisfy all the group members.
The existing methods are classified into two approaches. In the first, the
preferences of each group member are combined to create a virtual user; the
recommendations to the group are proposed w.r.t.\ to the virtual user. In the
second, individual recommendations for each member is computed; the
recommendations of all members are merged into a single recommendation.
A large number of group recommendation methods have been developed in several 
domains such as: music \cite{SWT08,CBH02,PiliponyteRK13,McCarthyA98,ChaoBF05,ZXD05}, movies \cite{CCKR01}, 
TV programs \cite{JM04,YZHG06,VKHA09,gb02}, restaurants \cite{PPC08,McCarthy2002}, 
sightseeing tours \cite{GSO11,ArdissonoGPST03,KW05},
vacation packages \cite{McCarthySCMSN06,Jameson04},
food \cite{ElahiGRMB14},
news \cite{PCCA05}, and online communities \cite{GXLBHMS10,KKYY10,BPD08}. 
Finally, several works study the problem of rank aggregation in
the context of group recommendations \cite{RACDY10,BMR10,BF10,CMS07,NSNK12b}.

Several methods to combine different ranked lists are presented in the IR
literature. There the \textit{data fusion} problem is defined. Given a set of
ranked lists of documents returned by different search engines, construct a
single ranked list combining the individual rankings \cite{DKNS01}. Data
fusion techniques can be classified based on whether they require knowledge of
the relevance scores \cite{AM01}. The simplest method based solely on the
documents' ranks is the Borda-fuse model. It assigns as score to each document
the summation of its rank in each list. The Condorcet-fuse method \cite{MA02}
is based on a majoritarian voting algorithm, which specifies that a document
$d_1$ is ranked higher in the fused list than another document $d_2$ if $d_1$
is ranked higher than $d_2$ more times than $d_2$ is ranked higher than $d_1$.
The approach in \cite{FV07}, assumes that a document is ranked better than
another if the majority of input rankings is in concordance with this fact and
at the same time only a few input rankings refute it. When the relevance
scores are available, other fusion techniques, including CombSUM, CombANZ and
CombMNZ, can be applied \cite{FS93}. In CombSUM, the fused relevance score of
a document is the summation of the scores assigned by each source. In CombANZ
(resp. CombMNZ), the final score of a document is calculated as that of
CombSUM divided (resp. multiplied) by the number of lists in which the
document appears.

\subsection{Algorithms for Pareto Aggregation}

The work of \cite{BKS01} rekindled interest in the problem of finding the
maximal objects \cite{Kung1975} and re-introduces it as the skyline operator. An object is
dominated if there exists another object before it according to the partial
order enforced by the Pareto-based aggregation. The maximal objects are
referred to as the skyline. The authors propose several external memory
algorithms.
The most well-known method is Block Nested Loops (BNL) \cite{BKS01}, 
which checks each point for dominance against the entire dataset.

The work in \cite{CGGL03} observes that examining
points according to a monotone (in all attributes) preference function reduces the average number of
dominance checks. Based on this fact, the Sort-first Skyline algorithm (SFS) is introduces;
including some variations (i.e., LESS \cite{GSG07}, and SaLSa \cite{BCP08}) 
belonging in the class of sort-based skyline algorithms,  that improve performance (see \cite{BikakisSS14} for more details).

In \cite{SLNX09} the multi-pass randomize algorithm RAND is proposed.
Initially, RAND selects a random sample; then, multiple passes over the dataset 
is performed in order to prune points and find the skyline.

In other approaches, multidimensional indexes are
used to guide the search for skyline points and prune large parts of the space. The most well-known
algorithms is the Branch and Bound Skyline (BBS) method \cite{PTFS05}, which uses an R-tree, and is
shown to be I/O optimal with respect to this index. 
Similarly, the Nearest Neighbor algorithm (NN)
\cite{KRR02} also uses an R-tree performing multiple nearest neighbor searches
to identify skyline objects. A bitmap structure is used by Bitmap \cite{TEO01}
algorithm to encode the input data. In the Index \cite{TEO01} algorithm,
several B-trees are used to index the data, one per dimension. Other methods,
e.g., \cite{LZLL07,LC10}, employ a space-filling curve, such as the
\mbox{Z-order} curve, and use a single-dimensional index. The Lattice Skyline
(LS) algorithm \cite{MPJ07} builds a specialized data structure for low-cardinality 
domains.

In {partitioning-based} approaches, the algorithms divide the initial
space into  several partitions. The first algorithm in this category, D\&C
\cite{BKS01} computes the skyline objects adopting the divide-and-conquer
paradigm.  A similar approach with stronger theoretical guarantees is presented
in \cite{ST12}. 
Recently, partitioning-based skyline algorithms are proposed in \cite{ZMC09,JS10}.
OSP \cite{ZMC09} attempts to reduce the number of checks 
between incomparable points by recursively partition the skyline points.
BSkyTree \cite{JS10}  enhances \cite{ZMC09} by considering both 
the notions of dominance and incomparability while partitioning the space.

Finally, specific algorithms are proposed to efficiently compute the skyline over partially ordered domains
\cite{CET05,WFP+08,SPP09,ZMKC10}, metric spaces \cite{CL09}, non-metric spaces \cite{PDMK09}, or anticorrelated distributions \cite{SK13}.

Several lines of research attempt to address the issue that the size of skyline cannot be
controlled, by introducing new concepts and/or ranking the skyline (see \cite{LB13} for a survey). 
\cite{YM07} ranks tuples based
on the number of records they dominate. \cite{CJT+06} deals with high-dimensional skylines, and
relaxes the notion of dominance to $k$-dominance, according to which a record is $k$-dominated if it
is dominated in a subspace of $k$ dimensions. \cite{LJZ11} uses a skyline-based partitioning to rank
tuples. The $k$ most representative skyline operator is proposed in \cite{LYZZ07}, which selects a
set of $k$ skyline points, so that the number of points dominated by at least one of them is
maximized. In a similar spirit, \cite{TDLP09} tries to select the $k$ skyline points that best
capture the trade-offs among the parameters.
Finally,  \cite{LYHSB12}   attempts to find a small and focused skyline set.
The size of the skyline is reduced by asking from users to state additional preferences.

%% file: define.tex

\section{Group Categorical Preferences}
\label{sec:define}

\autoref{tab:notation} shows the most important symbols and their definition.
Consider a set of $d$ categorical \emph{attributes} $\A = \{A_1, \dots,$ $ A_d \}$. The domain of
each attribute $A_k$ is a \emph{hierarchy} $\H(A_k)$. A hierarchy $\H(A_k)$ defines a tree, where a
leaf corresponds to a lowest-level value, and an internal node corresponds to a category, i.e.,
a set, comprising all values within the subtree rooted at this node. The root of a hierarchy
represents the category covering all lowest-level values. We use the symbol $|A_k|$ (resp.
$|\H(A_k)|$) to denote the number of leaf (resp. all hierarchy) nodes. With reference to
\autoref{fig:hierarchies}, consider the  ``Cuisine'' attribute. The node ``Eastern'' is a category
and is essentially a shorthand for the set $\{$``Greek'', ``Austrian''$\}$, since it contains the
two leaves, ``Greek'' and ``Austrian''.

Assume a set of \emph{objects} $\O$. An object $o_i \in \O$ is defined over \emph{all} attributes,
and the value of attribute $o_i.A_k$ is one of the nodes of the hierarchy $\H(A_k)$. For instance,
in \autoref{tab:rest}, the value of the ``Cuisine'' attribute of object $o_1$, 
is the node ``Eastern'' in the hierarchy of \autoref{fig:hierarchies}.

\begin{table}[t]
\caption{Notation}
\label{tab:notation}
\begin{tabular}{cl}
\tline
\textbf{Symbol} & \textbf{Description}\\ \dline
$\A$, $d$ & Set of attributes, number of attributes ($|\A|$)\\
$A_k$, $|A_k|$ & Attribute, number of distinct values in $A_k$\\
$\H(A_k)$, $|\H(A_k)|$ & Hierarchy of $A_k$, number of hierarchy nodes\\
$\O$, $o_i$ & Set of objects, an object\\
$\U$, $u_j$ & Set of users, a user\\
$o_i.A_k$, $u_j.A_k$ & Value of attribute $A_k$ in object $o_i$, user $u_j$\\
$o_i.I_k$, $u_j.I_k$ & Interval representation of the value of $A_k$ in $o_i$, $u_j$\\
$m_i^j$ & Matching vector of object $o_i$ to user $u_j$\\
$m_i^j.A_k$ & Matching degree of $o_i$ to user $u_j$ on attribute $A_k$\\
$o_a \succ o_b$ & Object $o_a$ is collectively preferred over $o_b$\\
$\T$ & The R$^*$-Tree that indexes the set of objects\\
$N_i$, $e_i$ & R$^*$-Tree node, the entry for $N_i$ in its parent node\\
$e_i.ptr$, $e_i.mbr$ & The pointer to node $N_i$, the MBR of $N_i$\\
$M_i^j$ & Maximum matching vector of entry $e_i$ to user $u_j$\\
$M_i^j.A_k$ & Maximum matching degree of $e_i$ to user $u_j$ on $A_k$\\
\bline
\end{tabular}
 \end{table}

Further, assume a set of \emph{users} $\U$. A user $u_i \in \U$ is defined
over a \emph{subset} of the attributes, and for each \emph{specified}
attribute $u_i.A_j$, its value in one of the hierarchy $\H(A_j)$ nodes. For
all unspecified attributes, we say that user $u_i$ is \emph{indifferent} to
them.  Note that, an object (resp.\ a user) may has (resp.\ specify) multiple values for each attribute (see \autoref{sec:multi}).

Given an object $o_i$, a user $u_j$, and a specified attribute $A_k$, 
the \emph{matching degree} of $o_i$ to $u_j$ with respect to $A_k$, denoted as $m_i^j.A_k$, is specified by a \textit{matching function}  
$\textbf{M} \colon dom(A_k) \times dom(A_k) \to [0,1]$. 
The matching function defines the \textit{relation} between the user's preferences and the objects attribute values.
For an indifferent attribute $A_k$ of a user $u_j$, we define $m_i^j.A_k= 1$.

Note that, different matching functions can be defined per attribute and
user; for ease of presentation, we assume a single matching function. Moreover, note that this
function can be \emph{any user defined function} operating on the
cardinalities of intersections and unions of hierarchy attributes. For
example, it can be the Jaccard coefficient, i.e.,
$m_i^j.A_k = \frac{|o_i.A_k \cap u_j.A_k|}{|o_i.A_k \cup u_j.A_k|}$.
The numerator counts the number of leaves
in the intersection, while the denominator counts the number of leaves in the union, of the
categories $o_i.A_k$ and $u_j.A_k$. Other popular choices are 
the Overlap coefficient:  $\frac{|o_i.A_k \cap u_j.A_k|}{ \min{(|o_i.A_k|, |u_j.A_k|)}}$, and
the Dice coefficient: $2\frac{|o_i.A_k \cap u_j.A_k|}{|o_i.A_k| +|u_j.A_k|}$.

In our running example, we assume the Jaccard coefficient. Hence, the matching degree of
restaurant $o_1$ to user $u_1$ w.r.t. ``Attire'' 
is $\frac{|\text{``Business casual''} \cap \text{``Casual''}|}{|\text{``Business casual''} \cup \text{``Casual''}|} = $  \linebreak $\frac{|\{\text{``Business casual''}\}|}{|\{\text{``Business casual''}, \text{``Smart casual''}|} =
\frac{1}{2}$, where we substituted ``Casual'' with the set 
$\{$``Business casual'', ``Smart casual''$\}$.

\begin{table}[]
\caption{Matching vectors}
\label{tab:match}
\begin{tabular}{cccc}
\tline
 &\multicolumn{3}{c}{\textbf{User}}\\ 
\cmidrule{2-4}
\specialcell[t]{\textbf{Restaurant}} & $u_1$ & $u_2$ & $u_3$\\ \dline
$o_1$ &  $\mat{1/2, 1/2, 1/6, 1, 1}$ & $\mat{0, 1, 1, 1, 0}$ & $\mat{0,1,0,1,1}$\\
$o_2$ & $\mat{1/4, 0, 0, 0, 0}$ & $\mat{1, 1,1, 1, 1}$ & $\mat{1/2,1,1,1,1}$\\
$o_3$  & $\mat{0, 1/2, 0, 0, 0}$ & $\mat{0, 1, 1, 1, 0}$ & $\mat{0,1,0,1,1}$\\
$o_4$  & $\mat{0, 0, 0, 0 ,0}$ & $\mat{0, 1, 1 ,1, 0}$ & $\mat{0,1,0,1,1}$\\
\bline
\end{tabular}
 \end{table}

Given an object $o_i$ and a user $u_j$, the \emph{matching vector} of $o_i$ to $u_j$, denoted as
$m_i^j$, is a $d$-dimensional point in $[0,1]^d$, where its $k$-th coordinate is the matching degree
with respect to attribute $A_k$. Furthermore, we define the norm of the matching vector to be
$\|m_i^j\| = \sum_{A_k \in \A} m_i^j.A_k$. In our example, the matching vector of restaurant $o_1$ to user
$u_1$ is $\langle 1/2, 1/2, 1/6, 1, 1\rangle$. All matching vectors of this example are shown in
\autoref{tab:match}.

%% file: algo.tex

\section{The Group-Maximal Categorical Objects (GMCO) Problem}
\label{sec:algo}

\autoref{sec:gmco_problem} introduces the GMCO problem, and
\autoref{sec:basic_algo} describes a straightforward baseline approach. Then,
\autoref{sec:transform} explains a method to convert categorical values into
intervals, and \autoref{sec:ics} introduces our proposed index-based solution.

\vspace{-5mm}

\subsection{Problem Definition}
\label{sec:gmco_problem}

We first consider a particular user $u_j$ and examine the matching vectors.
The \emph{first Pareto-based aggregation} across the attributes of the
matching vectors, induces the following partial and strict partial
``preferred'' orders on objects. An object $o_a$ is \emph{preferred} over
$o_b$, for user $u_j$, denoted as $o_a \succeq^j o_b$ iff for every specified
attribute $A_k$ of the user it holds that $m_a^j.A_k \geq m_b^j.A_k$.
Moreover, object $o_a$ is \emph{strictly preferred} over $o_b$, for user
$u_j$, denoted as $o_a \succ^j o_b$ iff $o_a$ is preferred over $o_b$ and
additionally there exists a specified attribute $A_k$ such that ${m_a^j.A_k >
m_b^j.A_k}$. Returning to our example, consider user $u_1$ and its matching
vector $\langle 0, 1/2, 0, 0, 0 \rangle$ for $o_3$, and $\langle 0, 0, 0, 0,
0\rangle$ for $o_4$. Observe that $o_3$ is strictly preferred over $o_4$.

We now consider all users in $\U$. The \emph{second Pareto-based aggregation}
across users, induces the following strict partial ``collectively preferred''
order on objects. An object $o_a$ is \emph{collectively preferred} over $o_b$,
if $o_a$ is preferred over $o_b$ for all users, and there exists a user $u_j$
for which $o_a$ is strictly preferred over $o_b$. 
From \autoref{tab:match},
it is easy to see that restaurant $o_1$ is collectively preferred over $o_3$, 
because $o_1$ is preferred by all three users, and strictly preferred
by user $u_1$.

Given the two Pareto-based aggregations, we define the \emph{collectively
maximal objects} in $\O$ with respect to users $\U$, as the set of objects for
which there exists no other object that is collectively preferred over them.
In our running example, $o_1$ and $o_2$ objects are both collectively
preferred over $o_3$ and $o_4$. There exists no object which is collectively
preferred over $o_1$ and $o_2$, and thus are the collectively maximal
objects. We next formally define the GMCO problem.

\begin{myprob}[GMCO]
\label{prob:GMCO}
Given a set of objects $\O$ and a set of users $\U$ defined over a set of
categorical attributes $\A$, the \emph{Group-Maximal Categorical Objects}
(GMCO) problem is to find the collectively maximal objects of $\O$ with
respect to $\U$.
\end{myprob}

\vspace{-3mm}
\subsection{A Baseline Algorithm (BSL)}
\label{sec:basic_algo}

The GMCO problem can be transformed to a maximal elements problem, or a
skyline query, where the input elements are the matching vectors. Note,
however, that the GMCO problem is different than computing the conventional
skyline, i.e., over the object's attribute values.

The \emph{Baseline} (BSL) method, whose pseudocode is depicted in
\autoref{algo:bcs}, takes advantage of this observation. The basic idea
of BSL is for each object $o_i$ (\textit{loop in line 1}) and for all users (\textit{loop in line 2}),
to compute the matching vectors $m_i^j$ (\textit{line 3}). 
Subsequently, BSL constructs a \mbox{$|\U|$-dimensional} 
tuple $r_i$ (\textit{line 4}), so that its $j$-th entry
is a composite value equal to the matching vector $m_i^j$ of object $o_i$ to
user $u_j$. When all users are examined, tuple $r_i$ is inserted in the set
$\R$ (\textit{line 5}).

The next step is to find the maximal elements, i.e., compute the skyline over
the records in $\R$. It is easy to prove that tuple $r_i$ is in the skyline of
$\R$ iff object $o_i$ is a collectively maximally preferred object of $\O$ w.r.t. $\U$.
Notice, however, that due to the two Pareto-based aggregations, each attribute
of a record $r_i \in \R$ is also a record that corresponds to a matching
vector, and thus is partially ordered according to the preferred orders
defined in \autoref{sec:define}. Therefore, in order to compute the
skyline of $\R$, we need to apply a skyline algorithm (\textit{line 6}), such as
\cite{BKS01,PTFS05,GSG07}.

\stitle{Computational Complexity.} 
The computational cost of BSL is the sum of two parts. The first is computing
the matching degrees, which takes $O(|\O| \cdot |\U|)$ time. The second is
computing the skyline, which requires ${O(|\O|^2 \cdot |\U| \cdot d)}$
comparisons, assuming a quadratic time skyline algorithms is used. Therefore,
BSL takes $O(|\O|^2 \cdot |\U| \cdot d)$ time.

\begin{algorithm}[t]
\SetCommentSty{textsf}
\DontPrintSemicolon
\KwIn{objects $\O$, users $\U$}
\KwOut{$CM$ the collectively maximal}
\SetKwInput{KwVar}{Variables}
\KwVar{$\R$ set of intermediate records}
\SetKwFunction{sky}{SkylineAlgo}
\vspace{1mm}
\ForEach{$o_i \in \O$}{
	\ForEach{$u_j \in \U$} {
		compute $m_i^j$\;
		$r_i[j] \gets m_i^j$\;
	}
	insert $r_i$ into $\R$\;
}
$CM \gets$ \sky($\R$)\;
\caption{BSL}
\label{algo:bcs}
\end{algorithm}

\subsection{Hierarchy Transformation}
\label{sec:transform}

This section presents a simple method to transform the hierarchical domain of a categorical
attribute into a numerical domain. The rationale is that numerical domains can be ordered, and thus
tuples can be stored in multidimensional index structures. The index-based algorithm of
\autoref{sec:ics} takes advantage of this transformation.

Consider an attribute $A$ and its hierarchy $\H(A)$, which forms a tree. We assume that any internal
node has at least two children; if a node has only one child, then this node and its child are
treated as a single node. Furthermore, we assume that there exists an ordering, e.g., the
lexicographic, among the children of any node that totally orders all
leaf nodes.

The hierarchy transformation assigns an interval to each node, similar to labeling schemes such as
\cite{ABJ89}. The $i$-th leaf of the hierarchy (according to the ordering) is assigned the interval
$[i-1, i)$. Then, each internal node is assigned the smallest interval that covers the intervals of
its children. \autoref{fig:hierarchies} depicts the assigned intervals for all nodes in the two
car hierarchies.

Following this transformation, the value on the $A_k$ attribute of an object $o_i$ becomes an interval
$o_i.I_k = [o_i.I_k^-, o_i.I_k^+)$. The same holds for a user $u_j$. Therefore, the transformation
translates the hierarchy $H(A_k)$ into the numerical domain $[0, |A_k|]$.

An important property of the transformation is that it becomes easy to compute
matching degrees for metrics that are functions on the cardinalities of
intersections or unions of hierarchy attributes. This is due to the following
properties, which use the following notation: for a closed-open interval
$I=[\alpha,\beta)$, define $\| I \| = \beta-\alpha$. \alt{}{Note that the
proofs can be found in the full version of the paper \cite{BBS13}.}

\begin{myproposition}
\label{propos:intervals}
For objects/users $x$, $y$, and an attribute $A_k$, let $x.I_k$, $y.I_k$ denote the
intervals associated with the value of $x$, $y$ on $A_k$. Then the following hold:
\begin{enumerate}[(1)]
\setlength{\itemindent}{15pt}

\item $|x.A_k| = \|x.I_k\|$

\item $|x.A_k \cap y.A_k| = \| x.I_k \cap y.I_k \| $

\item $|x.A_k \cup y.A_k| = \|x.I_k\| + \|y.I_k\| - \| x.I_k \cap y.I_k \| $

\end{enumerate}

\end{myproposition}

 \begin{myproof}
For a leaf value $x.A_k$, it holds that $|x.A_k|=1$. By construction of the
transformation, $\|x.I_k\| = 1$. For a non-leaf value $x.A_k$, $|x.A_k|$ is
equal to the number of leaves under $x.A_k$. Again, by construction of the
transformation, $\|x.I_k\|$ is equal to the smallest interval that covers the
intervals of the leaves under $x.A_k$, and hence equal to $|x.A_k|$. Therefore
for any hierarchy value, it holds that $x.A_k = \|x.I_k\|$.

Then, the last two properties trivially follow. The third holds since \linebreak ${|x.A_k \cup y.A_k| =}$ $|x.A_k| + |y.A_k| - |x.A_k \cap y.A_k|$. \qed
\end{myproof}

\vspace{-3mm}
\subsection{An Index-based Algorithm (IND)}
\label{sec:ics}

This section introduces the \emph{Index-based} GMCO (IND) algorithm. The key ideas of IND
are: (1) apply the hierarchy transformation, previously described, and index the resulting
intervals, and (2) define upper bounds for the matching degrees of a group of objects, so as to
guide the search and quickly prune unpromising objects.

We assume that the set of objects $\O$ and the set of users $\U$ are transformed so that each
attribute $A_k$ value is an interval $I_k$. Therefore, each object (and user) defines a \mbox{(hyper-)rectangle} on the $d$-dimensional cartesian product of the numerical domains, i.e., $[0, |A_1|)
\times \dots \times [0, |A_d|)$.

\begin{figure}
\includegraphics[scale=0.27]{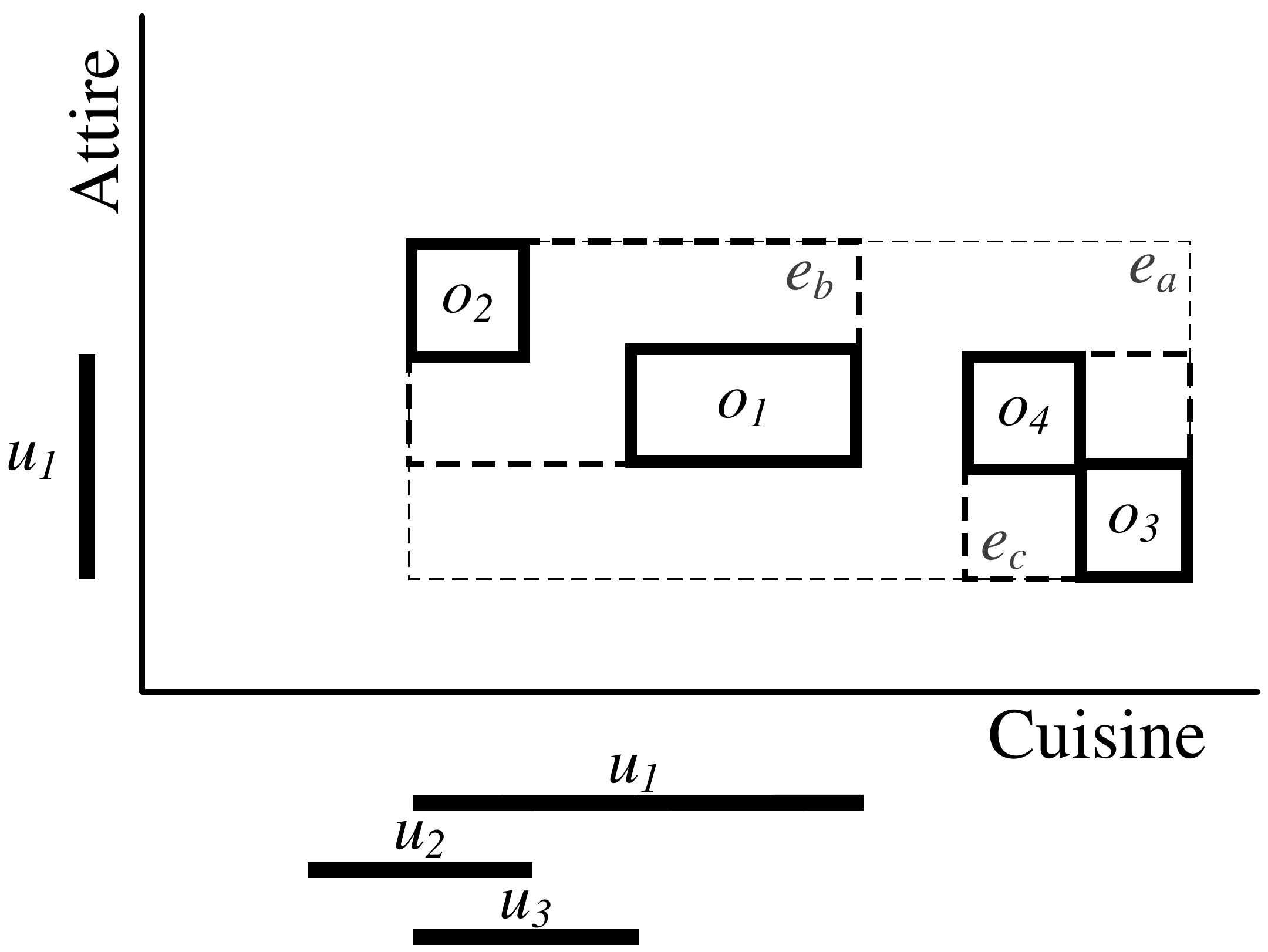}
\caption{Transformed objects and users}
\label{fig:trans}
\end{figure}

\autoref{fig:trans} depicts the transformation of the objects and users shown in
Tables~\ref{tab:rest}~\&~\ref{tab:users}, considering only the attributes Cuisine and Attire. For
instance, object $o_1$ is represented as the rectangle $[4,6) \times [2,3)$ in the
``Cuisine''$\times$``Attire'' plane. Similarly, user $u_1$ is represented as two intervals, $[2,6)$,
$[1,3)$, on the transformed ``Cuisine'', ``Attire'' axes, respectively.

The IND algorithm indexes the set of objects in this $d$-dimensional space. In particular,
IND employs an R$^*$-Tree $\T$ \cite{BKSS90}, which is well suited to index rectangles. Each $\T$
node corresponds to a disk page, and contains a number of entries. Each entry $e_i$ comprises (1) a
pointer $e_i.ptr$, and (2) a Minimum Bounding Rectangle (MBR) $e_i.mbr$. A leaf entry $e_i$
corresponds to an object $o_i$, its pointer $o_i.ptr$ is \textit{null}, and $e_i.mbr$ is the rectangle
defined by the intervals of $o_i$. A non-leaf entry $e_i$ corresponds to a child node $N_i$, its
pointer $e_i.ptr$ contains the address of $N_i$, and $e_i.mbr$ is the MBR of (i.e., the tightest
rectangle that encloses) the MBRs of the entries in $N_i$.

Due to its enclosing property, the MBR of an entry $e_i$ encloses all
objects that are stored at the leaf nodes within the $\T$ subtree rooted at node $N_i$. It is often
helpful to associate an entry $e_i$ with all the objects it encloses, and thus treat $e_i$ as a
group of objects.

Consider a $\T$ entry $e_i$ and a user $u_j \in \U$. Given only the information within entry $e_i$,
i.e., its MBR, and not the contents, i.e., its enclosing objects, at the subtree rooted at $N_i$, it
is impossible to compute the matching vectors for the objects within this subtree. However, it is
possible to derive an \emph{upper bound} for the matching degrees of any of these objects.

We define the \emph{maximum matching degree} $M_i^j.A_k$ of entry $e_i$ on
user $u_j$ w.r.t. specified attribute $A_k$ as the highest attainable matching
degree of any object that may reside within $e_i.mbr$. To do this we first
need a way to compute lower and upper bounds on unions and intersections of a
user interval with an MBR.

\begin{myproposition}
\label{lem:mmd}
Fix an attribute $A_k$. Consider an object/user $x$, and let $I_x$,
denote the interval associated with its value on $A_k$.
Also, consider another object/user $y$ whose interval $I_y$ on
$A_k$ is contained within a range $R_y$. 
Given an interval $I$, $\delta(I)$ returns $0$ if $I$ is empty, and $1$ otherwise.
Then the following hold:

\begin{enumerate}[(1)]
\setlength{\itemindent}{15pt}
\vspace{-5pt}
\item $1 \ \leq |y.A_k| \leq \|R_y\| $

\item $\delta(I_x \cap R_y) \ \leq |x.A_k \cap y.A_k| \leq \| I_x \cap R_y \| $

\item $\|I_x\| + 1 - \delta(I_x \cap R_y) \ \leq |x.A_k \cup y.A_k| \leq \|I_x\| + \|R_y\| - \delta(I_x \cap R_y) $
\end{enumerate}
\end{myproposition}

 \begin{myproof}
Note that for the object/user $y$ with interval $I_y$ on $A_k$, it holds that
$I_y \subseteq R_y$.

(1) 
For the left inequality of the first property, observe that value $y.A_k$ is a node that contains at
least one leaf, hence $1 \leq |y.A_k|$. Furthermore, for the right inequality, $|y.A_k| =
\| I_y\| \leq \| R_y \|$.

(2) For the left inequality of the second property, observe that the value ${x.A_k \cap y.A_k}$ contains either at least one leaf when the intersection is not
empty, and no leaf otherwise. The right inequality follows from the fact that \linebreak
${I_x \cap I_y \subseteq I_x \cap R_y}$.

(3)
For the left inequality of the third property, assume first that $I_x \cap I_y
= \varnothing$; hence $\delta(I_x \cap R_y) = 0$. In this case, it holds that
$\|I_x \cup I_y \| = \|I_x\| + \|I_y \|$. By the first property, we obtain $1
\leq \|I_y\|$. Combining the three relations, we obtain the left inequality.
Now, assume that $I_x \cap I_y \neq \varnothing$; hence $\delta(I_x \cap R_y)
= 1$. In this case, it also holds that $\|I_x\| \leq \|I_x \cup I_y \|$, and
the left inequality follows.

For the right inequality of the third property, observe that \linebreak$\|I_x \cup I_y \| = $  ${\|I_x\| + \|I_y\| - \|I_x \cap I_y\|}$. By the first property, we obtain \linebreak
$\|I_y\| \leq \|R_y\| $, and ${-\|I_x \cap I_y\|}$ $\leq -\delta(I_x \cap R_y) $,
by the second. The right inequality follows from combining these three
relations.  \qed
 \end{myproof}

Then, defining the maximum matching degree reduces to appropriately selecting
the lower/upper bounds for the specific matching function used. For example,
consider the case of the Jaccard coefficient, $\frac{|o_i.A_k \cap
u_j.A_k|}{|o_i.A_k \cup u_j.A_k|}$. Assume $e_i$ is a non-leaf entry, and let
$e_i.R_k$ denote the range of the MBR on the $A_k$ attribute. We also assume
that $u_j.I_k$ and $e_i.R_k$ overlap. Then, we define $M_i^j.A_k =
\frac{\|e_i.R_k \cap u_j.I_k\|}{\|u_j.I_k\|}$, where we have used the upper
bound for the intersection in the enumerator and the lower bound for the union
in the denominator, according to \autoref{lem:mmd}. For an indifferent to
the user attribute $A_k$, we define $M_i^j.A_k= 1$. Now, assume that $e_i$ is
a leaf entry, that corresponds to object $o_i$. Then the maximum matching
degree $M_i^j.A_k$ is equal to the matching degree $m_i^j.A_k$ of $o_i$ to
$u_j$ w.r.t. $A_k$.

Computing maximum matching degrees for other metrics is straightforward. In
any case, the next lemma shows that an appropriately defined maximum matching
degree is an upper bound to the matching degrees of all objects enclosed in
entry $e_i$.

\begin{myproposition} \label{lem:max}
The maximum matching degree $M_i^j.A_k$ of entry $e_i$ on user $u_j$ w.r.t. specified attribute
$A_k$ is an upper bound to the highest matching degree in the group that
$e_i$ defines.
\end{myproposition}

 \begin{myproof}
The maximum matching degree is an upper bound from \autoref{lem:mmd}. \qed
\end{myproof}

In analogy to the matching vector, the \emph{maximum matching vector} $M_i^j$ of entry $e_i$ on user
$u_j$ is defined as a $d$-dimensional vector whose $k$-th coordinate is the maximum matching degree
$M_i^j.A_k$. Moreover, the norm of the maximum matching vector is $\|M_i^j\| = \sum_{A_k \in \A}
M_i^j.A_k$.

Next, consider a $\T$ entry $e_i$ and the entire set of users $\U$. We define the \emph{score} of an
entry $e_i$ as $score(e_i) = \sum_{u_j \in \U} \| M_i^j \|$. This score quantifies how well the
enclosed objects of $e_i$ match against all users' preferences. Clearly, the higher the score, the
more likely that $e_i$ contains objects that are good matches to users.

\begin{algorithm}[]
\SetKw{Break}{break}
\SetCommentSty{textsf}
\DontPrintSemicolon
\KwIn{R$^*$-Tree $\T$, users $\U$}
\KwOut{$CM$ the collectively maximal}
\SetKwInput{KwVar}{Variables}
\KwVar{$H$ a heap with $\T$ entries sorted by $score()$}
\vspace{1mm}
$CM \gets \varnothing$\;
read $\T$ root node\;
insert in $H$ the root entries\;
\While{$H$ is not empty}{
	$e_x \gets$ pop $H$\;
	\eIf{$e_x$ is non-leaf}{
		$N_x \gets$ read node $e_x.ptr$\;
		\ForEach{$e_i \in N_x$}{
			$pruned \gets false$\;
			\ForEach{$u_j \in \U$} {
				compute $M_i^j$\;
			}
			\ForEach{$o_a \in CM$} {
				\If{$\forall A_j\!: m_a^j \!\succeq\! M_i^j \land \exists A_k\!: m_a^k \!\succ\! M_i^k$}{
					$pruned \gets true$\;
					\Break\;
				}
			}
			\If{not $pruned$} {
				insert $e_i$ in $H$ \;
			}
		}
	}{
		$o_x \gets e_x$ \;
		$result \gets true$\;
		\ForEach{$o_a \in CM$} {
			\If{$o_a \succ o_x$} {
				$result \gets false$\;
				\Break\;
			}
		}
		\If{$result$} {
			insert $o_x$ in $CM$ \;
		}
	}
}
\caption{IND}
\label{algo:ics}
\end{algorithm}
 

\stitle{Algorithm Description.}
\autoref{algo:ics} presents the pseudocode for IND. The algorithm maintains two data
structures: a heap $H$ which stores $\T$ entries sorted by their score, and a list $CM$ of
collectively maximal objects discovered so far. Initially the list $CM$ is empty (\textit{line 1}), and the
root node of the R$^*$-Tree is read (\textit{line 2}). The score of each root entry is computed and all
entries are inserted in $H$ (\textit{line 3}). Then, the following process (\textit{loop in line 4}) is repeated as
long as $H$ has entries.

The $H$ entry with the highest score, say $e_x$, is popped (\textit{line 5}). If $e_x$ is a non-leaf entry
(\textit{line 6}), it is \emph{expanded}, which means that the node $N_x$ identified by $e_x.ptr$ is read
(\textit{line 7}). For each child entry $e_i$ of $N_x$ (\textit{line 8}), its maximum matching degree $M_i^j$ with
respect to every user $u_j \in \U$ is computed (\textit{lines 10--11}). Then, the list $CM$ is scanned (\textit{loop
in line 12}). If there exists an object $o_a$ in $CM$ such that (1) for each user $u_j$, the matching
vector $m_a^j$ of $o_a$ is better than $M_i^j$, and (2) there exists a user $u_k$ so that the
matching vector $m_a^k$ of $o_a$ is strictly better than $M_i^k$, then entry $e_i$ is discarded
(\textit{lines 13--15}). It is straightforward to see (from \autoref{lem:max}) that if this condition
holds, $e_i$ cannot contain any object that is in the collectively maximal objects, which guarantees IND'
correctness. When the condition described does not hold (\textit{line 16}), the score of $e_i$ is computed
and $e_i$ is inserted in $H$ (\textit{line 17}).

Now, consider the case that $e_x$ is a leaf entry (\textit{line 18}), corresponding to object $o_x$ (\textit{line
19}). The list $CM$ is scanned (\textit{loop in line 21}). If there exists an object that is collectively preferred over
$o_x$ (\textit{line 22}), it is discarded. Otherwise (\textit{line 25--26}), $o_x$ is inserted in $CM$.

The algorithm terminates when $H$ is empty (\textit{loop in line 4}), at which time the list $CM$ contains
the collectively maximal objects.

\stitle{Computational Analysis.}
IND performs object to object comparisons as well as object to non-leaf
entries. Since there are at most $|\O|$ non-leaf entries, IND performs
$O(|\O|^2 \cdot |\U| \cdot d)$ comparisons in the worst case. Further it
computes matching degrees on the fly at a cost of $O(|\O| \cdot |\U|)$.
Overall, IND takes ${O(|\O|^2 \cdot |\U| \cdot d)}$ time, the same as BSL.
However, in practice IND is more than an order of magnitude faster than BSL
(see \autoref{sec:exp}).
 

\stitle{Example.}
We demonstrate IND, using our running example, as depicted in \autoref{fig:trans}. The four
objects are indexed by an R$^*$-Tree, whose nodes are drawn as dashed rectangles. Objects $o_1$,
$o_2$ are grouped in entry $e_b$, while $o_3$, $o_4$ in entry $e_c$. Entries $e_b$ and $e_c$ are the
entries of the root $e_a$. Initially, the heap contains the two root entries, $H = \{ e_b, e_c\}$.
Entry $e_b$ has the highest score (the norm of its maximum matching vector is the largest), and is
thus popped. The two child entries $o_1$ and $o_2$ are obtained. Since the list $CM$ is empty, no
child entry is pruned and both are inserted in the heap, which becomes $H = \{ o_1, o_2, e_c\}$. In
the next iteration, $o_2$ has the highest score and is popped. Since this is a leaf entry, i.e., an
object, and $CM$ is empty, $o_2$ is inserted in the result list, $CM = \{ o_2\}$.
 Subsequently, $o_1$ is popped and since $o_2$ is not collectively preferred over it, 
 $o_1$ is also placed in the result list, $CM = \{ o_2, o_1\}$.
In the final iteration, entry $e_c$ is popped, but the
objects in $CM$ are collectively preferred over both $e_c$ child.
Algorithm IND concludes, finding the collectively maximal 
$CM = \{ o_2, o_1\}$.

%% file: rank.tex

\section{The {\large\textit{{p}}}-Group-Maximal Categorical Objects ({\large\textit{{p}}}-GMCO) Problem}
\label{sec:pGMCO}

\autoref{sec:pgmco_pro} introduces the $p$-GMCO problem, and
\autoref{sec:pbsl} presents an adaptation of the BSL method, while
\autoref{sec:pind} introduces an index-based approach.

\subsection{Problem Definition}
\label{sec:pgmco_pro}

As the number of users increases, it becomes more likely that the users
express very different and conflicting preferences. Hence, it becomes
difficult to find a pair of objects such that the users unanimously agree that
one is worst than the other. Ultimately, the number of maximally preferred
objects increases. This means that the answer to an GMCO problem with a large
set of users becomes less meaningful.

The root cause of this problem is that we require unanimity in deciding
whether an object is collectively preferred by the set of users. The following
definition relaxes this requirement. An object $o_a$ is \emph{$p$-collectively
preferred} over $o_b$, denoted as $o_a \succ_p o_b$, iff there exist a subset
$\U_p \subseteq \U$ of at least $\lceil\frac{p}{100}\cdot|\U|\rceil$ users such that for each user
$u_i \in \U_p$ $o_a$ is preferred over $o_b$, and there exists a user $u_j \in
\U_p$ for which $o_a$ is strictly preferred over $o_b$. In other words, we
require only $p\%$ of the users votes to decide whether an object is
universally preferred. Similarly, the \emph{$p$-collectively maximal objects}
of $\O$ with respect to users $\U$, is defined as the set of objects in $\O$
for which there exists no other object that is $p$-collectively preferred over
them. The above definitions give rise to the $p$-GMCO problem.

\begin{myprob}[$p$-GMCO]
\label{prob:pGMCO}
Given a set of objects $\O$ and a set of users $\U$ defined over a set of
categorical attributes $\A$, the \emph{$p$-Group-Maximal Categorical Objects}
($p$-GMCO) problem is to find the $p$-collectively maximal objects of $\O$
with respect to $\U$.
\end{myprob}
\vspace{3pt}

Following the definitions, we can make a number of important observations,
similar to those in   the $k$-dominance notion \cite{CJT+06}. First, if an
object is collectively preferred over some other object, it is also
$p$-collectively preferred over that same object for any $p$. As a result, an
object that is $p$-collectively maximal is also collectively maximal for any
$p$. In other words, the answer to the $p$-GMCO problem is a
\emph{subset} of the answer to the corresponding GMCO.

Second, consider an object $o$ that is not $p$-collectively maximal. Note that
it is possible that no $p$-collectively maximal object is $p$-collectively
preferred over $o$. As a result checking if $o$ is a result by considering
only the $p$-collectively maximal objects may lead to false positives.
Fortunately, it holds that there must exist a collectively maximal object that
is $p$-collectively preferred over $o$. So it suffices to check $o$ against
the collectively maximal objects only (and not just the subset that is
$p$-collectively maximal).

\stitle{Example.} Consider the example in
Tables~\ref{tab:rest}~\&~\ref{tab:users}. If we consider $p=100$, we require
all users to agree if an object is collectively
preferred. So, the $100$-collectively maximal objects are the same as the
collectively maximal objects (i.e., $o_1$, $o_2$). Let's assume that $p=60$; i.e., $\lceil\frac{60}{100}\cdot 3\rceil
= 2$ users. In this case, only the restaurant $o_2$ is $60$-collectively
maximal, since, $o_2$ is $60$-collectively preferred over $o_1$, if we
consider the set of users $u_2$ and $u_3$. Finally, if $p=30$, we consider
only one user in order to decide if an object is collectively preferred. In
this case, the $30$-collectively maximal is an empty set, since $o_2$ is
$30$-collectively preferred over $o_1$, if we consider either user $u_2$ or
$u_3$, and also $o_1$ is $30$-collectively preferred over $o_2$, if we consider
user $u_1$.


\begin{algorithm}[]
\SetCommentSty{textsf}
\DontPrintSemicolon
\LinesNotNumbered
\SetKw{Break}{break}
\KwIn{objects $\O$, users $\U$}
\KwOut{$p$-$CM$ the $p$-collectively maximal}
\SetKwInput{KwVar}{Variables}
\KwVar{$CM$ the collectively maximal}

\vspace{5pt}
$\vdots$\;
\nlset{7} 
%
\ForEach{$o_i \in CM$} {
			\nlset{8} $inpCM \gets true$\;
	\nlset{9}\ForEach{$o_j \in CM\backslash o_i$ } {
		\nlset{10}\If{$o_j \succ_p o_i$} {
			\nlset{11} $inpCM \gets false$\;
			\nlset{12}\Break;
		}
	}
	\nlset{13}\If{$inpCM$}{
				\nlset{14} insert $o_i$ to $p\text{-}CM$\;				
	}
}
\caption{$p$-BSL}
\label{algo:pbsl}
\end{algorithm}

\subsection{A Baseline Algorithm ({\large\textit{{p}}}-BSL)}
\label{sec:pbsl}
Based on the above observations, we describe a baseline algorithm for the
$p$-GMCO problem, based on BSL. \autoref{algo:pbsl} shows the changes with
respect to the BSL algorithm; all omitted lines are identical to those in
\autoref{algo:bcs}. The $p$-BSL algorithm first computes the collectively
maximal objects applying BSL (\textit{lines 1--6}). Then, each collectively
maximal object, is compared with all other collectively maximal objects
(\textit{lines 7--14}). Particularly, each object $o_i$ is checked whether
there exists another object in $CM$ that is $p$-collectively preferred over
$o_i$ (\textit{lines 10--12}). If there is no such object, object $o_i$ is
inserted in $p$-$CM$ (\textit{line 14}). When the algorithm terminates, the
set $p$-$CM$ contains the $p$-collectively maximal objects.

\stitle{Computational Analysis.} 
Initially, the algorithm is computing the collectively maximal set using the
BSL algorithm (\textit{lines 1--6}), which requires $O(|\O|^2 \cdot |\U| \cdot
d)$. Then, finds the $p$-collectively maximal objects (\textit{lines 7--14}),
performing in the worst case $O(|\O|^2)$ comparisons. 
Since, in worst case we have that $|CM|=|\O|$.
Therefore, the
computational cost of \autoref{algo:pbsl} is $O(|\O|^2 \cdot |\U| \cdot d)$.

\subsection{An Index-based Algorithm ({\large\textit{{p}}}-IND)}
\label{sec:pind}

We also propose an extension of IND for the $p$-GMCO problem, termed $p$-IND.
\autoref{algo:pics} shows the changes with respect to the IND algorithm; all omitted lines are
identical to those in \autoref{algo:ics}.

\begin{algorithm}[]
\SetCommentSty{textsf}
\LinesNotNumbered
\SetKw{Break}{break}
\DontPrintSemicolon
\KwIn{R$^*$-Tree $\T$, users $\U$}
\KwOut{$p$-$CM$ the $p$-collectively maximal}
\SetKwInput{KwVar}{Variables}
\KwVar{$H$ a heap with $\T$ entries sorted by $score()$,
$CM$ the collectively maximal object 
}
\vspace{1mm}
\nlset{1} $CM \gets \varnothing$; $p$-$CM \gets \varnothing$\;
\vspace{-5pt}$\vdots$\;
\nlset{4} \While{$H$ is not empty}{
	\vspace{-5pt}$\vdots$\;
	\nlset{18} \Else{ 
		\nlset{19} $o_x \gets e_x$ \;
		\nlset{20} $inCM \gets true$; $inpCM \gets true$\;
		\nlset{21} \ForEach{$o_a \in CM$} {
			\nlset{22} \If{$o_a \succ o_x$} {
				\nlset{23} $inCM \gets false$\;
				\nlset{24} \Break\;
			}
			\nlset{25} \If{$inpCM$} {
				\nlset{26} \If{$o_a \succ_p o_x$} {
					\nlset{27} $inpCM \gets false$\;
				}
			}			
			\nlset{28} \If{$o_a \in $ $p$-$CM$} {	
			
					\nlset{29} \If{$o_x \succ_p o_a$} {
					\nlset{30} remove $o_a$ from $p\text{-}CM$\;
							
				}	
			}
			
		}
 		\nlset{31} \If{$inCM$} {
				\nlset{32} insert $o_x$ to $CM$\;				
	\nlset{33}	\If{$inpCM$} {
				\nlset{34} insert $o_x$ to $p\text{-}CM$\;				
		
		} 
		} 
 	}
}
\caption{$p$-IND}
\label{algo:pics}
\end{algorithm}

In addition to the set $CM$, $p$-IND maintains the set $p$-$CM$ of
$p$-collectively maximal objects discovered so far (\textit{line 1}). It holds
that $p$-$CM \subseteq CM$; therefore, an object may appear in both sets. When
a leaf entry $o_x$ is popped (\textit{line 19}), it is compared against each
object $o_a$ in $CM$ (\textit{lines 21--30}) in three checks. First, the
algorithm checks if $o_a$ is collectively preferred over $o_x$ (\textit{lines 22--24}).
In that case, object $o_x$ is not in the $CM$ and thus not in the $p$-$CM$.
Second, it checks if $o_a$ is $p$-collectively preferred over $o_x$
(\textit{lines 25--27}). In that case, object $o_x$ is not in the $p$-$CM$,
but is in the $CM$. Third, the algorithm checks if the object $o_x$ is
$p$-collectively preferred over $o_a$ (\textit{lines 28--30}). In that case,
object $o_a$ is removed from the $p$-collectively maximal objects
(\textit{line 30}), but remains in $CM$.

After the three checks, if $o_x$ is collectively maximal (\textit{line 31}) it
is inserted in $CM$ (\textit{line 32}). Further, if $o_x$ is p-collectively
maximal (\textit{line 33}) it is also inserted in $p$-$CM$ (\textit{line 34}).
When the $p$-IND algorithm terminates, the set $p$-$CM$ contains the answer to
the $p$-GMCO problem.

\stitle{Computational Analysis.} 
$p$-IND performs at most 3 times more object to object comparisons than IND.
Hence its running time complexity remains $O(|\O|^2 \cdot |\U| \cdot d)$.

\section{The Group-Ranking Categorical Objects (GRCO) Problem}
\label{sec:rank}

\autoref{sec:rank_method} introduces the GRCO problem, and \autoref{sec:rank_algo} describes an algorithm for GRCO. Then, \autoref{sec:rank_properties} discusses some theoretical properties of our proposed ranking scheme.

\vspace{-6mm}
\subsection{Problem Definition}
\label{sec:rank_method}

\eat{
A na\"ive
approach is to assume the ``composition'' is obtained by summing the
individual degrees of match for all users and all attributes. That is, for an 
object $o_i$, the rank is defined as: $\rank(o_i) =\sum_{u_j \in \U} \sum_{ A_k \in \A}
m_i^j.A_k$.
}

\eat{
For each object we
determine an integer number called the \emph{tier}. Higher tiers are
better, and thus objects are ranked in decreasing order of their tier.

We define the \emph{tier} of an object $o_i$ to be the largest integer $\tau$,
where $0 \leq \tau < |\U|$, such that $o_i$ is $p$-collectively maximal for
any $p \ge \frac{|\U| -\tau}{|\U|}\cdot100$. The non-collectively maximal
objects are assigned the lowest possible tier $-1$ \note[dsachar]{we should keep this... does it mess with the properties?}.
}

As discussed in \autoref{sec:intro}, it is possible to define a ranking among
objects by ``composing'' the degrees of match for all users.
However, any ``compositing'' ranking function is unfair, as there is no objective
way to aggregate individual degrees of match.
In contrast, we propose an objective ranking method based on the concept of
$p$-collectively preference. The obtained ranking is a weak order, meaning
that it is possible for objects to share the same rank (ranking with ties).
We define the \emph{rank} of an object $o$ to be the smallest integer $\tau$,
where $1 \le \tau \le |\U|$, such that $o$ is $p$-collectively maximal for any
$p \ge \frac{\tau}{|\U|}\cdot100$. The non-collectively maximal objects are
assigned the lowest possible rank $|\U|+1$. Intuitively, rank $\tau$ for an
object $o$ means that any group $\U' \subseteq \U$ of at least $\tau$ users
(i.e., $|\U'| \ge \tau$) would consider $o$ to be preferable, i.e., $o$ would
be collectively maximal for these $\U'$ users. At the highest rank $1$, an
object $o$ is preferred by each user individually, meaning that $o$ appears in
all possible $p$-collectively maximal object sets.

\begin{myprob}[GRCO]
\label{prob:GRCO}
Given a set of objects $\O$ and a set of users $\U$ defined over a set of
categorical attributes $\A$, the \emph{Group-Ranking Categorical Objects}
(GRCO) problem is to find the rank of all collectively maximal objects of $\O$
with respect to $\U$.
\end{myprob}

\stitle{Example.} Consider the restaurants and the users presented in
Tables~\ref{tab:rest}~\&~\ref{tab:users}. In our example, the collectively
maximals are the restaurants $o_1$ and $o_2$. As described in the previous
example (\autoref{sec:pGMCO}), the restaurant $o_2$ is collectively maximal
for any group of two users. Hence, the rank for the restaurant $o_2$ is equal
to two. 
In addition, $o_1$ requires all the three users in order to be considered
as collectively maximal; so its rank is equal to three. Therefore, the
restaurant $o_2$ is ranked higher than $o_1$.

\subsection{A Ranking Algorithm (RANK-CM)}
\label{sec:rank_algo}

The RANK-CM algorithm (\autoref{algo:rcs}), computes the rank for all collectively maximal objects. 
The algorithm takes as input, the collectively maximal objects $CM$,
as well as the number of users $|\U|$. 
Initially, in each object is assigned the highest rank; i.e., $rank(o_i)\gets1$ (\textit{line 2}). 
Then, each object is compared against all other objects in $CM$ (\textit{loop in line 3}).
Throughout the objects comparisons, we 
increase $\tau$ (\textit{lines 5--11}) from the current rank (i.e., $\rank(o_x)$) (\textit{line 4}) up to $|\U|$.
If $o_i$ is not $p$-collectively maximal (\textit{line 7}), 
for $p=\frac{\tau}{|\U|} \cdot 100$ (\textit{line 6}), then $o_x$ cannot be in the $p$-$CM$
and can only have rank at most $\tau+1$ (\textit{line 8}).
Finally, each object is inserted in the $rCM$ based on its rank (\textit{line 12}).

\stitle{Computational Analysis.} 
The algorithm compares each collective maximal object with all other collective maximal objects. 
Between two objects the algorithm performs at most 
$|\U|-1$ comparisons. 
Since, in worst case we have that ${|CM|=|\O|}$, 
the computational cost of \autoref{algo:rcs} is $O(|\O|^2 \cdot |\U|)$.



\begin{algorithm}[]
\SetCommentSty{textsf}
\SetKw{Break}{break}
\DontPrintSemicolon
\KwIn{$CM$ the collectively maximal objects, $|\U|$ the number of users }
\KwOut{$rCM$ the ranked collectively maximal objects}
 \vspace{1mm} 
 
		 \ForEach{$o_i \in CM$} {
			$\rank(o_i) \gets 1$\;
			 \ForEach{$o_j \in CM\backslash o_i$} {
					$\tau \gets \rank(o_i)$\;
				\While{$\tau \leq |\U| - 1$}{
				 	$p \gets \frac{ \tau}{|\U|}\cdot100$ \;
					 \eIf{ $o_j \succ_p o_i$ } {
							$rank(o_i) = \tau+1$\;
					}{
\Break;					
					}
					$\tau \gets \tau+1$
			}
		}
		insert $o_i$ in $rCM$ at $\rank(o_i)$ \;			
	}
\caption{RANK-CM}
\label{algo:rcs}
\end{algorithm}

 \input{prop}

%% file: prop.tex

\vspace{-5mm}
\subsection{Ranking Properties}
\label{sec:rank_properties}

In this section, we discuss some theoretical properties in the context of the
rank aggregation problem. These properties have been widely used in voting
theory as evaluation criteria for the fairness of a voting system
\cite{Taylor05,A63,R88}. We show that the proposed ranking scheme satisfies several of
these properties.

 \begin{myproperty}[Majority] 
If an object is strictly preferable over all other objects by the 
\textit{majority} of the users, then this object is ranked above all other objects.
\end{myproperty}

\begin{myproof} 
Assume that $k_a $ users strictly prefer $o_a$ over all other objects, where ${k_a > \frac{|\U|}{2}}$.
We will prove that the rank $r_a$ of the object $o_a$ is lower than the rank of any other object.
 
Since, $k_a$ users strictly prefer $o_a$ over all other objects, 
any group of at least $|\U|- k_a+1$ users, will consider $o_a$ as collectively maximal. 
This holds since, any group of at least $|\U|- k_a+1$ users, 
contains at least one user which strictly prefers $o_a$ over all other objects.
Note that, $|\U|- k_a+1$ may not be the smallest group size. 
That is, it may hold that, for any group of less than $|\U|- k_a+1$ 
users, $o_a$ is collectively maximal.
 
Recall the definition of the ranking scheme, if the rank of an object $o$ is $\tau$, 
then $\tau$ is the smallest integer that, for any group of at least $\tau$ users, $o$ will be collectively maximal (for this group).
Therefore, in any case we have that, 
the rank $r_a$ of $o_a$ is at most $|\U|-k_a+1$, i.e., $r_a \leq |\U|-k_a+1$ \myFontH{(1)}.
 
On the other hand, let an object $o_i \in \O \backslash o_a$. 
Then, $o_i$ is not collectively maximal, for any group with $|\U|- k_a+1$ users. 
This holds since, we have that $k_a > \frac{|\U|}{2}$. 
So, there is a group of $|\U|- k_a+1$ users, 
for which, each user strictly preferred $o_a$ over $o_i$.
As a result, in order for $o_i$ to be considered as collectively maximal 
for any group of a specific size, we have to consider groups with more than $|\U|- k_a+1$ users. 
From the above, it is apparent that, in any case, the rank 
$r_i$ for an object $o_i$ is greater than $|\U|-k_a +1$, i.e., $r_i >|\U|- k_a + 1$ \myFontH{(2)}.

Therefore, from \myFontH{(1)} and \myFontH{(2)}, in any case the rank 
of the object $o_a$ will be lower than the rank of any other object. This concludes the proof of the property. \qed
\end{myproof}

\begin{myproperty}[Independence of Irrelevant Alternatives]
\label{pr:irr_alt}
The rank of each object is not affected if non-collectively maximal objects are inserted or removed.
\end{myproperty}

\begin{myproof}
According to the definition of the ranking scheme, if the rank of an object $o$ is $\tau$, 
then $\tau$ is the smallest integer that, for any group of at least $\tau$ users, $o$ will be collectively maximal (for this group).

As a result, the rank of an object is specified from the minimum group size, 
for which, for any group of that size, the object is collectively maximal.
Therefore, it is apparent that, the rank of each object is not affected by the non-collectively maximal objects.
To note that, the non-collectively maximal objects are ranked with the lowest possible rank, i.e., $|\U|+1$. \qed
\end{myproof}

\begin{myproperty}[Independence of Clones Alternatives]
The rank of each object is not affected if {non-collectively maximal objects similar to an existing object} are inserted.
\end{myproperty}

\begin{myproof}
Similarly to the \autoref{pr:irr_alt}. 
Based on the ranking scheme definition, the non-collectively maximal objects 
do not affect the ranking. \qed
\end{myproof}

 \begin{myproperty}[Users Equality]
The result will remain the same if two users switch their preferences.
This property is also know as \textit{Anonymity}.
\end{myproperty}

\begin{myproof}
According to the definition of the ranking scheme, if the rank of an object $o$ is $\tau$, 
then $\tau$ is the smallest integer that, for any group of at least $\tau$ users, $o$ will be collectively maximal (for this group). 

As a result, the rank of an object is specified from the minimum group size, 
for which, for any group of that size, the object is collectively maximal.
Hence, if two users switch preferences, it is apparent that, 
the minimum group of any users, for which an object is collectively maximal, remains the same, for all objects.
Therefore, the rank of all objects remains the same.  \qed
\end{myproof}

Let an object $o_i \in \O$ and a user $u_j \in \U$. Also, let $m_i^j$ be the matching vector between $u_j$ and $o_i$.
We say that the user $u_j$ \textit{increases his interest} over $o_i$, if
$\exists A_k: m_i^j.A_k < {\grave{m}}_{i}^{j}.A_k$, where 
$\grave{m}_{i}^{j}$ is the matching degree resulted by the interest change.

\begin{myproperty}[Monotonicity]
If an object $o_a$ is ranked above an object $o_b$,
and a user {increases his interest} over $o_a$, 
then $o_a$ maintains its position above $o_b$.
\end{myproperty}

\begin{myproof}
Let $r_{a}$ and $r_{b}$ be the rank of objects $o_a$ and $o_b$, respectively.
Since, $o_a$ is ranked above the object $o_b$, we have that $r_{a} < r_{b}$.

According to the definition of the ranking scheme, if the rank of an object $o$ is $\tau$, 
then $\tau$ is the smallest integer that, 
for any group of at least $\tau$ users, 
$o$ will be collectively maximal (for this group).
So, we have that for any group of at least 
$r_{a}$ and $r_{b}$ members, 
$o_a$ and $o_b$ will be collectively maximal.

Assume a user $u_j \in \U$ {increases his interest} over the object $o_a$.
Further, assume that $r'_{a}$ and $r'_{b}$ are the new ranks of 
the objects $o_a$ and $o_b$, resulting from the interest change.
 We show that in any case $r'_a \leq r_a$ and $r'_b \geq r_b$.

First let us study what holds for the new rank of the object $o_a$.
After the interest change, $r'_{a}$ is the smallest group size 
that $o_a$ is collectively maximal for any group of that size.
We suppose for the sake of contradiction that $r'_a > r_a$. 
Hence, after the interest change, we should consider larger group sizes
in order to ensure that $o_a$ will be collectively maximal for any group of that size.
This means that, after the interest change, there is a group of $r_{a}$ users 
for which $o_a$ is not collectively maximal.
Hence, since $o_a$ is not collectively maximal, 
there must exist an object $o_i \in \O \backslash o_a$ 
that is collectively preferred over $o_a$.
To sum up, considering $r_{a}$ users, we have that:
before the interest change, there is no object that is collectively preferred over $o_a$; 
and, after the interest change, there is an object that is collectively preferred over $o_a$.
This cannot hold, since the matching degrees between all other users and objects remain the same, 
while some matching degrees between $o_a$ and $u_j$ have increased (due to interest change).
So, for any group of $r_{a}$ users, there cannot exist an object $o_i$ which is collectively preferred over $o_a$.
Hence, we proved by contradiction that in any case $r'_a \leq r_a$.

Now, let us study what holds for the new rank of the object $o_b$.
After the interest change, $r'_{b}$ is the smallest group size, 
that, for any group of that size, $o_b$ is collectively maximal.
For the sake of contradiction, we assume that $r'_b < r_b$.
Hence, after the interest change, we should consider smaller group sizes,
in order to ensure that $o_b$ will be collectively maximal for any group of that size.
This means that, before the interest change, there is a group of $r'_{b}$ users, 
for which $o_b$ is not collectively maximal.
Hence, since $o_b$ is not collectively maximal, 
there must be an object $o_i \in \O \backslash o_b$ 
that is collectively preferred over $o_b$.
To sum up, considering $r'_{b}$ users, we have that:
before the interest change, there is an object that is collectively preferred over $o_b$; 
and, after the interest change, there is no object that is collectively preferred over $o_b$.
It is apparent that this also cannot hold. 
So, we proved by contradiction, that in any case $r'_b \geq r_b$.

We show that, $r'_a \leq r_a$ and $r'_b \geq r_b$. 
Since, $r_{a} < r_{b}$, in any case the object $o_a$ will be ranked above $o_b$. This concludes the proof. \qed
\end{myproof}

For some user $u$, the following property ensures that the result when $u$ participates is the same or better (w.r.t.\ $u$'s preferences) compared to that 
when $u$  does not participate.

 \begin{myproperty}[Participation]
\label{pr:particip}
 \textit{Version 1:} 
If the object $o_a$ is ranked above the object $o_b$, 
then after adding one or more users, which strictly prefer $o_a$ over all other objects, 
object $o_a$ maintains its position above $o_b$.

\vspace{1mm}
\noindent
\textit{Version 2:} 
Assume an object $o_a$ that is ranked above the object $o_b$, 
and that there is at least one user $u \in \U$ which has not stated any preferences; 
then if $u$ expresses that strictly prefers $o_a$ over all other objects, object $o_a$ maintains its position above $o_b$.
 \end{myproperty}
 
 \begin{myproof}
 \textit{Version 1:} 
Let $r_{a}$ and $r_{b}$ be the ranks of the objects $o_a$ and $o_b$, respectively.
Since, $o_a$ is ranked above the object $o_b$, we have that $r_{a} < r_{b}$.
 
According to the definition of the ranking scheme, if the rank of an object $o$ is $\tau$, 
then $\tau$ is the smallest integer that, 
for any group of at least $\tau$ users, 
$o$ will be collectively maximal (for this group).
Hence, we have that for any group of at least $r_{a}$ and $r_{b}$ 
members, $o_a$ and $o_b$ will be collectively maximal, respectively.

We assume a new user $u_n$, where $u_n \cap \U = \varnothing$.
The new user $u_n$ strictly prefers $o_a$ 
over all other objects $\O\backslash o_a$.
For the sake of simplicity, we consider a singe new user;
the proof for more users is similar. 
The new user set $\U_n$ is 
generated by adding the new user $u_n$ to the user set $\U$, i.e., $\U_n = \U \cup u_n$.

Let $r'_a$ and $r'_b$ be the ranks for the objects $o_a$ and $o_b$, 
respectively, for the new user set $\U_n$.
We show that, in any case, rank $r'_a$ is lower than $r'_b$.

First let us study what holds for the new rank of the object $o_a$.
We show that for any group of $r_{a}$ members from the new user set $\U_n$, 
$o_a$ will be collectively maximal. 
We assume a set $S$ of $r_{a}$ members from $\U_n$; 
i.e., $S \subseteq \U_n$ and $|S|=r_a$. 
Then, based on the users contained in $S$, we have two cases: 
(a) All users from $S$ initially belong to $\U$; 
i.e., $S \subseteq \U_n$. 
In this case $o_a$ is collectively maximal based on the initial hypothesis.
(b) The new user $u_n$ is included to $S$; i.e., $u_n \in S$.
Also in this case $o_a$ is collectively maximal,
since for the user $u_n$, $o_a$ is strictly preferred over all other objects.

Hence, in any case for any group of $r_{a}$ members from $\U_n$, 
$o_a$ will be collectively maximal. 
Also, depending on $\U$, the minimum size of any group of $\U_n$
for which $o_a$ is collectively maximal, may be smaller than $r_{a}$; i.e., $r'_a \leq r_a$. 
Therefore, we have that in any case $r'_a \leq r_a$~\myFontH{(1)}.

Now, let's determine the new rank for the object $o_b$.
It easy to verify that,
 if we consider groups of less than $r_{b}$ users from $\U_n$, 
 then $o_b$ cannot be collectively maximal for any group of that size.
Therefore, we have to select groups with equal to or greater than $r_{b}$ users from $\U_n$, 
 in order for any group of that size to consider $o_b$ as collectively maximal.
Hence, we have that in any case $r'_b \geq r_b$~\myFontH{(2)}.

Since, $r_a < r_b$, for \myFontH{(1)} and \myFontH{(2)} we have that $r'_a < r'_b$.
This concludes the proof of Version 1.

\vspace{2mm}
\noindent
\textit{Version 2:} 
The second version can be proved in similar way, 
since it can be ``transformed" into the first version. 

Assume we have a user $u_j \in \U$ that has not expressed any preferences.
Note that the following also holds if we have more than one users that have not expressed any preferences.

In this case, it is apparent that the ranking process ``ignores" the user $u_j$. 
In other words:
let $r_a$ and $r_b$ be the rank for the objects $o_a$ and $o_b$, respectively, when we consider the set of users $\U$.
In addition, let $r'_a$ and $r'_b$ be the ranks if we consider the users $\U \backslash u_j$. 
Based on our ranking scheme, it is apparent that, if $r_a > r_b$, then $r'_a > r'_b$.

In this version of the property, we assume that a user $u_j$ has not initially expressed any preferences. Afterwards, $u_j$ states that he strictly prefers $o_a$ over any other object. 
This scenario is equivalent to the following. 

Since, as described above, the rankings are not effected if we remove $u_j$; 
we initially consider the users $\U \backslash u_j$. 
Afterwards, a user that strictly prefers $o_a$ over all other objects is inserted in the users set 
$\U \backslash u_j$. 
This is the same as the first version of our property.

Note that, in order for the second version to be considered in our implementation, 
we have to modify the initialization of matching vector for the 
indifferent attributes. 
Particularly, the matching vector for indifferent attributes should be setting to 0, instead of 1.  \qed
\end{myproof}

The following property ensures a low possibility of objects being ranked in the same position.
\begin{myproperty}[Resolvability]
\textit{Version 1:} If two objects are ranked in the same position, adding a new user 
can cause an object to be ranked above the other.

\vspace{1mm}
\noindent
\textit{Version 2:} Assume that two objects are ranked in the same position, and that
there is at least one user $u$ which has not stated any preferences; 
if $u$ expresses preferences, then this can cause an object to be ranked above the other.
 \end{myproperty}

\begin{myproof}
\textit{Version 1:} 
Assume that we have the objects $o_a$ and $o_b$.
Let $r_{a}$ and $r_{b}$ be the rank of objects $o_a$ and $o_b$, respectively.
Initially, the objects are ranked in the same position, so we have that $r_{a} = r_{b}$.
In order to prove this property, we consider the following example. 

Assume that we have an object set $\O$ and four users $\U$ (i.e., $|\U|=4$).
For each of the first two users (i.e., $u_1 $ and $u_2$) the object $o_a$ is strictly preferred over all other objects in $\O$.
 
On the other hand, for each of the users $u_3$ and $u_4$, the object $o_b$ is strictly preferred over the all other objects in $\O$.

So, for the object $o_a$, we have that, for any group of three members,
$o_a$ will be collectively maximal. 
This holds, since at least one of the three members 
is one of the first two users ($u_1$ or $u_2$), for which $o_a$ is strictly preferred over all other objects. 
In addition, it is apparent that three is the smallest size for which, for any group of that size,
 $o_a$ will be collectively maximal.

According to the definition of the ranking scheme, if the rank of an object $o$ is $\tau$, 
then $\tau$ is the smallest integer that, for any group of at least $\tau$ users, 
$o$ will be collectively maximal (for this group).

As a result, for the rank of $o_a$ we have that $r_a=3$.
Using similar reasoning, for the rank of $o_b$ we have that $r_b=3$.
Hence, we have that in our example both objects $o_a$ and $o_b$ have the same rank, i.e., $r_a=r_b=3$. 

Now lets assume that we add a new user $u_5$, 
for which the object $o_a$ is strictly preferred over the all other objects in $\O$.
 So, for the following, we consider the new users set 
$\U'$ that includes the new user $u_5$, i.e., $\U'=\U \cup u_5$. 
We show that, considering the new users set $\U'$, 
the new rank $r'_b$ of object $o_b$ will be greater than the initial rank $r_b$; 
and for $o_a$ its new rank $r'_a$ will be the same as the initial $r_a$ rank.
Hence, in any case, if we also consider a new user $u_5$, the objects $o_a$ and $o_b$ will have different ranks. 

Considering the new users $\U'$, there is a group with three 
users for which $o_b$ is not collectively maximal. 
For example, if we select the users $u_1$, $u_2$ and $u_5$, 
then $o_b$ is not collectively maximal.
Hence, in order for $o_b$ to be collectively maximal, we have to select a larger group (at least four users) from $\U'$.
So, four users is the smallest group, for which for any group of that size, $o_b$ will be collectively maximal.
As a result, $r'_b = 4$. 
Hence, the new rank of the object $o_b$ is greater than the initial rank.

Regarding the object $o_a$, for any group of three 
users from $\U'$, $o_a$ will be collectively maximal. 
This hold since, for the three out of the five users (i.e., $u_1$, $u_2$, $u_5$), 
the object $o_a$ is strictly preferred over all other objects.
In addition, three users is the smallest group, for which for any group of that size, 
$o_a$ will be collectively maximal.
Therefore, the new rank of $o_a$ is $r'_a = 3$. 

So, the new ranks after the addition of user $u_5$ will be $r'_a = 3$ and $r'_b =4$, i.e., the objects $o_a$ and $o_b$ will have different ranks.
This concludes the proof of Version 1.

\vspace{2mm}
\noindent
\textit{Version 2:} 
The second version can be proved in similar way, 
since it can be ``transformed" to the first version as in the proof of \autoref{pr:particip}.  \qed
\end{myproof}

 \begin{myproperty}[Users' Preferences Neutrality]
Users with different number of preferences, or different preference granularity, are equally important.
\end{myproperty}

\begin{myproof}
It is apparent from the ranking scheme definition that this property holds. 
\qed
\end{myproof}

 \begin{myproperty}[Objects' Description Neutrality]
Objects with different description (i.e., attributes values) granularity are equal important.
\end{myproperty}
 
\begin{myproof}
It is apparent from the ranking scheme definition that this property holds. 
\qed
\end{myproof}

%
%
%
%
%


\eat{ antiparadeigma 
4 users.
1os stricly prefers to b.
oloi oi aloi 3 stricly prefer a.
ara gia to a object opia groups me 2 users pareis to a ine preferable ..ara to rank tu a ine 4-2 =2.
eno gia to b prepei na paris 4 users gia na ine preferable ara to rank tu b ine 4 -4 =0.
ara to rank tu a megalitero tu b.
pernume 2 sets pu to ena set ine o 1os users kai to 2o oi 3ois aloi.. 
ara sto proto group to rank tu b ine 1 eno to rank tu b ine 0.
 }

{}

%% file: ext.tex
 
\section{Extensions}
\label{sec:ext}

\autoref{sec:multi} discusses the case of multi-valued attributes and
\autoref{sec:nontree} the case of non-tree hierarchies.
\autoref{sec:subspace} presents an extension of IND (and thus of $p$-IND) for
the case when only a subset of the attributes is indexed.
\autoref{sec:objattr} discusses semantics of objective attributes. 

\subsection{Multi-valued Attributes}
\label{sec:multi}

There exist cases where objects have, or users specify, multiple values for an attribute.
Intuitively, we want the matching degree of an object to a user w.r.t.\ a multi-valued attribute to
be determined by the \emph{best possible match} among their values.
Note that, following  a similar approach, different semantics can be adopted for the matching degree  of multi-valued attributes. For example, the matching degree of multi-valued attributes may be defined as the average or the minimum match among their values.

Consider an attribute $A_k$, an
object $o$ and a user $u$, and also let $\{o.A_k[i]\}$, $\{u.A_k[j]\}$ denote the set of values for
the attribute $A_k$ for object $o$, user $u$, respectively. We define the matching degree of $o$ to
$u$ w.r.t.\ $A_k$ to be the largest among matching degrees computed over pairs of $\{o.A_k[i]\}$,
$\{u.A_k[j]\}$ values. For instance, in case of Jaccard coefficient we have, $m.A_k = \max_{i,j} \frac{|o.A_k[i] \cap u.A_k[j]|}{|o.A_k[i] \cup
u.A_k[j]|}$.

In order to extend IND to handle multi-valued attributes, we make the
following changes. We can relate an object $o_x$ to multiple virtual objects
$\{o_x[i]\}$, corresponding to different values in the multi-valued
attributes. Each of these virtual objects correspond to different rectangles
in the transformed space. For object $o_x$, the R$^*$-Tree $\T$ contains a
leaf entry $e_x$ whose MBR is the MBR enclosing all rectangles of the virtual
objects $\{o_x[i]\}$. The leaf entry $e_x$ also keeps information on how to
re-construct all virtual objects. During execution of IND, when leaf entry
$e_x$ is de-heaped, all rectangles corresponding to virtual objects
$\{o_x[i]\}$ are re-constructed. Then, object $o_x$ is collectively maximal,
if there exists no other object which is collectively preferred over all
virtual objects. If this is the case, then all virtual objects are inserted in
the list $CM$, and are used to prune other entries. Upon termination, the
virtual objects $\{o_x[i]\}$ are replaced by object $o_x$.

\subsection{Non-Tree Hierarchies}
\label{sec:nontree}

We consider the general case where an attribute hierarchy forms a directed acyclic graph (dag),
instead of a tree. The distinctive property of such a hierarchy is that a category is allowed to
have multiple parents. 
For example, consider an Attire attribute hierarchy 
slightly different than that of \autoref{fig:hierarchies}, which also has an new attire category 
``Sport casual". In this case, the ``Sport casual" category will have two parents, ``Street wear" and ``Casual". 

In the following, we extend the hierarchy transformation to handle dags. The extension follows the
basic idea of labeling schemes for dags, as presented in \cite{ABJ89}. First, we obtain a spanning
tree from the dag by performing a depth-first traversal. Then, we assign intervals to nodes for the
obtained tree hierarchy as in \autoref{sec:transform}. Next for each edge, i.e., child to parent
relationship, not included in the spanning tree, we propagate the intervals associated with a child
to its parent, merging adjacent intervals whenever possible. In the end, each node might be
associated with more than one interval.

The IND algorithm can be adapted for multi-interval hierarchy nodes similar to how it can handle
multi-valued attributes (\autoref{sec:multi}). That is, an object may be related to multiple
virtual objects grouped together in a leaf entry of the R$^*$-Tree.

The following properties extend \autoref{propos:intervals} for the general case
of non-tree hierarchies.

\begin{myproposition}
\label{lem:intervals}
For objects/users $x$, $y$, and an attribute $A_k$, let $\{x.I_k\}$, $\{y.I_k\}$ denote the set of
intervals associated with the value of $x$, $y$ on $A_k$. Then it holds that:

\newpage
\begin{enumerate}[(1)]
\setlength{\itemindent}{15pt}



\item $\displaystyle |x.A_k| = \hspace{-0pt} \underset{ I_x \in \{x.I_k\} } { \sum } \hspace{0pt}\|I_x\| $ 

\item $\displaystyle |x.A_k \cap y.A_k| =    \hspace{-0pt}  \underset{\substack{I_x \in \{x.I_k\}\\ I_y \in \{y.I_k\}}} {\sum}  \hspace{-0pt}  \| I_x \cap I_y \| $

\item $ \displaystyle |x.A_k \cup y.A_k| =  \hspace{-0pt}  \underset{ I_x \in \{x.I_k\} }  { \sum } \hspace{-0pt} \|I_x\| +  \underset{ I_y \in \{y.I_k\} } { \sum } \hspace{-0pt}   \|I_y\| - \hspace{-0pt}     \underset{\substack{I_x \in \{x.I_k\}\\ I_y \in \{y.I_k\}}} {\sum} \hspace{-0pt}  \| I_x \cap I_y \| $

\end{enumerate}

\end{myproposition}

\begin{myproof}
Regarding the first property, observe that  $ |x.A_k| =$ $\big\| \bigcup_{I_x \in \{x.I_k\}} I_x \big\|$ $= \sum_{I_x \in \{x.I_k\}} \|I_x\| $, since the intervals $I_x$ are disjoint.

Also, $ |x.A_k \cap y.A_k| = $ $\big\| \big( \bigcup_{I_x \in \{x.I_k\}} I_x \big) \cap \big( \bigcup_{I_y \in \{y.I_k\}} I_y \big) \big\| =$ \linebreak
$\big\| \bigcup_{I_x \in \{x.I_k\}, I_y \in \{y.I_k\}} I_x \cap I_y \big\|$ $ = \sum_{I_x \in \{x.I_k\}, I_y \in \{y.I_k\}} \| I_x \cap I_y \| $, since the intervals $I_x \cap I_y$ are disjoint.

Finally, the third property holds since $|x.A_k \cup y.A_k| = |x.A_k| + |y.A_k| -$  ${|x.A_k \cap y.A_k|}$. \qed
\end{myproof}

\subsection{Subspace Indexing}
\label{sec:subspace}

This section deals with the case that the index on the set of objects is built on a subset of the
object attributes. Recall that R$^*$-Tree indices are efficient for small dimensionalities, e.g.,
when the number of attributes is less than 10. Therefore, to improve performance, it makes sense to
build an index only on a small subspace containing the attributes most frequently occuring in users'
preferences. In the following, we present the changes to the IND algorithm necessary to handle this
case.

First, a leaf R$^*$-Tree entry $e_i$ contains a pointer to the disk page storing the non-indexed
attributes of the object $o_i$ corresponding to this entry. Second, given a non-leaf R$^*$-Tree
entry $e_i$, we define its maximum matching degree on user $u_j$ to be $M_i^j.A_k =
\frac{\|e_i.mbr.I_k \cap u_j.I_k\|}{\|u_j.I_k\|}$ with respect to an indexed attribute $A_k$ (as in
regular IND), and $M_i^j.A_{k'}=1$ with respect to a non-indexed attribute $A_{k'}$.   
Third, for a
leaf entry $e_i$ corresponding to object $o_i$, its maximum matching degree is equal to the matching
degree of $o_i$ to $u_j$ w.r.t.\@ $A_k$, as in regular IND. Note that in this case an additional I/O
operation is required to retrieve the non-indexed attributes.

It is easy to see that the maximum matching degree $M_i^j.A_k$ of entry $e_i$ on user $u_j$ w.r.t.\
specified attribute $A_k$ is an upper bound to the highest matching degree among all objects in the
group that $e_i$ defines. However, note that it is not a \emph{tight} upper bound as in the case of
the regular IND (\autoref{lem:max}).

The remaining definitions, i.e., the maximum matching vector and the score of an entry, as well as
the pseudocode are identical to their counterparts in the regular IND algorithm.

\subsection{Objective Attributes}
\label{sec:objattr}
In this section we describe \textit{objective attributes}. As objective
attributes we refer to the attributes that the order of their values is the
same for all users. Hence, in contrast to the attributes considered before,
the users are not expressing any preferences over the  objective attributes.
Particularly, in objective attributes, their preference relation is derived
from the attributes' semantics and it is the same for all users. For instance,
in our running example  in addition to  the restaurants' attributes in which
different users may have  different preferences (i.e., subjective attributes);
we can assume an objective attribute ``Rating", representing the restaurant's
score. Attribute Rating is totally ordered, and higher rated
restaurants are more preferable from all users.

Based on the attributes' semantics, we categorized attributes into two groups: 
(1) \textit{objective attributes}, and (2) \textit{subjective attributes}.
Let $\A_o \in \A$ and $\A_s \in \A$ denote the objective and subjective attributes respectively,
where   $\A_o \cup \A_s =\A$ and $\A_o \cap \A_s  = \varnothing$.

Let two objects $o_a$ and $o_b$, having an objective attribute $A_k \in \A_o$. 
As $o_a.A_k$ we denote the value of the attribute $A_k$ for the object $o_a$. 
Here, without loss of generality, we assume that objective attributes are single-value numeric attributes, and the object $o_a$ is \textit{better} than another object $o_b$ on the objective attribute $A_k$,  iff  $o_a.A_k > o_b.A_k$.

Considering objects with both objective and subjective attributes, 
the preferred and strictly preferred relations presented in \autoref{sec:define}, are defined as follows.

An object $o_a$ is \emph{preferred} over $o_b$, for user $u_j$, denoted as $o_a \succeq^j o_b$ iff 
(1) for every specified subjective attribute $A_h \in \A_s$ it holds that $m_a^j.A_h \geq m_b^j.A_h$, and 
(2) for each objective attribute $A_k \in \A_o$ hold that $o_a.A_k \geq o_b.A_k$.
Moreover, object $o_a$ is \emph{strictly preferred} over $o_b$, for user
$u_j$, denoted as $o_a \succ^j o_b$ iff 
(1) $o_a$ is preferred over $o_b$,  
(2) there exists a specified subjective attribute $A_h \in \A_s$
such that $m_a^j.A_h > m_b^j.A_h$, and
(3) there exists  an  objective attribute $A_k \in \A_o$ such that 
$o_a.A_k > o_b.A_k$.

%% file: exp.tex

\section{Experimental Analysis}
\label{sec:exp}

\autoref{sec:datasets} describes the datasets used for the evaluation. 
Sections~\ref{sec:e_mcp} and \ref{sec:e_pmcp} study the efficiency of the GMCO
and $p$-GMCO algorithms, respectively. Finally, \autoref{sec:r_mcp}
investigates the effectiveness of the ranking in the GRCO problem.

\subsection{Datasets \& User preferences}
\label{sec:datasets}

We use five datasets in our experimental evaluation, one synthetic and four real. 
The first is \textsf{Synthetic},
where objects and users are synthetically generated. All attributes have the same hierarchy, a
binary tree of height $\log|A|$, and thus all attributes have the same number of leaf hierarchy
nodes $|A|$. To obtain the set of objects, we fix a level, $\ell_o$ (where $\ell_o=1$ corresponds to
the leaves), in all attribute hierarchies. Then, we randomly select nodes from this level to obtain
the objects' attribute value. The number of objects is denoted as $|\O|$, while the number of
attributes for each object is denoted as $d$. Similarly, to obtain the set of users, we fix a level,
$\ell_u$, in all hierarchies. 
The group size (i.e., number of users) is denoted as $|\U|$.

\begin{table}[t!]
\caption{Parameters (\textsf{Synthetic})}
\label{tab:param_S}
\begin{tabular}{lcc}
\tline
 \textbf{Description} &\textbf{Symbol} & \textbf{Values} \\ \dline
Number of objects & $|\O|$ & 50K, 100K, \textbf{500K}, 1M, 5M \\
Number of attribute & $d$ & 2, 3, \textbf{4}, 5, 6 \\
Group size& $|\U|$ & 2, 4, \textbf{8}, 16, 32 \\
Hierarchy height & $\log|A|$ & 4, 6, \textbf{8}, 10, 12 \\
Hierarchy level for objects & $\ell_o$ & \textbf{1}, 2, 3, 4, 5 \\
Hierarchy level for users & $\ell_u$ & \textbf{2}, 3, 4, 5, 6 \\
\bline
\end{tabular}
\end{table}

The second dataset is \textsf{RestaurantsF}, which contains $85,681$ US
restaurant retrieved from Factual\footnote{www.factual.com}. We consider
\textit{three} categorical attributes, \textit{Cuisine}, \textit{Attire} and
\textit{Parking}. The hierarchies of these attributes are presented in
\autoref{fig:hierarchies} (the figure only depicts a subset of the hierarchy
for Cuisine). Particularly, for the attributes Cuisine, Attire and Parking,
we have $6$, $3$, $3$ levels and $126$, $5$, $5$ leaf hierarchy nodes, respectively.

The third dataset is \textsf{ACM}, which contains $281,476$ research
publications from the ACM, obtained from
datahub\footnote{datahub.io/dataset/rkb-explorer-acm}. The
\textit{Category} attribute is categorical and is used by the ACM in order to
classify research publications. The hierarchy for this attributed is defined
by the ACM Computing Classification
System\footnote{www.acm.org/about/class/ccs98-html}, and is organized in $4$
levels and has $325$ leaf nodes.

The fourth dataset is \textsf{Cars}, containing a set of $30,967$ car
descriptions retrieved from the Web\footnote{www.epa.gov}. We consider
\textit{three} attributes, \textit{Engine}, \textit{Body} and
\textit{Transmission}, having $3$, $4$, $3$ levels, and $11$, $23$, $5$ leaf hierarchy
nodes, respectively. We note that this is not the same dataset used in
\cite{BBS14}.

The fifth dataset is \textsf{RestaurantsR}, obtained from a recommender system
prototype\footnote{archive.ics.uci.edu/ml/datasets/Restaurant+\&+consum
er+data}. This dataset contains a set of $130$ restaurants descriptions and a
set of $138$ users along with their preferences. For our purposes, we consider
\textit{four} categorical attributes,
\textit{Cuisine}, \textit{Smoke}, \textit{Dress}, and \textit{Ambiance},
having $5$, $3$, $3$, $3$ levels, and $83$, $3$, $3$, $3$ leaf hierarchy nodes, respectively. This
dataset is used in the effectiveness analysis of the GRCO problem, while the
other datasets are used in the efficiency evaluation of the GMCO algorithms.


For the efficiency evaluation, the user preferences for real datasets are
obtained following two different approaches. In the first approach, denoted as
\textit{Real preferences}, we attempt to simulate real user preferences.
Particularly, for the \textsf{RestaurantF} dataset, we use as user preferences
the restaurants' descriptions from the highest rated New York restaurant
list\footnote{www.yelp.com}. For the \textsf{Car} dataset, the user
preferences are obtained from the top rated
cars\footnote{www.edmunds.com/car-reviews/top-rated.html}. Finally, for the
\textsf{ACM} dataset, the user preferences are obtained by considering ACM
categories from the papers published within a research
group\footnote{www.dblab.ntua.gr/pubs}. In the second approach, denoted as
\textit{Synthetic preferences}, the user preferences are obtained using a
method similar to this followed in \textsf{Synthetic} dataset. Particularly,
the user preferences are specified by randomly selecting hierarchy nodes from
the second hierarchy level (i.e., $\ell_u=2$).
Table~\ref{tab:realdatachar} summarizes the basic characteristics of the employed real datasets.


\begin{table}[t!]
\caption{Real Datasets Basic Characteristics}
\label{tab:realdatachar}
\begin{tabular}{lcl}
\tline
 \textbf{Dataset} &\textbf{Number of Objects} & \textbf{Attributes (Hierarchy height)} \\ \dline
\textsf{RestaurantsF} & $85,691$ & \textit{Cuisine} (6), \textit{Attire} (3), \textit{Parking} (3)\\
\textsf{ACM} & $281, 476$ & \textit{Category} (4)\\
\textsf{Cars} & $30, 967$ & \textit{Engine} (3), \textit{Body} (4), \textit{Transmission} (3)\\
\textsf{RestaurantsR} & $130$ & \textit{Cuisine} (5), \textit{Smoke} (3), \textit{Dress} (3), \textit{Ambiance} (3)\\
\bline
\end{tabular}
\end{table}

\subsection{Efficiency of the GMCO algorithms}
\label{sec:e_mcp}

For the GMCO problem, we implement IND (\autoref{sec:algo}) and three flavors
of the BSL algorithm (\autoref{sec:basic_algo}), denoted BSL-BNL, BSL-SFS, and
BSL-BBS, which use the skyline algorithms BNL \cite{BKS01}, SFS \cite{CGGL03},
BBS \cite{PTFS05}, respectively.

To gauge the efficiency of all algorithms, we measure: 
(1) the number of disk I/O operations, denoted as I/Os; 
(2) the number of dominance checks, denoted as Dom. Checks; and 
(3) the total execution time, denoted as Total Time, and measured in secs. 
In all cases, the reported time values are the averages of 3 executions.
All algorithms were written in C++, compiled with gcc, and the experiments were performed on a 2GHz CPU.

\subsubsection{Results on Synthetic Dataset}

In this section we study the efficiency of the GMCO algorithms using the \textsf{Synthetic} dataset described in \autoref{sec:datasets}. 

\stitle{Parameters.} \autoref{tab:param_S} lists the parameters that we vary and the range of
values examined for \textsf{Synthetic}. To segregate the effect of each parameter, we perform six
experiments, and in each we vary a single parameter, while we set the remaining ones
to their default (bold) values.

\begin{figure}[t]
\hspace{-1.0 cm}
\subfloat[I/O Operations]{\includegraphics[width=2.1in]{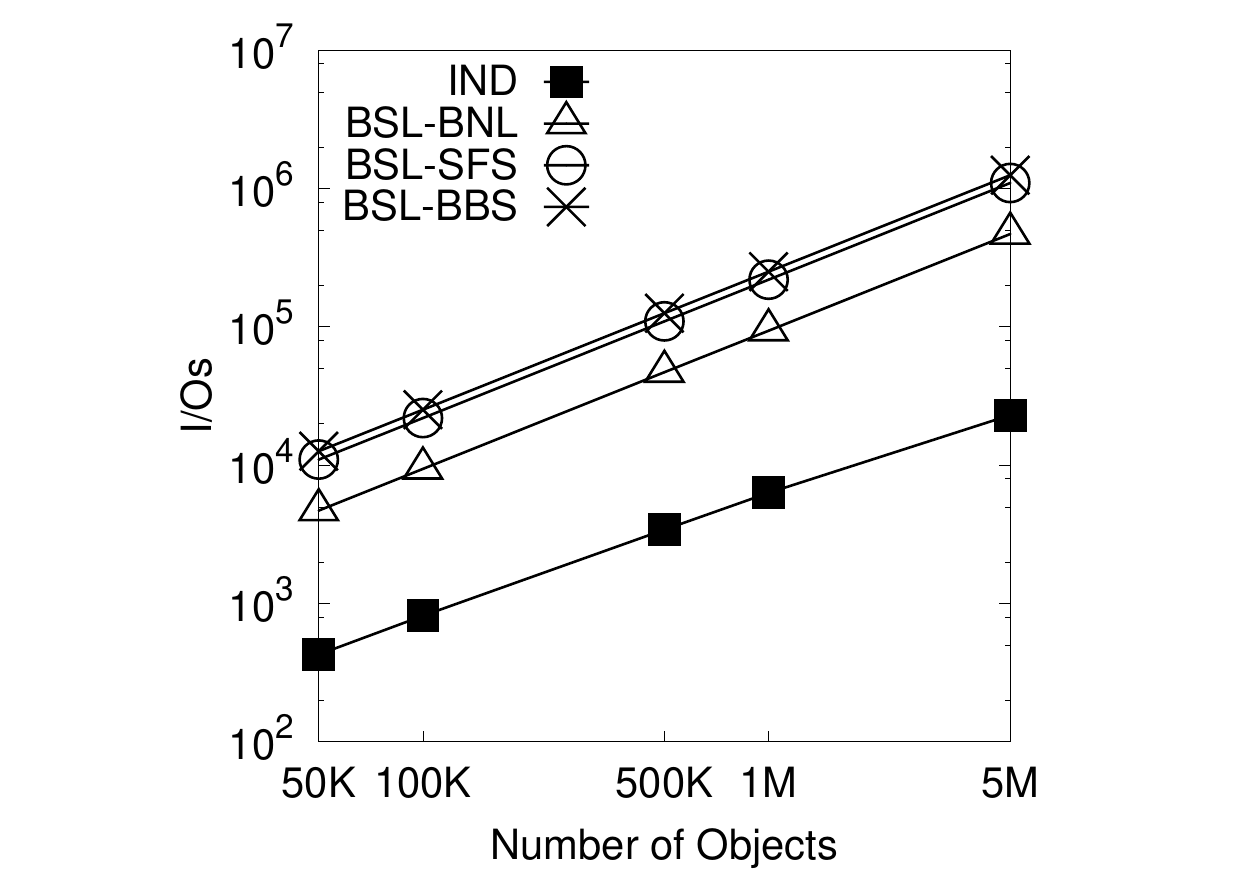}\label{fig:syn_o_io}}
\subfloat[Dom. Checks]{\hspace{-1.0 cm}\includegraphics[width=2.1in]{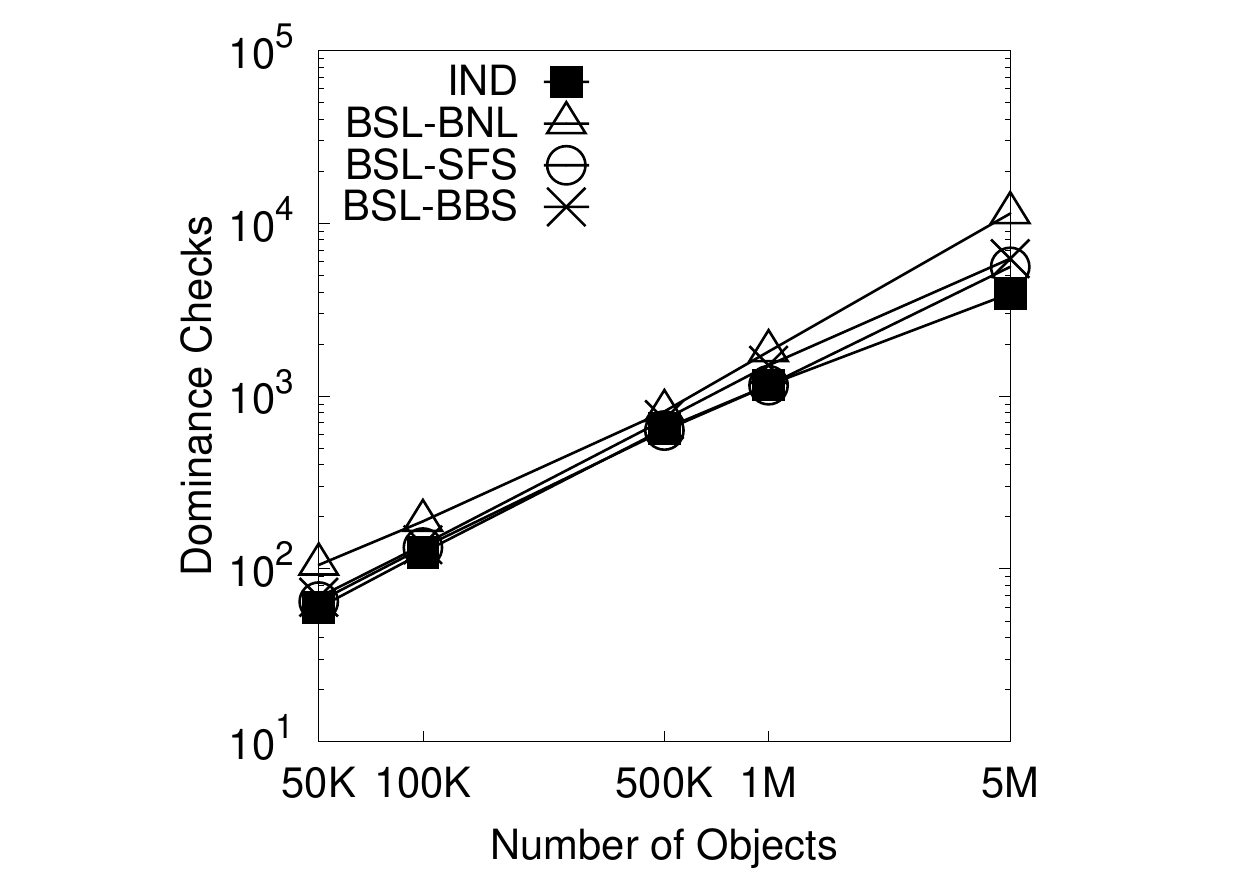}\label{fig:syn_o_dc}\hspace{-1.0 cm}}
\subfloat[Total Time]{\includegraphics[width=2.1in]{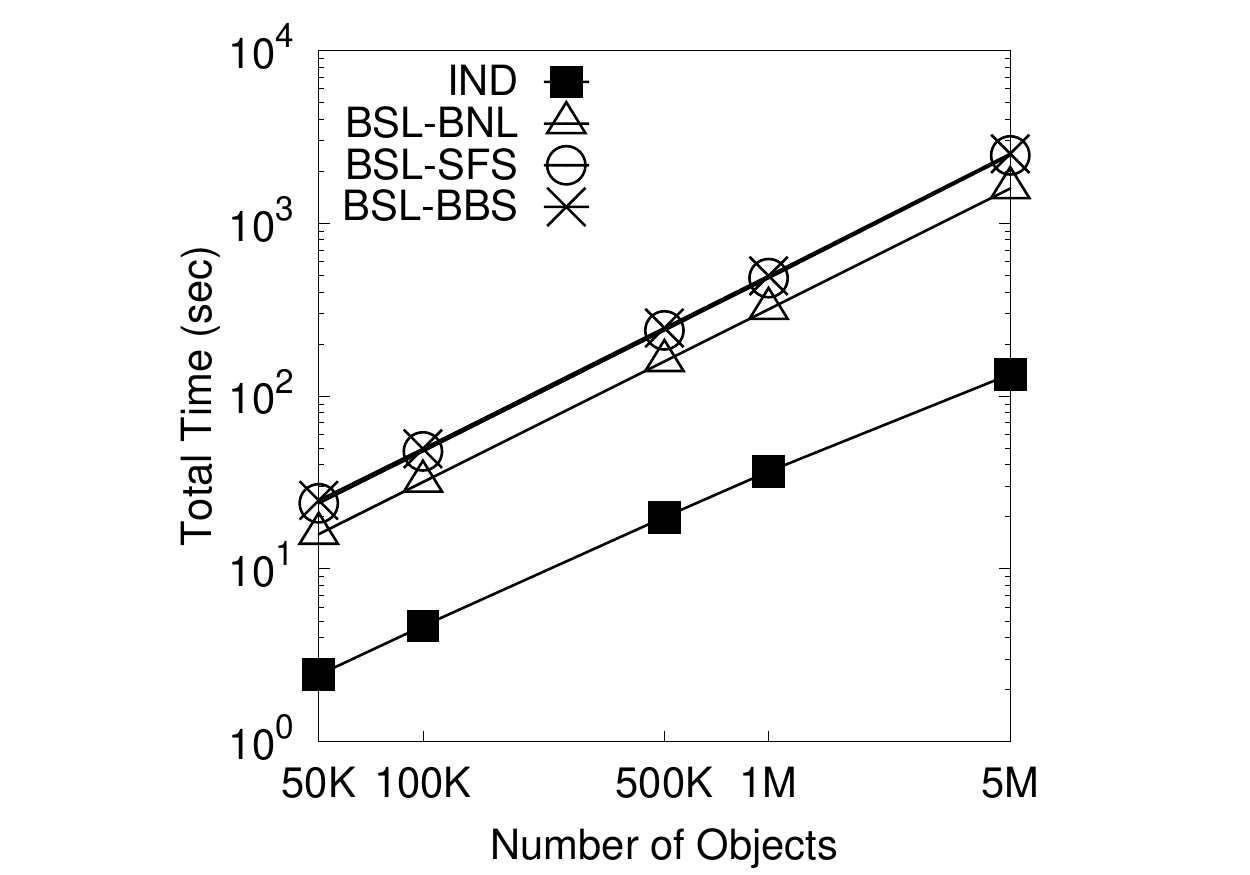}\label{fig:syn_o_time}}
\caption{GMCO algorithms, \textsf{Synthetic}: varying $|\O|$}
\label{fig:syn_o}
\vspace{-5pt}
\end{figure}

\stitle{Varying the number of objects.} In the first experiment, we study performance with respect
to the objects' set cardinality $|\O|$. Particularly, we vary the number of objects from 50K up to
5M and measure the number of I/Os, the number of dominance checks, and the total processing time, in
Figures~\ref{fig:syn_o_io}, \ref{fig:syn_o_dc} and \ref{fig:syn_o_time}, respectively.

When the number of objects increases, the performance of all methods deteriorates. The number of
I/Os performed by IND is much less than the BSL variants, the reason being BSL needs to construct a
file containing matching degrees. Moreover, the SFS and BBS variants have to preprocess this file, i.e.,
sort it and build the R-Tree, respectively. Hence, BSL-BNL requires the fewest I/Os among the BSL
variants.

\begin{figure*}[]
\hspace{-1.0 cm}
\subfloat[I/O Operations]{\includegraphics[width=2.1in]{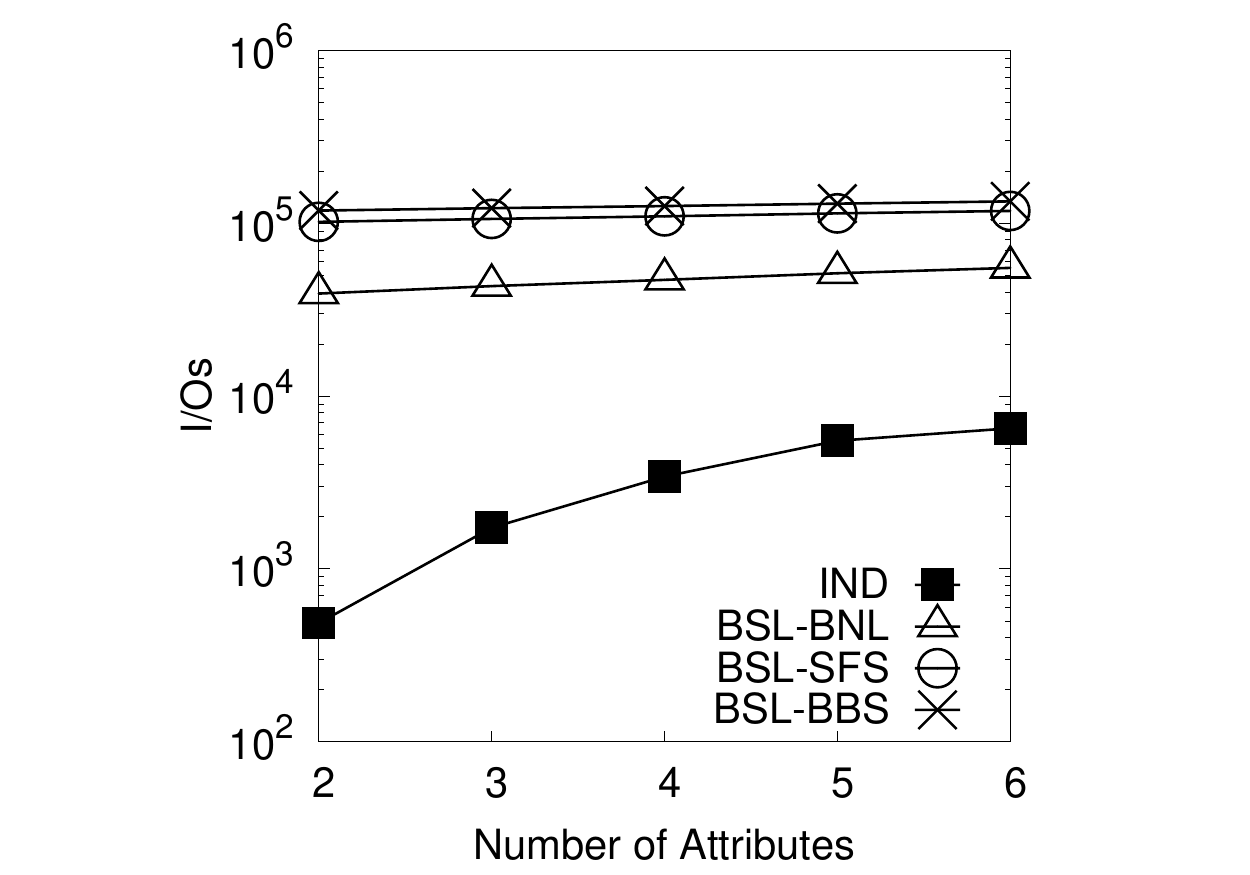}\label{fig:syn_d_io}}
\subfloat[Dom. Checks]{\hspace{-1.0 cm}\includegraphics[width=2.1in]{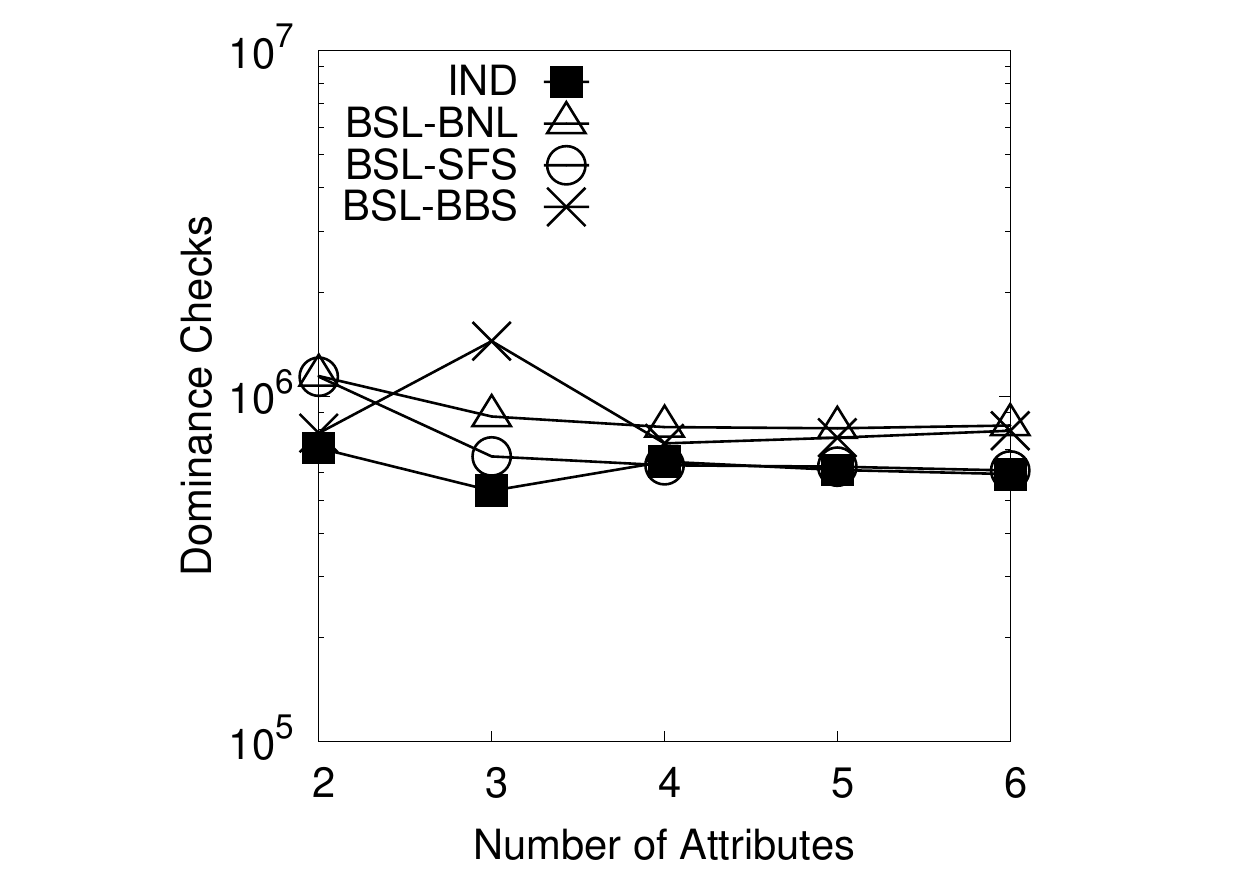}\label{fig:syn_d_dc}\hspace{-1.0 cm}}
\subfloat[Total Time]{\includegraphics[width=2.1in]{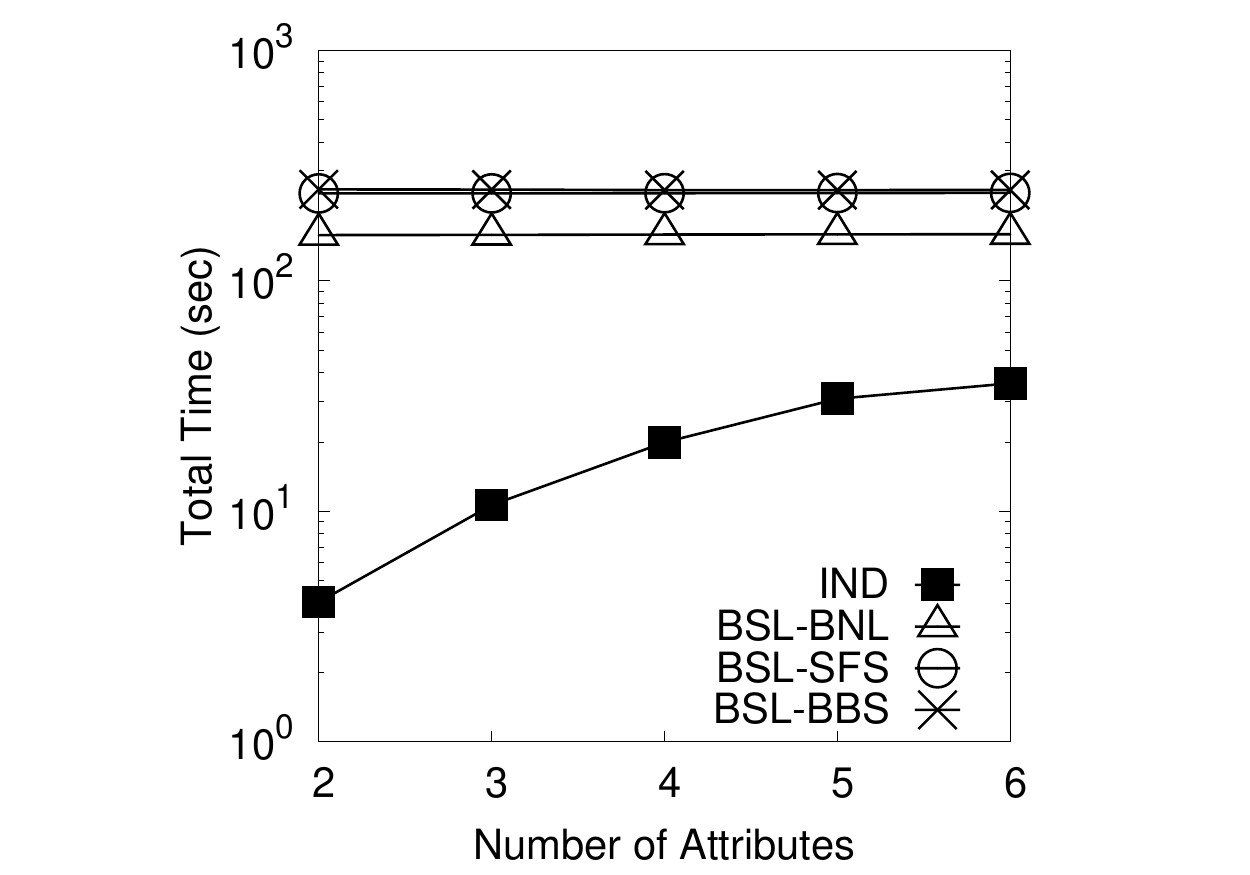}\label{fig:syn_d_time}}
\caption{GMCO algorithms, \textsf{Synthetic}: varying $d$}
\label{fig:syn_d}
\vspace{-5pt}\end{figure*}

All methods require roughly the same number of dominance checks as seen in
\autoref{fig:syn_o_dc}. IND performs fewer checks, while BSL-BNL the most. Compared to the other
BSL variants, BSL-BNL performs more checks because, unlike the others, computes the skyline over an
unsorted file. IND performs as well as BSL-SFS and BSL-BBS, which have the easiest task. Overall,
\autoref{fig:syn_o_time} shows that IND is more than an order of magnitude faster than
the BSL variants.

\stitle{Varying the number of attributes.} \autoref{fig:syn_d} investigates the effect as we
increase the number of attributes $d$ from 2 up to 6. The I/O cost, shown in
\autoref{fig:syn_d_io} of the BSL variants does not depend on $|\O|$ and thus remains roughly
constant as $d$ increases. On the other hand, the I/O cost of IND increases slightly with $d$. The
reason is that $d$ determines the dimensionality of the R-Tree that IND uses. Further, notice that
the number of dominance checks depicted in \autoref{fig:syn_d_dc} is largely the same across
methods. \autoref{fig:syn_d_time} shows that the total time of IND increases with $d$, but it is
still significantly smaller (more than 4 times) than the BSL methods even for $d=6$.

\begin{figure*}[]
\hspace{-1.0 cm}
\subfloat[I/O Operations]{\includegraphics[width=2.1in]{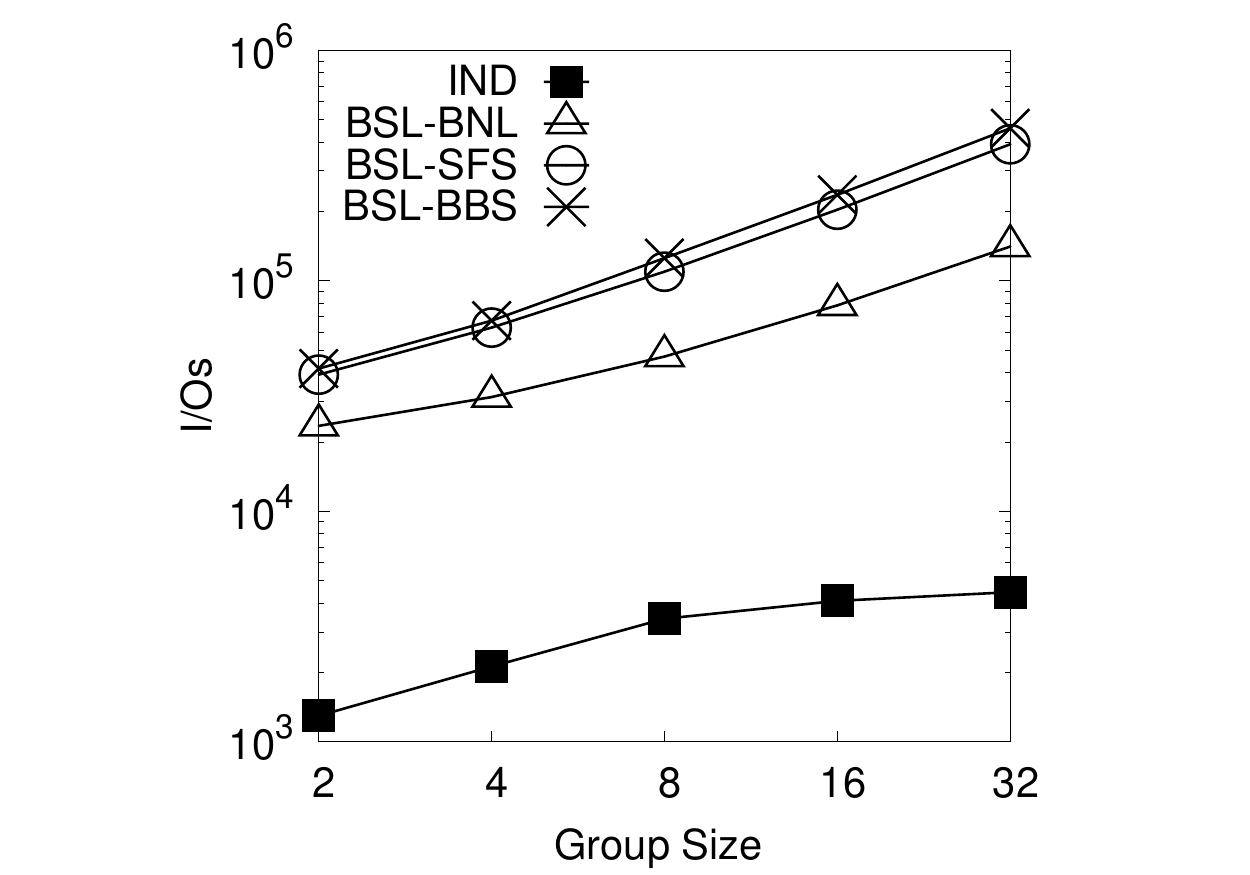}\label{fig:syn_u_io}}
\subfloat[Dom. Checks]{\hspace{-1.0 cm}
\includegraphics[width=2.1in]{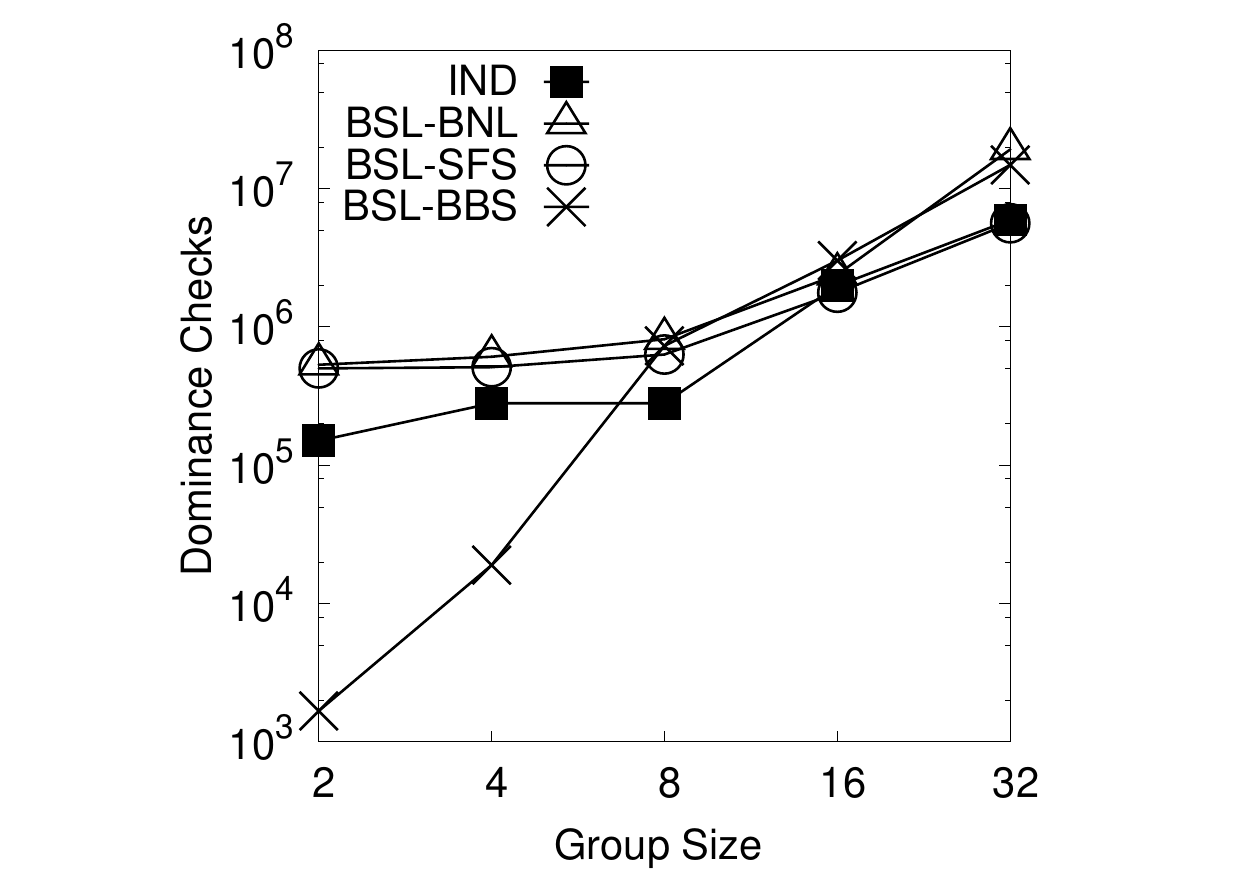}\label{fig:syn_u_dc}\hspace{-1.0 cm}}
\subfloat[Total Time]{\includegraphics[width=2.1in]{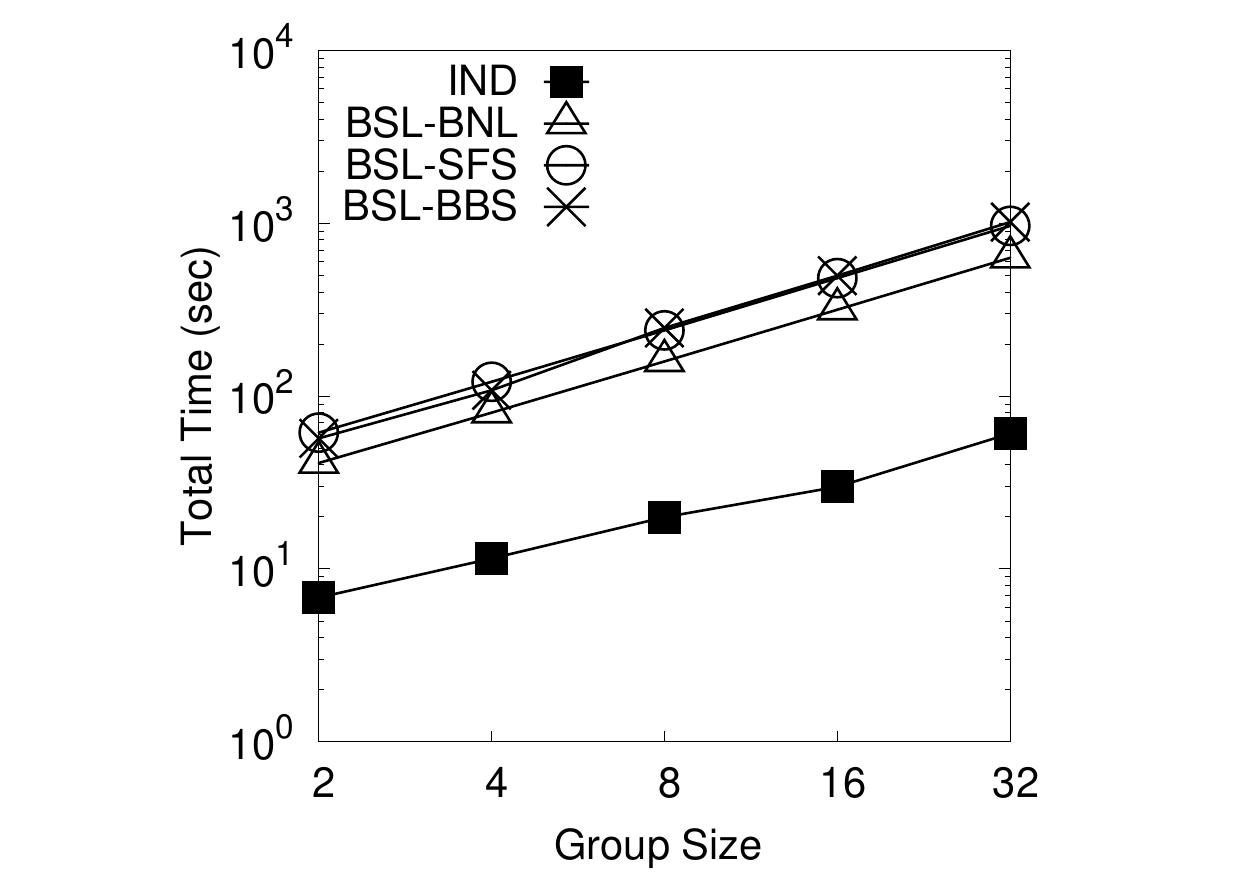}\label{fig:syn_u_time}}
\caption{GMCO algorithms, \textsf{Synthetic}: varying $|\U|$}
\label{fig:syn_u}
\vspace{-5pt}\end{figure*}

\stitle{Varying the group size.} In the next experiment, we vary the users' set cardinality
$|\U|$ from 2 up to 32; results are depicted in \autoref{fig:syn_u}. The performance of all
methods deteriorates with $|\U|$. The I/O cost for IND is more than an order of magnitude smaller
than the BSL variants, and the gap increases with $|\U|$, as \autoref{fig:syn_u_io} shows. As
before, BSL-BNL requires the fewest I/Os among the BSL variants.

Regarding the number of dominance checks, shown in \autoref{fig:syn_u_dc}, IND performs the
fewest, except for 2 and 4 users. In these settings, the BBS variant performs the fewest checks, as
it is able to quickly identify the skyline and prune large part of the space. Note that $|\U|$
determines the dimensionality of the space that BSL-BBS indexes. As expected, for more than 4
dimensions the performance of BBS starts to take a hit. Overall, \autoref{fig:syn_o_time} shows
that IND is more than an order of magnitude faster than all the BSL variants, among which BSS-BNL is
the fastest.

\begin{figure*}[]
\hspace{-1.0 cm}
\subfloat[I/O Operations]{\includegraphics[width=2.1in]{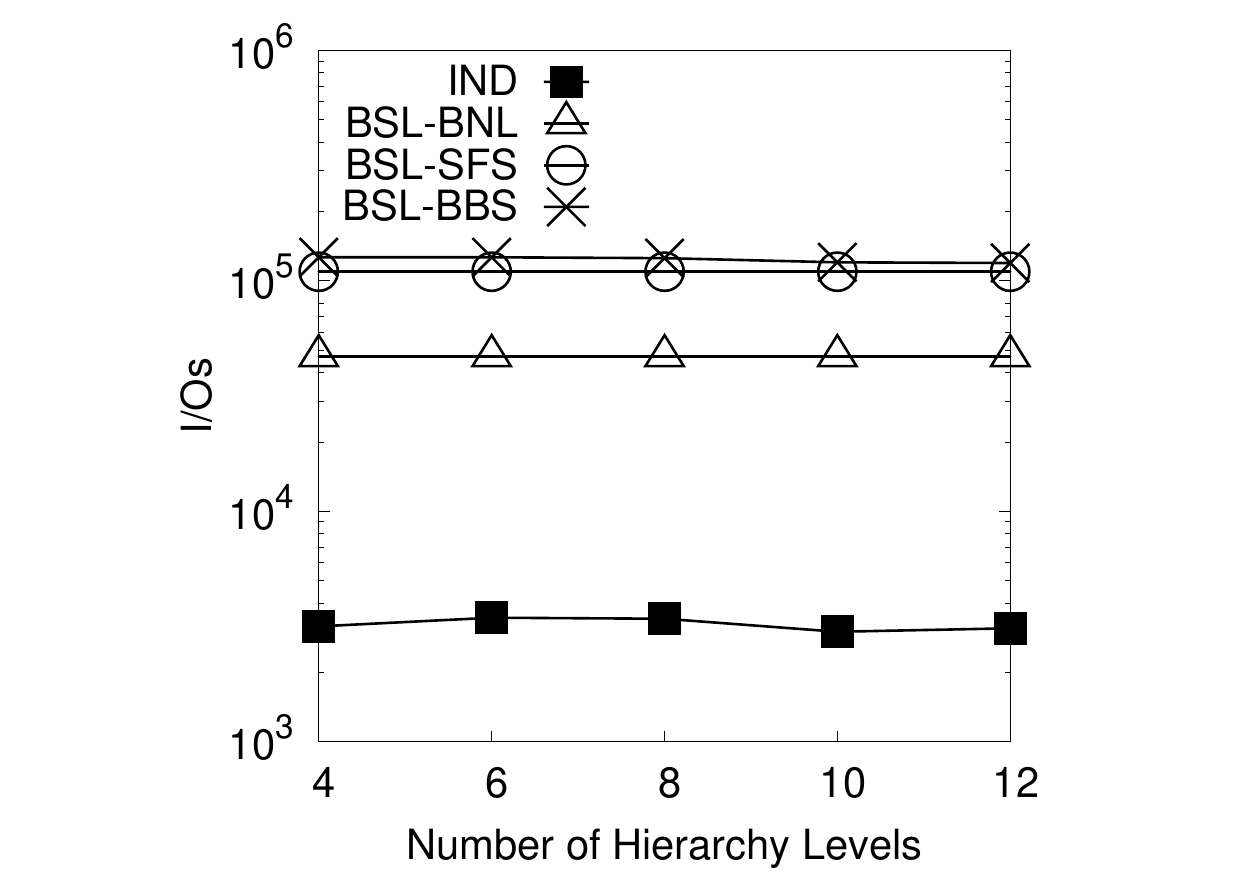}\label{fig:syn_a_io}}
\subfloat[Dom. Checks]{\hspace{-1.0 cm}\includegraphics[width=2.1in]{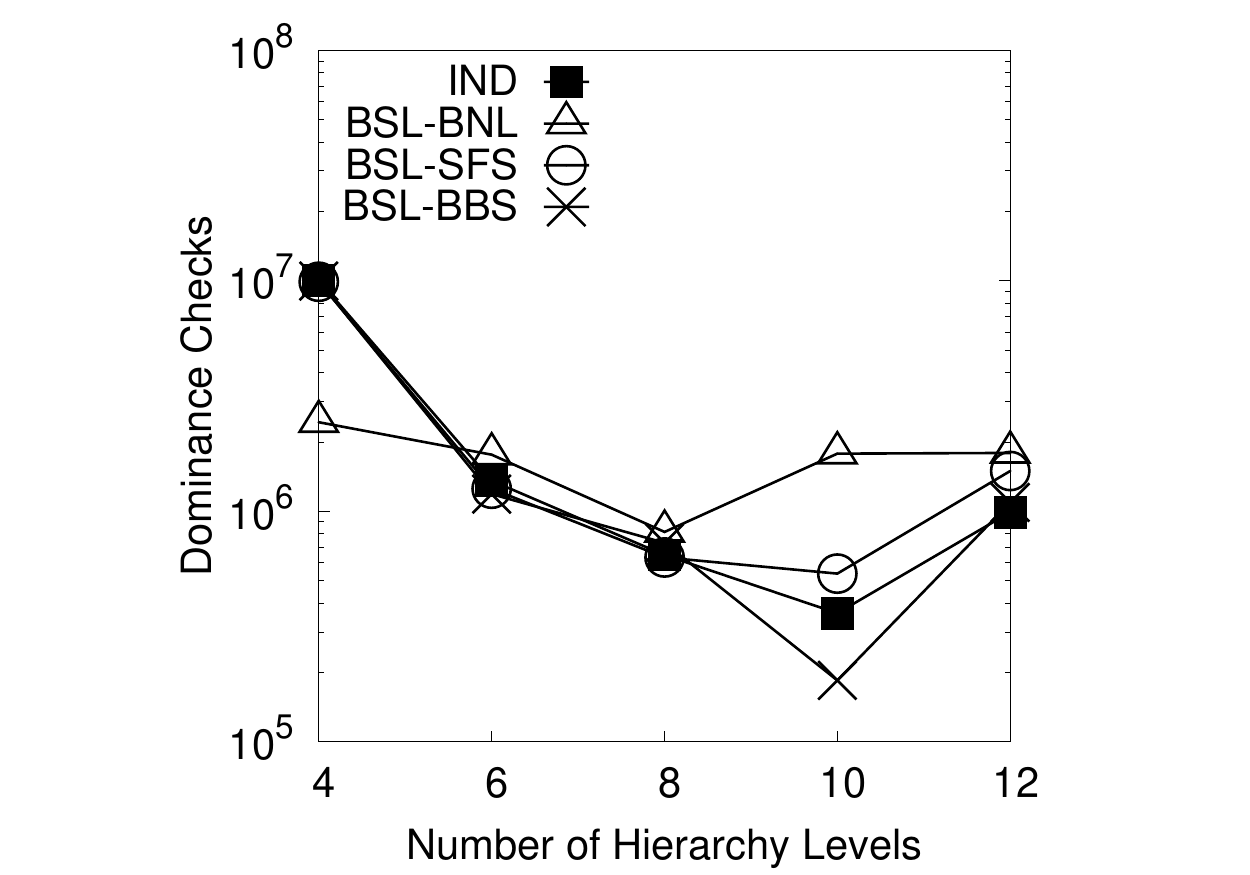}\label{fig:syn_a_dc}\hspace{-1.0 cm}}
\subfloat[Total Time]{\includegraphics[width=2.1in]{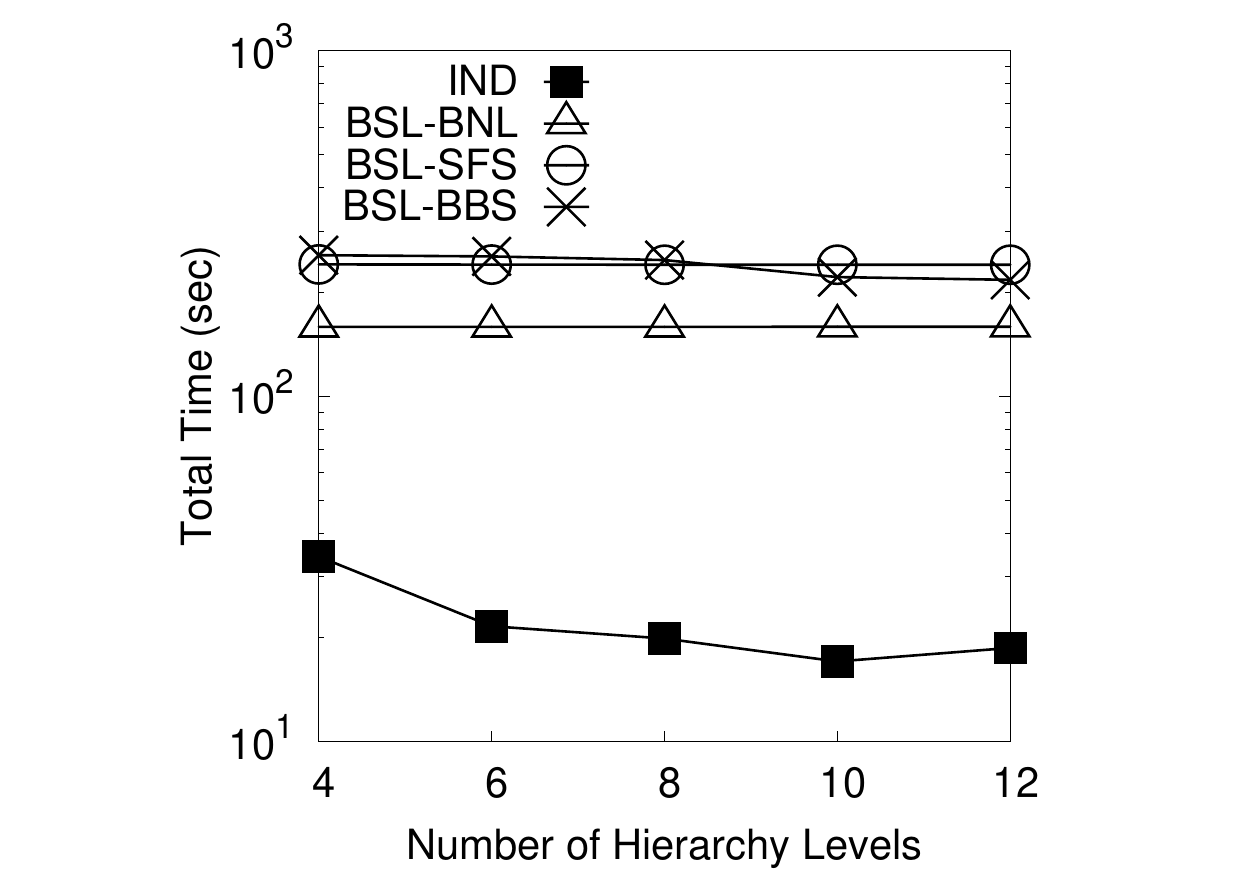}\label{fig:syn_a_time}}
\caption{GMCO algorithms, \textsf{Synthetic}: varying $\log|A|$}
\label{fig:syn_a}
\vspace{-5pt}\end{figure*}

\stitle{Varying the hierarchy height.} In this experiment, we vary the hierarchy height $\log |A|$
from 4 up to 12 levels. \autoref{fig:syn_a} illustrates the results. All methods are largely
unaffected by this parameter. Note that the number of dominance checks varies with $\log |A|$, and
IND performs roughly as many checks as the BSL variants which operated on a sorted file, i.e.,
BSL-SFS and BSL-BBS. Overall, IND is more than an order of magnitude faster than all BSL variants.

\begin{figure*}[]
\hspace{-1.0 cm}
\subfloat[I/O Operations]{\includegraphics[width=2.1in]{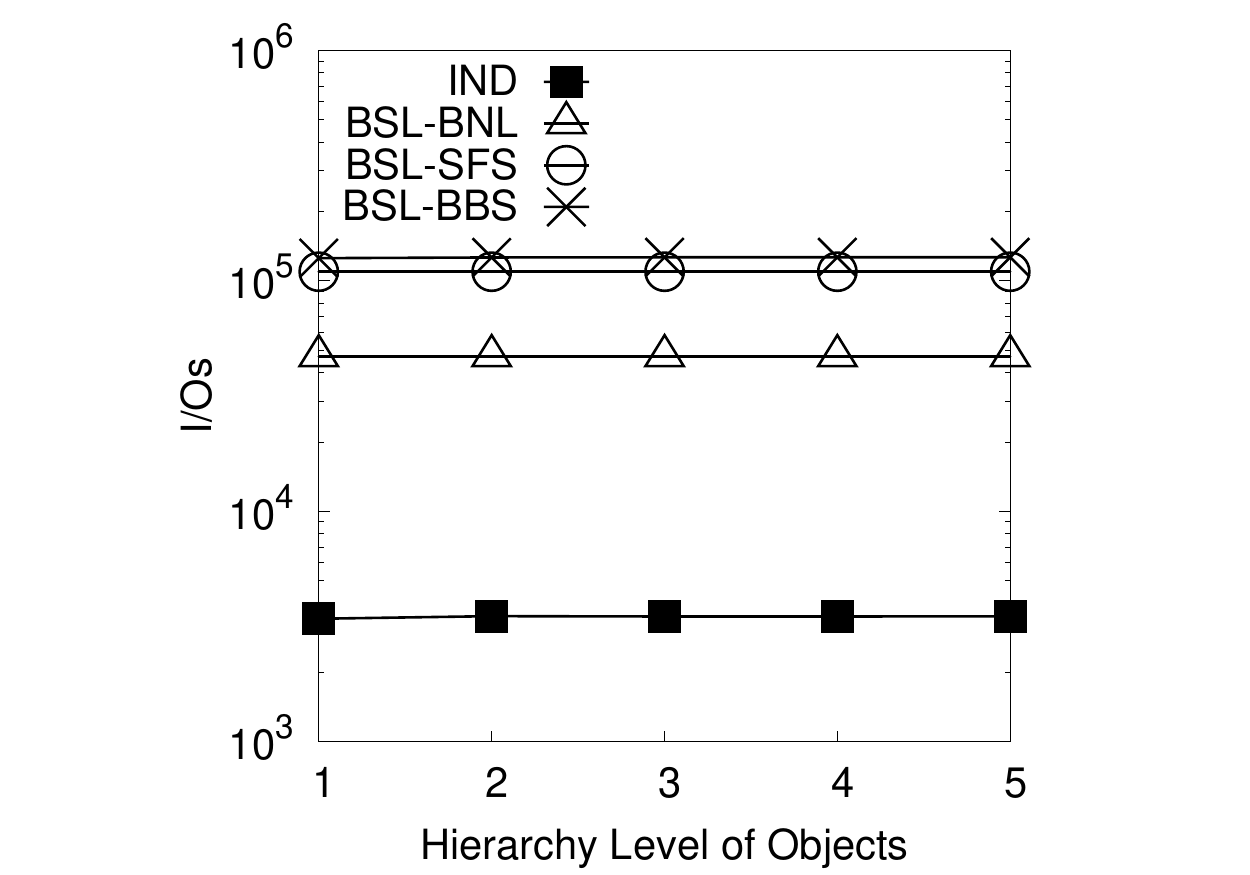}\label{fig:syn_Lo_io}}
\subfloat[Dom. Checks]{\hspace{-1.0 cm}\includegraphics[width=2.1in]{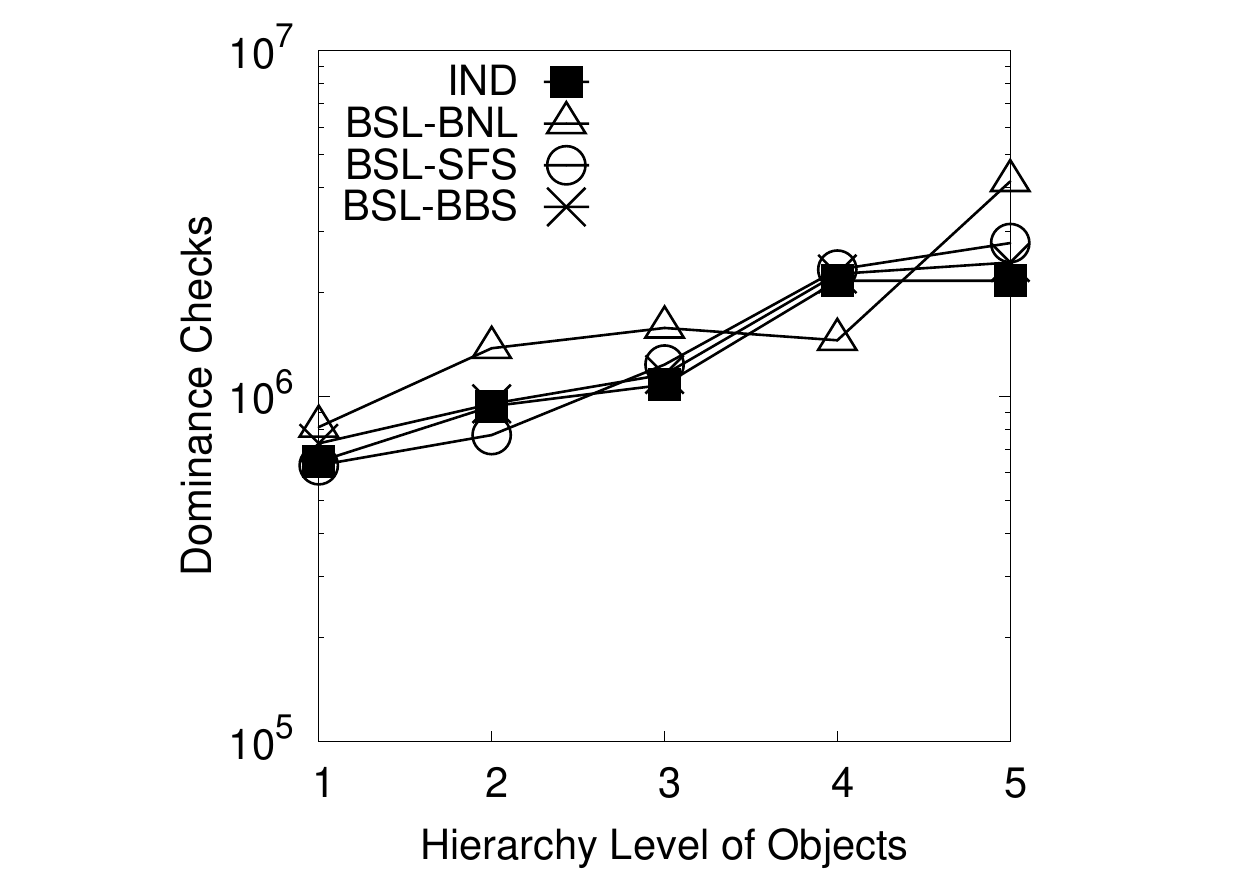}\label{fig:syn_Lo_dc}\hspace{-1.0 cm}}
\subfloat[Total Time]{\includegraphics[width=2.1in]{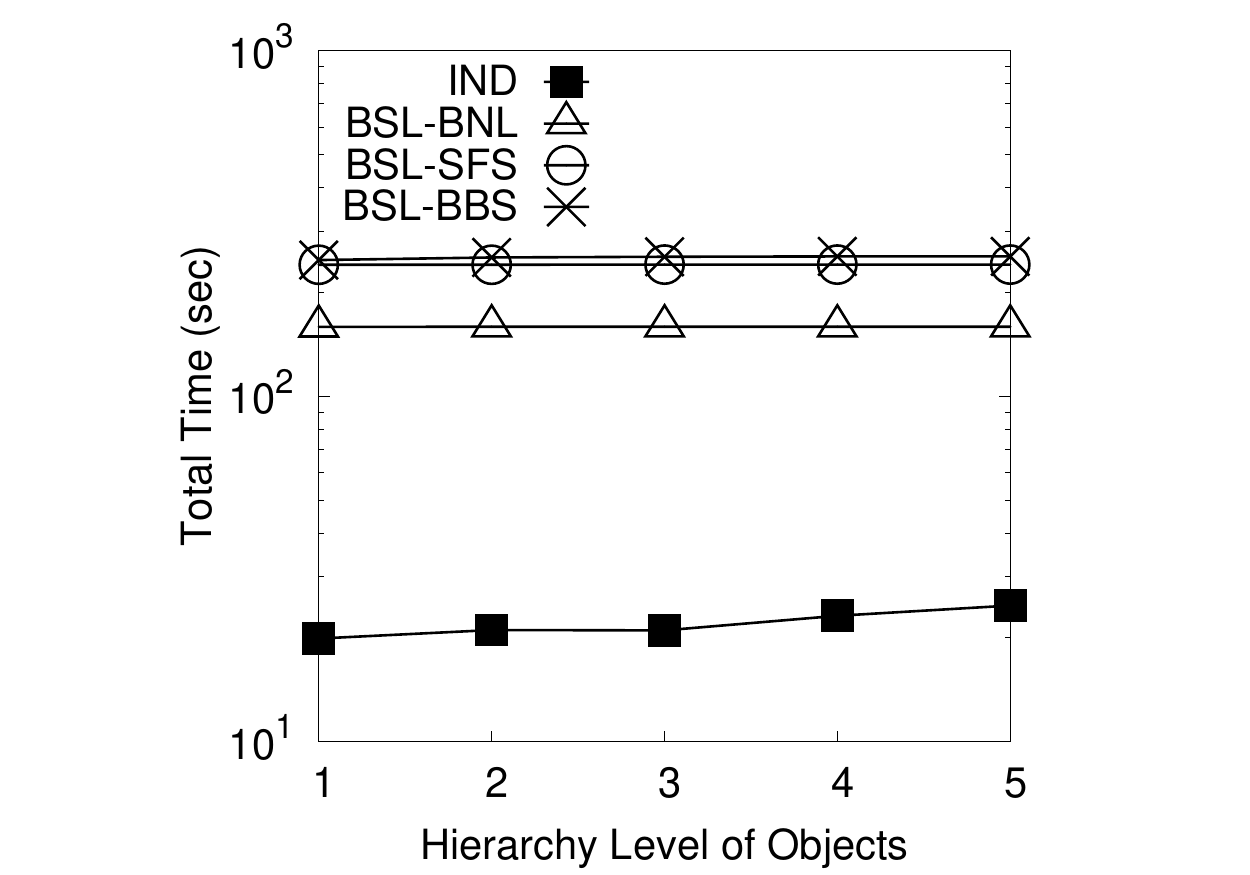}\label{fig:syn_Lo_time}}
\caption{GMCO algorithms, \textsf{Synthetic}: varying $\ell_o$}
\label{fig:syn_Lo}
\vspace{-5pt}\end{figure*}

\stitle{Varying the objects level.} \autoref{fig:syn_Lo} depicts the results of varying the level
$\ell_o$ from which we draw the objects' values. The performance of all methods is not significantly
affected by $\ell_o$. Note though that the number of dominance checks increases as we select
values from higher levels.

\begin{figure*}[]
\hspace{-1.0 cm}
\subfloat[I/O Operations]{\includegraphics[width=2.1in]{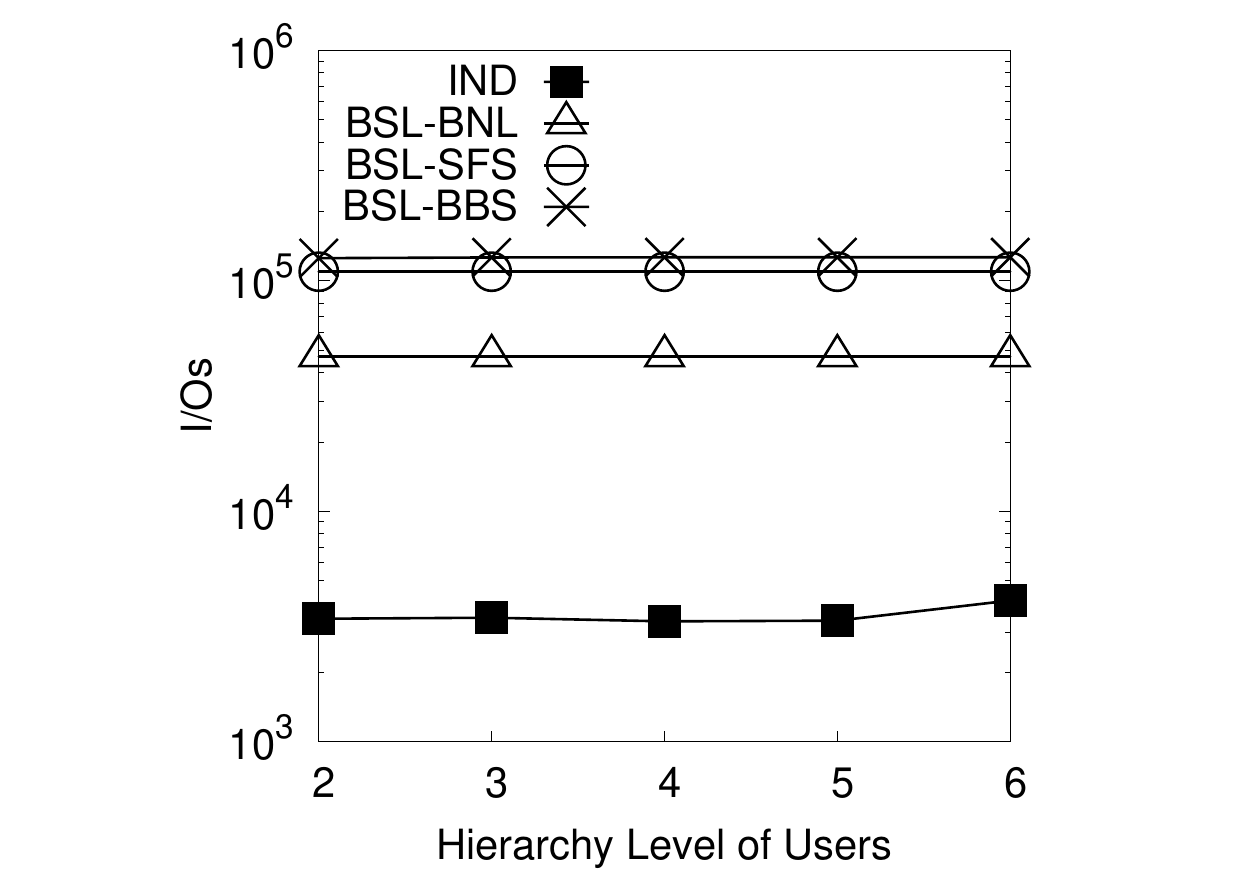}\label{fig:syn_Lu_io}}
\subfloat[Dom. Checks]{\hspace{-1.0 cm}\includegraphics[width=2.1in]{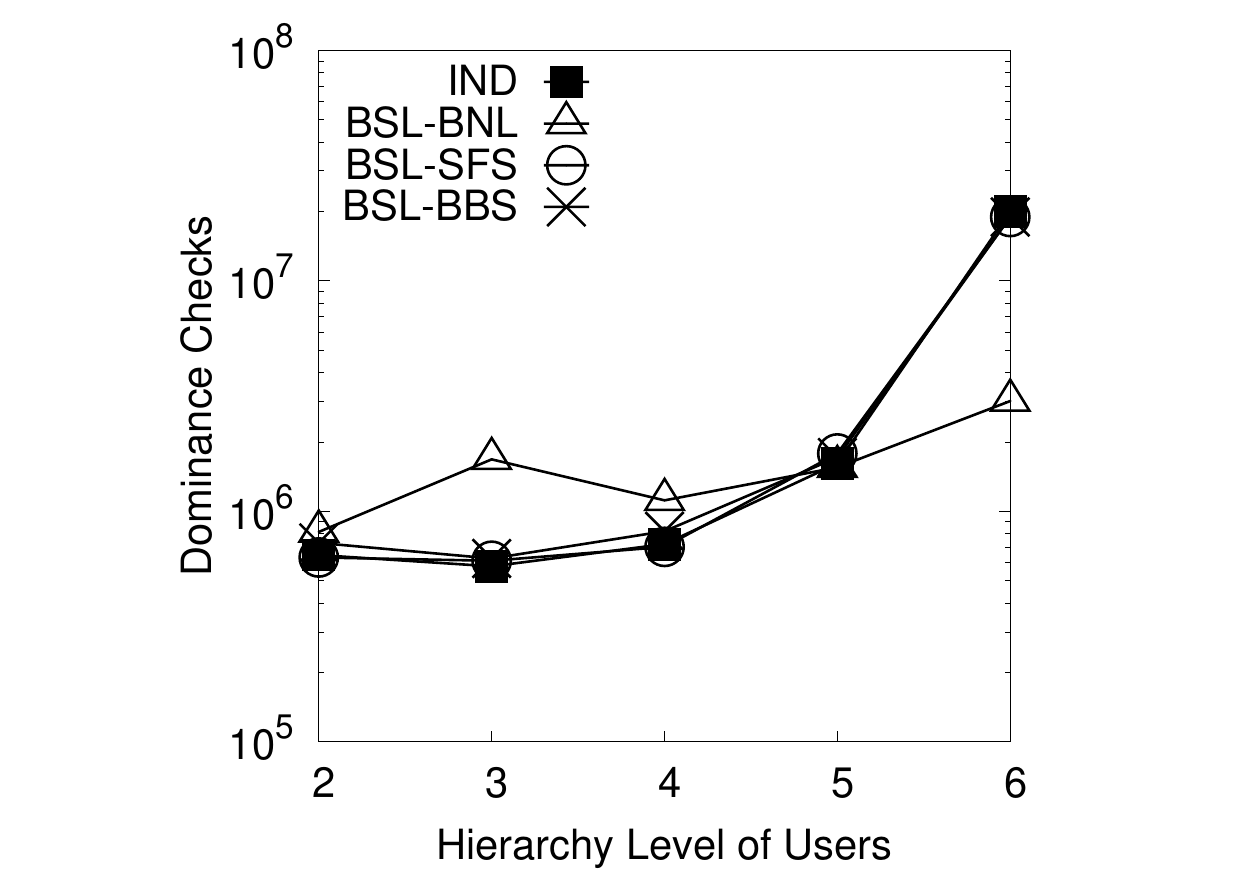}\label{fig:syn_Lu_dc}\hspace{-1.0 cm}}
\subfloat[Total Time]{\includegraphics[width=2.1in]{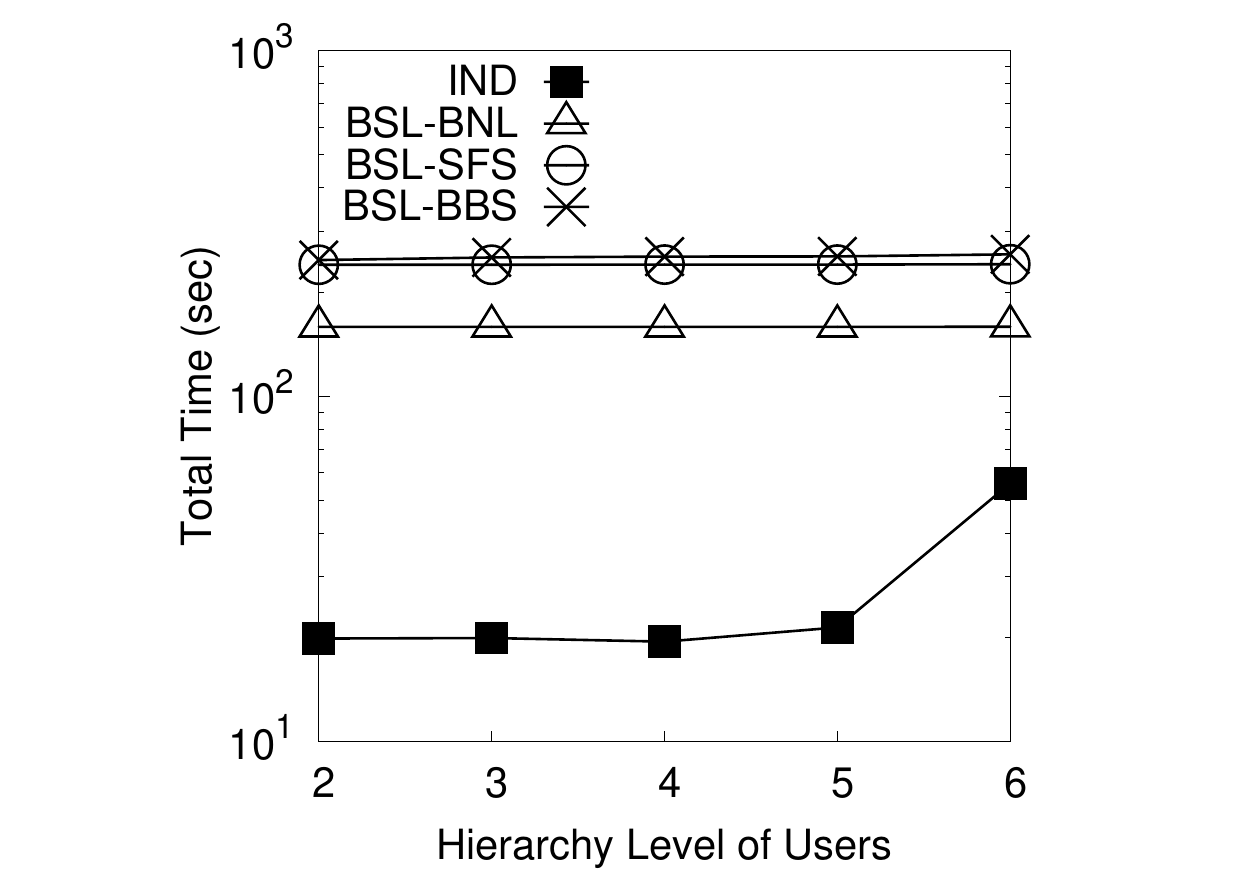}\label{fig:syn_Lu_time}}
\caption{GMCO algorithms, \textsf{Synthetic}: varying $\ell_u$}
\label{fig:syn_Lu}
\vspace{-5pt}\end{figure*}

\stitle{Varying the users level.} \autoref{fig:syn_Lu} depicts the results of varying the level
$\ell_u$ from which we draw the users' preference values. As with the case of varying $\ell_o$, the
number of dominance checks increases with $\ell_u$, while the performance of all methods remains
unaffected. The total time of IND takes its highest value of $\ell_u = 6$, as the number of required
dominance checks increases sharply for this setting. Nonetheless, IND is around 3 times faster than BSL-BNL.

\subsubsection{Results on Real Datasets}

In this section we study the efficiency of the GMCO algorithms using the three real datasets described in \autoref{sec:datasets}. 
For each dataset, we examine both real and synthetic preferences, obtained as described in \autoref{sec:datasets}.
Also, we vary the group size $|\U|$ from 2 up to 32 users.

\begin{figure*}[]
\hspace{-1.0 cm}
\subfloat[I/O Operations]{\includegraphics[width=2.1in]{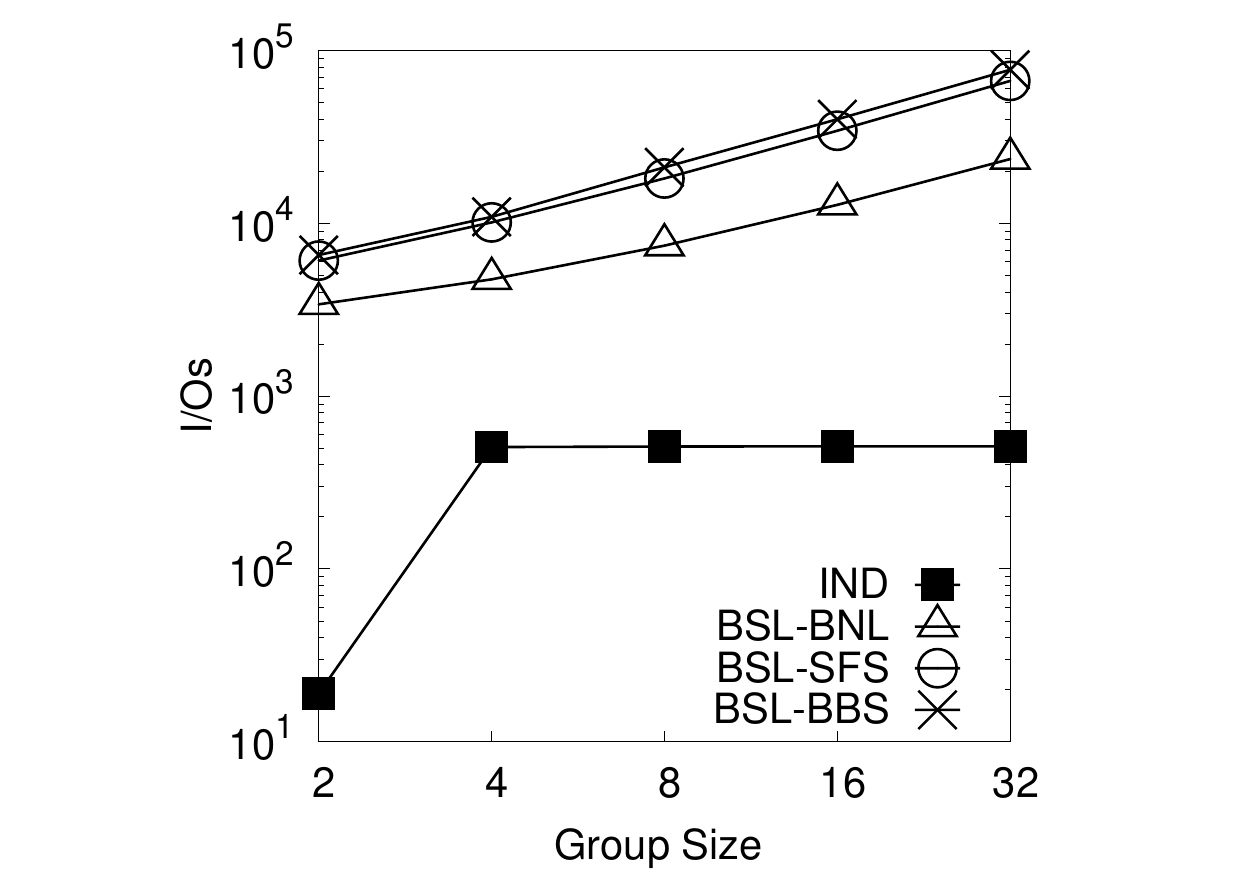}\label{fig:rest_r_io}}
\subfloat[Dom. Checks]{\hspace{-1.0 cm}\includegraphics[width=2.1in]{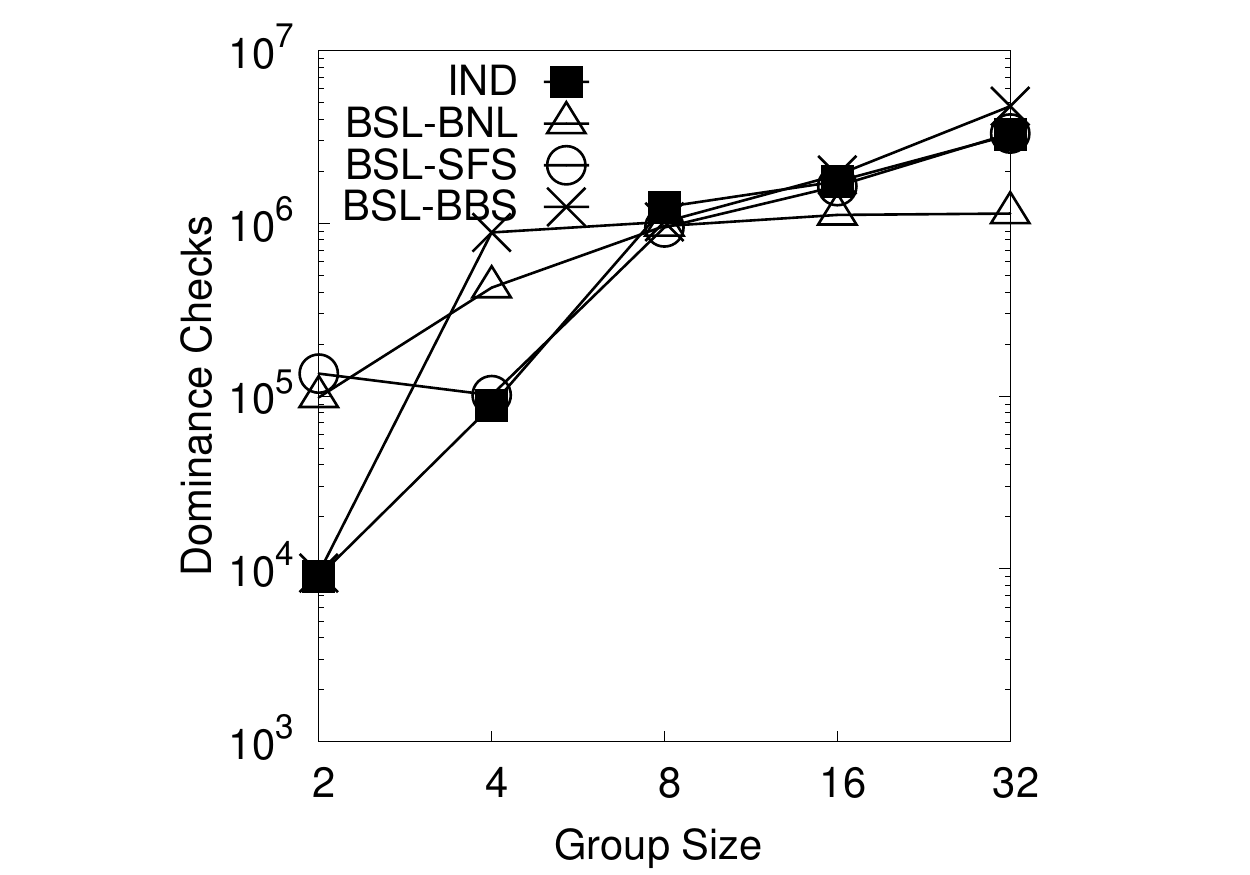}\label{fig:rest_r_dc}\hspace{-1.0 cm}}
\subfloat[Total Time]{\includegraphics[width=2.1in]{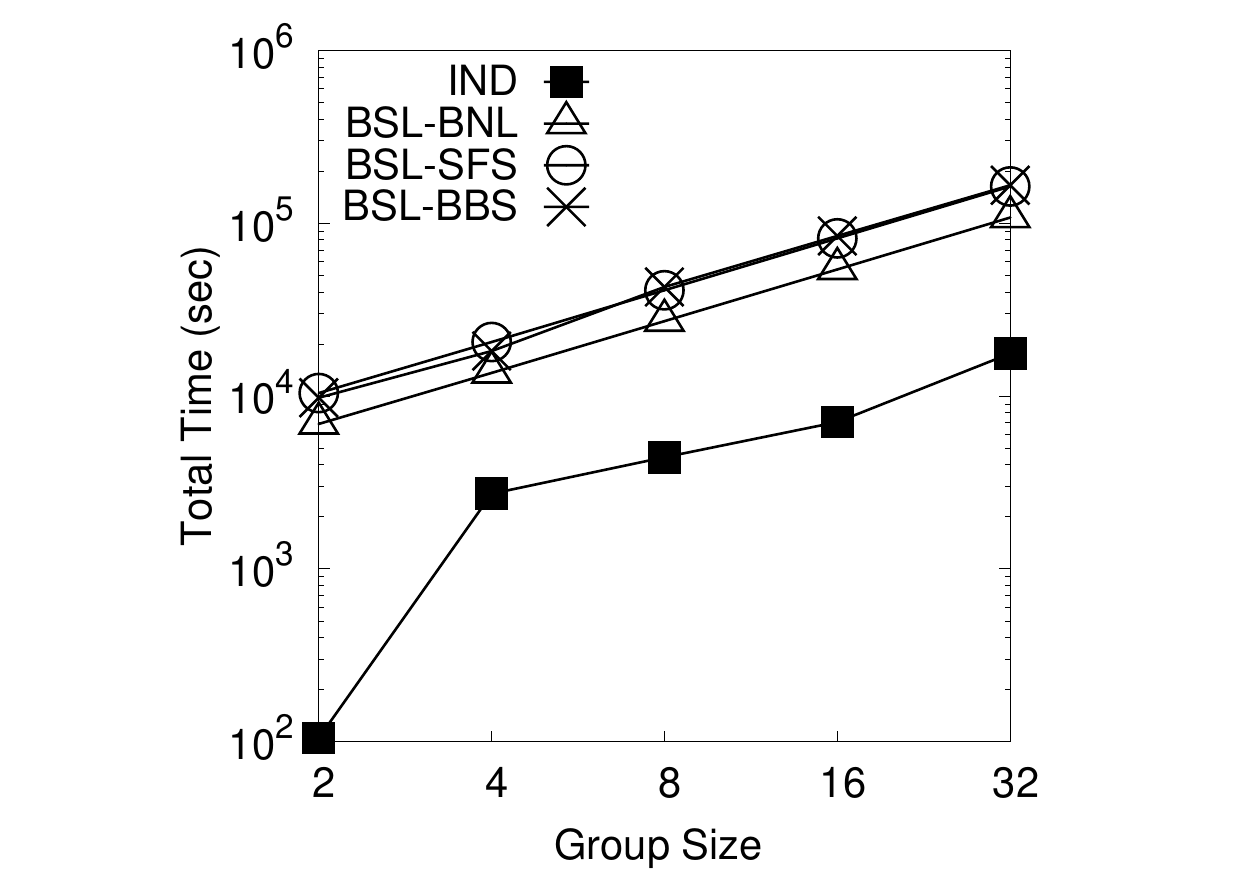}\label{fig:rest_r_time}}
\caption{GMCO algorithms, \textsf{RestaurantsF} (Real preferences): varying $|\U|$}
\label{fig:rest_r}
\end{figure*}

\begin{figure*}[]
\hspace{-1.0 cm}
\subfloat[I/O Operations]{\includegraphics[width=2.1in]{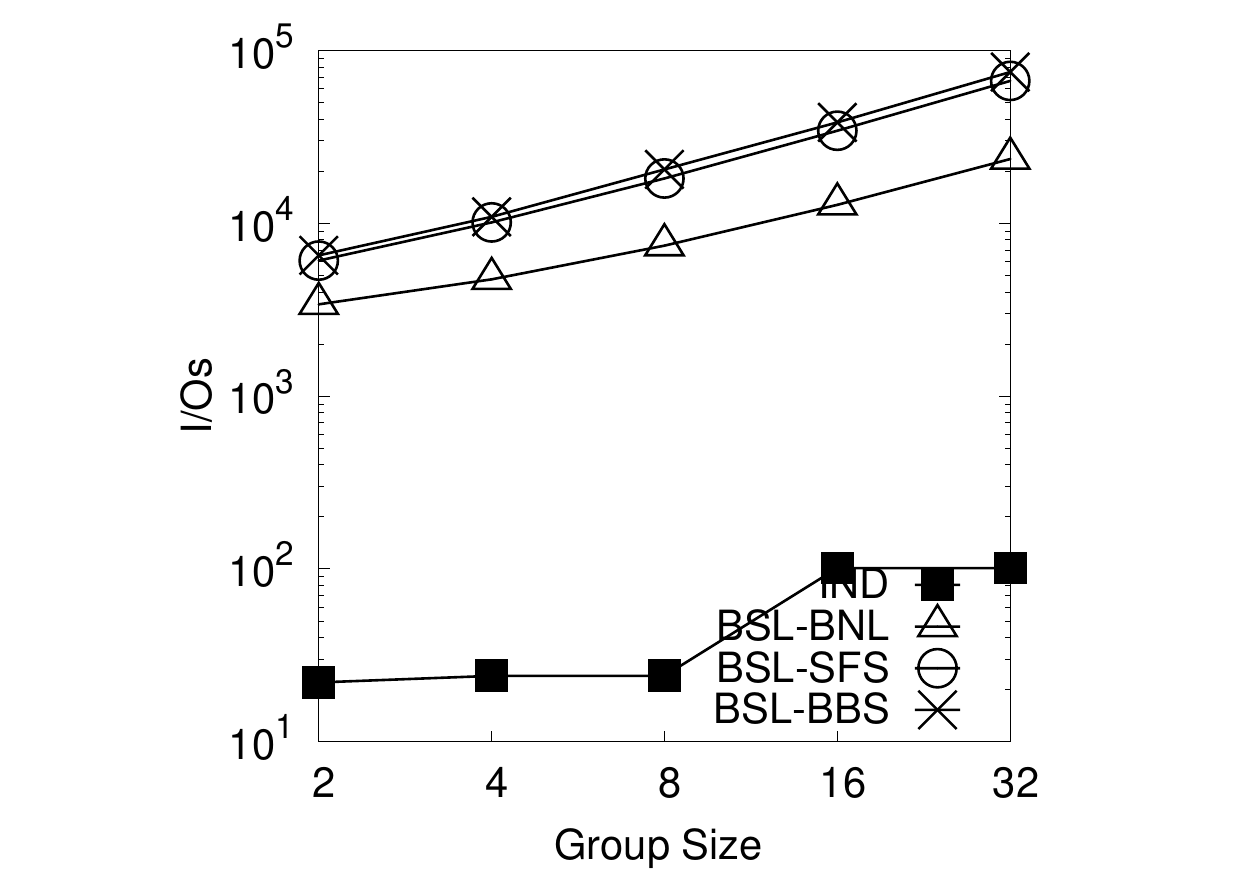}\label{fig:rest_s_io}}
\subfloat[Dom. Checks]{\hspace{-1.0 cm}\includegraphics[width=2.1in]{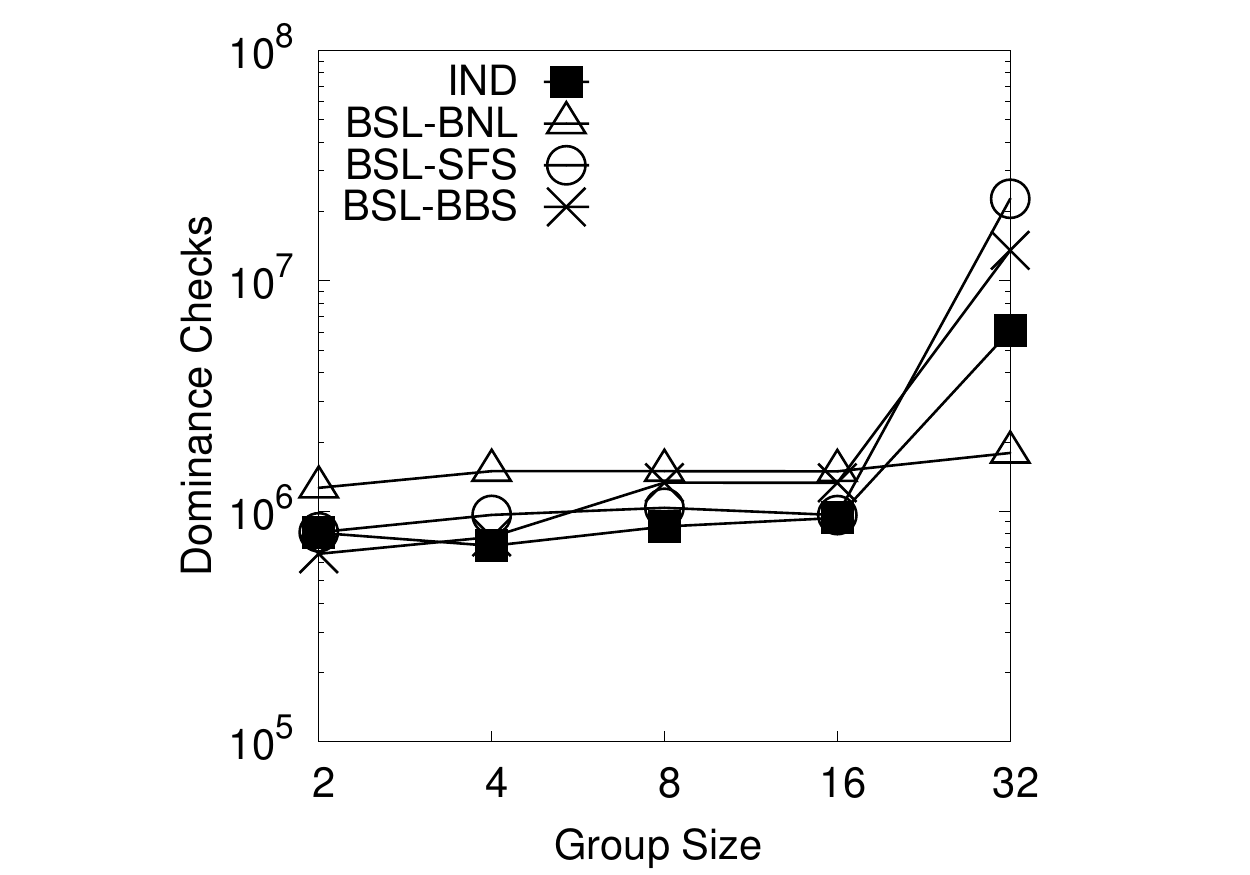}\label{fig:rest_s_dc}\hspace{-1.0 cm}}
\subfloat[Total Time]{\includegraphics[width=2.1in]{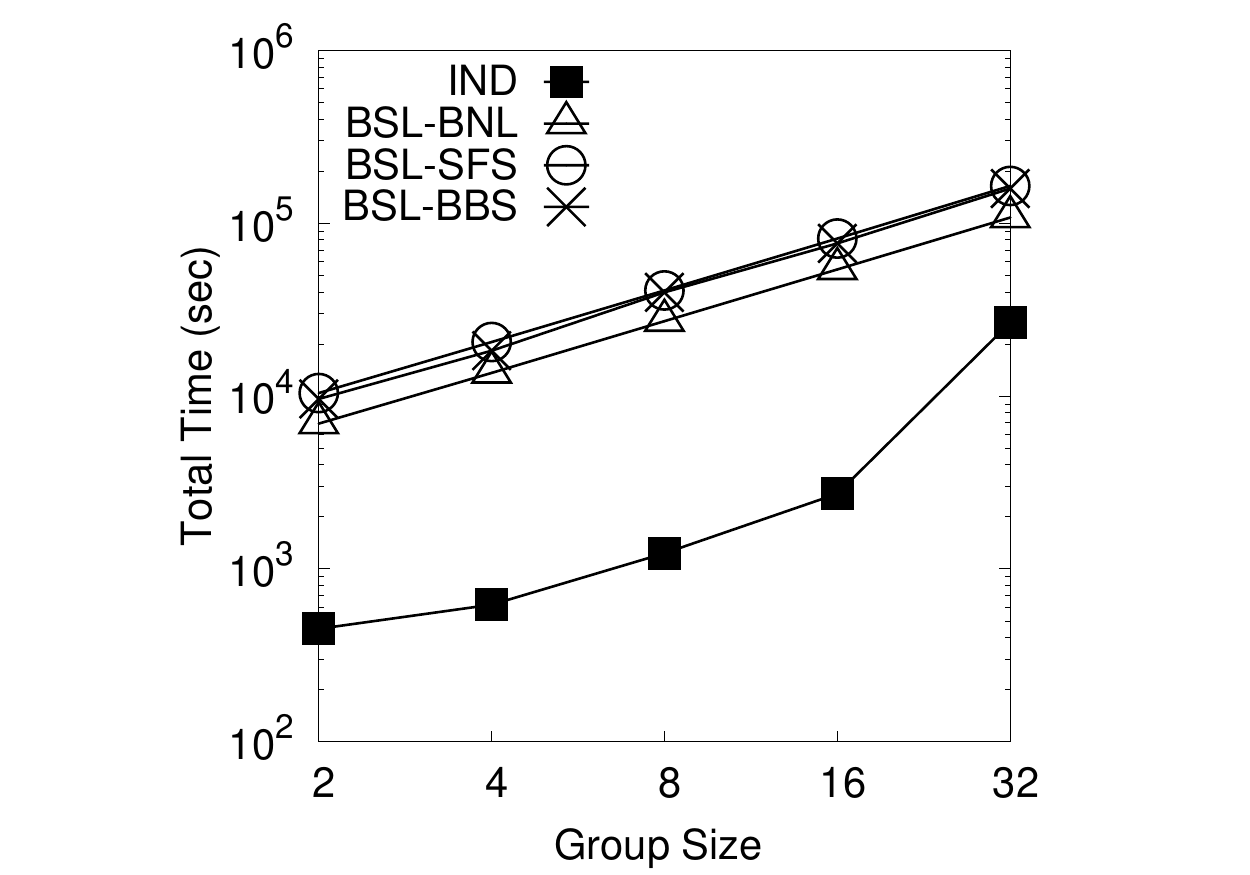}\label{fig:rest_s_time}}
\caption{GMCO algorithms, \textsf{RestaurantsF} (Synthetic preferences): varying $|\U|$}
\label{fig:rest_s}
\end{figure*}

\begin{figure*}[]
\hspace{-1.0 cm}
\subfloat[I/O Operations]{\includegraphics[width=2.1in]{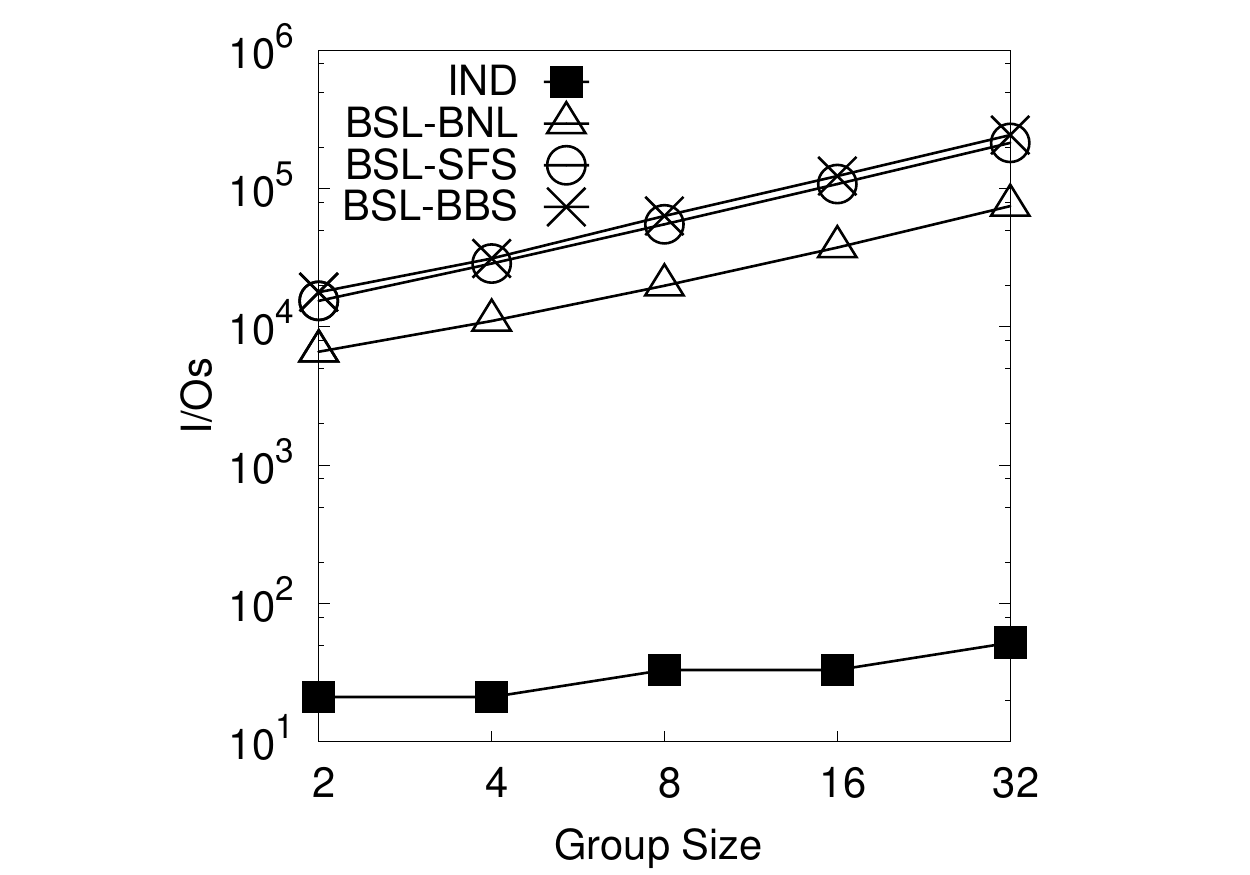}\label{fig:acm_r_io}}
\subfloat[Dom. Checks]{\hspace{-1.0 cm}\includegraphics[width=2.1in]{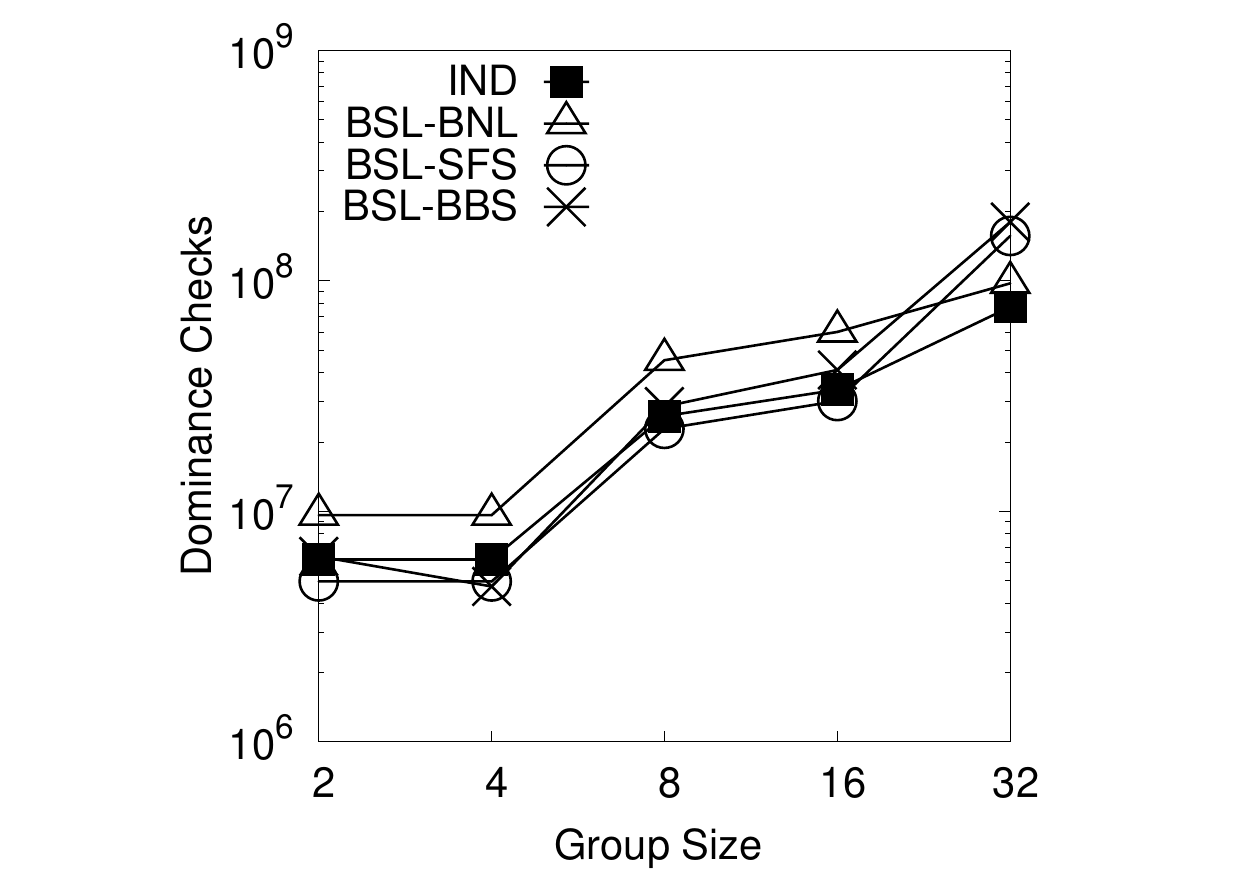}\label{fig:acm_r_dc}\hspace{-1.0 cm}}
\subfloat[Total Time]{\includegraphics[width=2.1in]{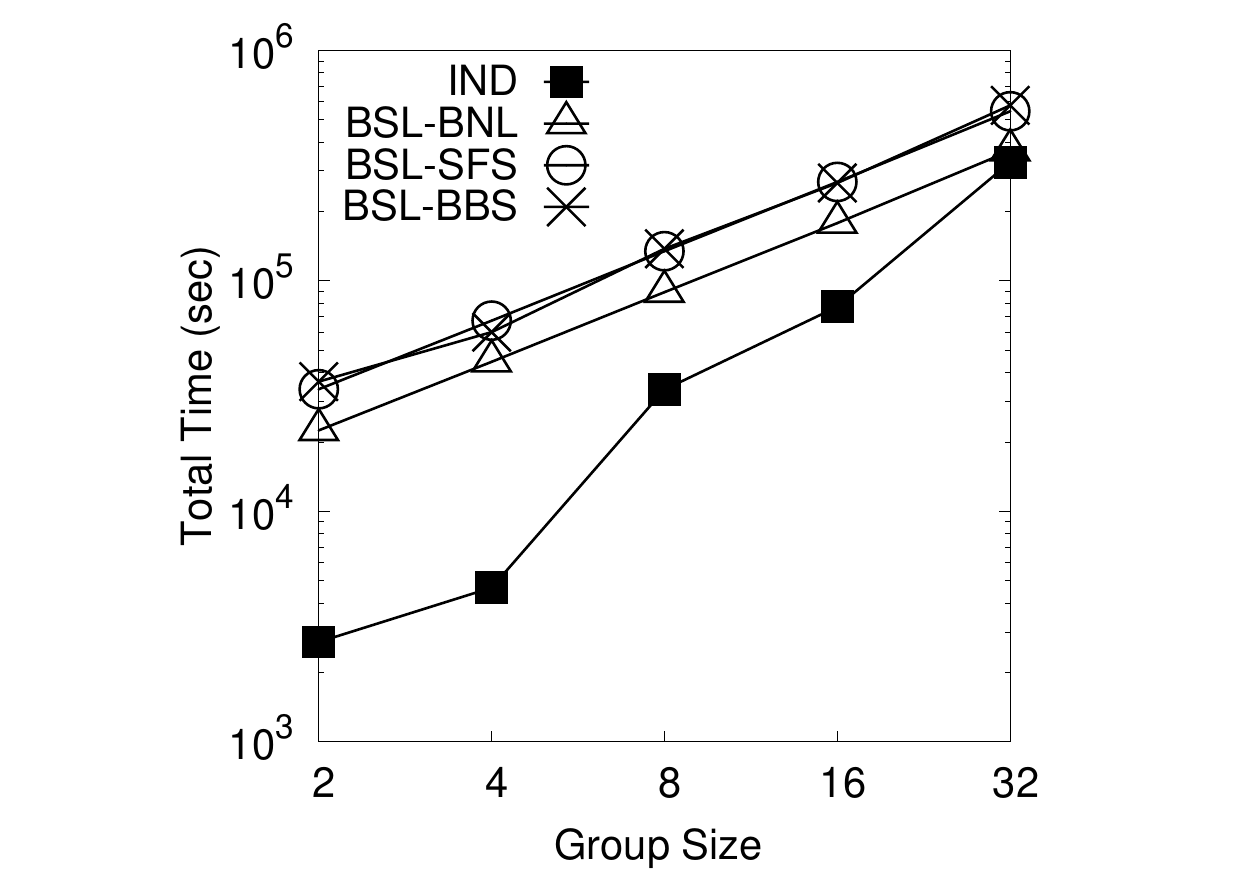}\label{fig:acm_r_time}}
\vspace{-5pt}
\caption{GMCO algorithms, \textsf{ACM} (Real preferences): varying $|\U|$}
\label{fig:acm_r}
\end{figure*}

\begin{figure*}[h!]
\hspace{-1.0 cm}
\subfloat[I/O Operations]{\includegraphics[width=2.1in]{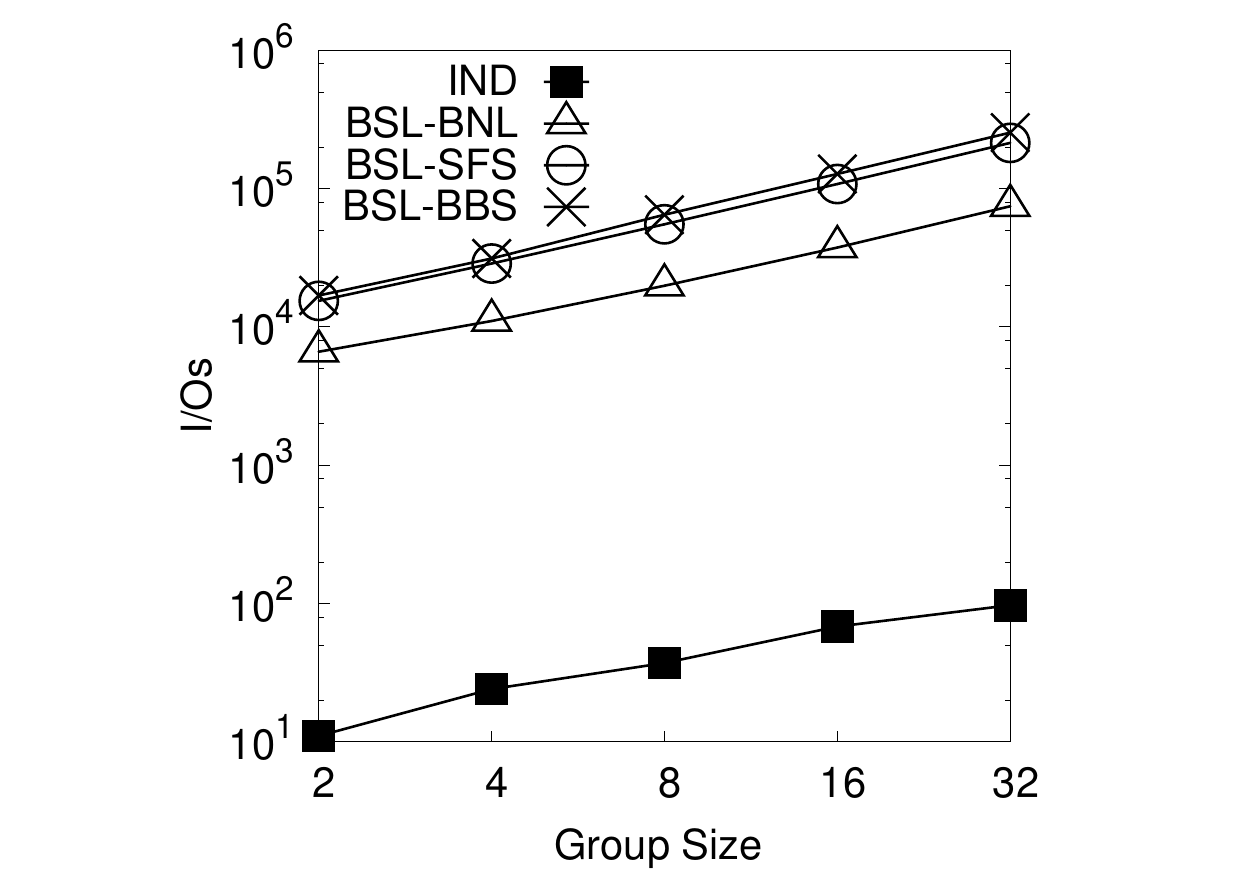}\label{fig:acm_s_io}}
\subfloat[Dom. Checks]{\hspace{-1.0 cm}\includegraphics[width=2.1in]{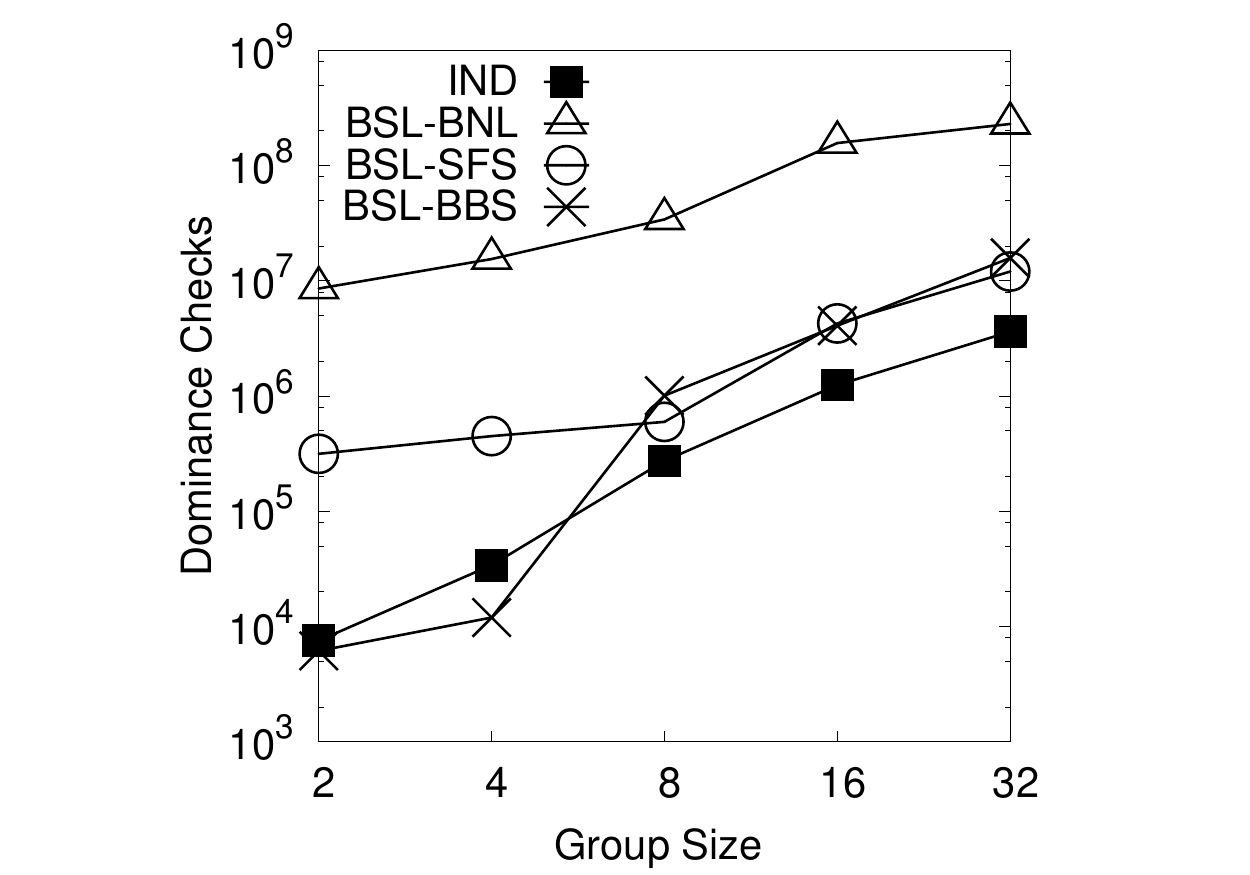}\label{fig:acm_s_dc}\hspace{-1.0 cm}}
\subfloat[Total Time]{\includegraphics[width=2.1in]{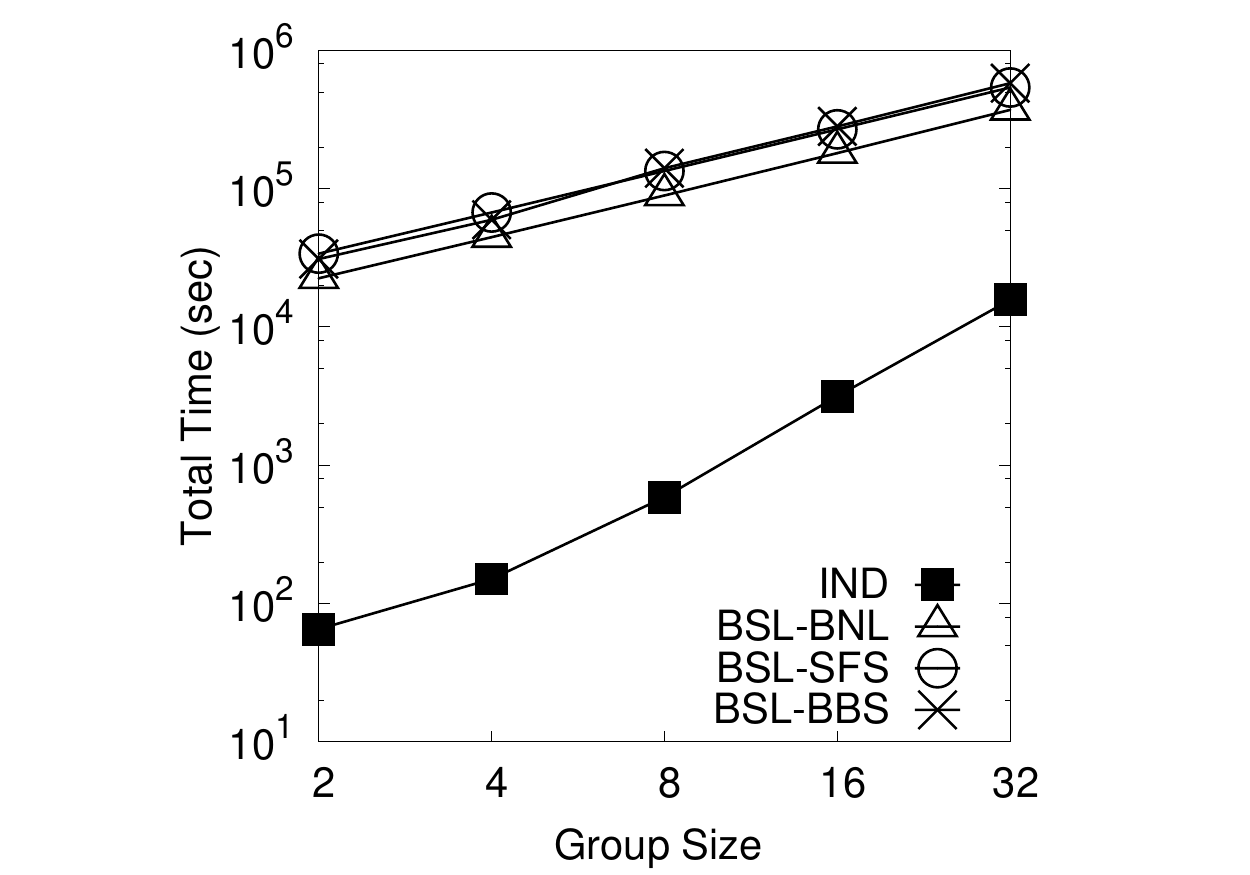}\label{fig:acm_s_time}}
\caption{GMCO algorithms, \textsf{ACM} (Synthetic preferences): varying $|\U|$}
\label{fig:acm_s}
\end{figure*}

\begin{figure*}[]
\hspace{-1.0 cm}
\subfloat[I/O Operations]{\includegraphics[width=2.1in]{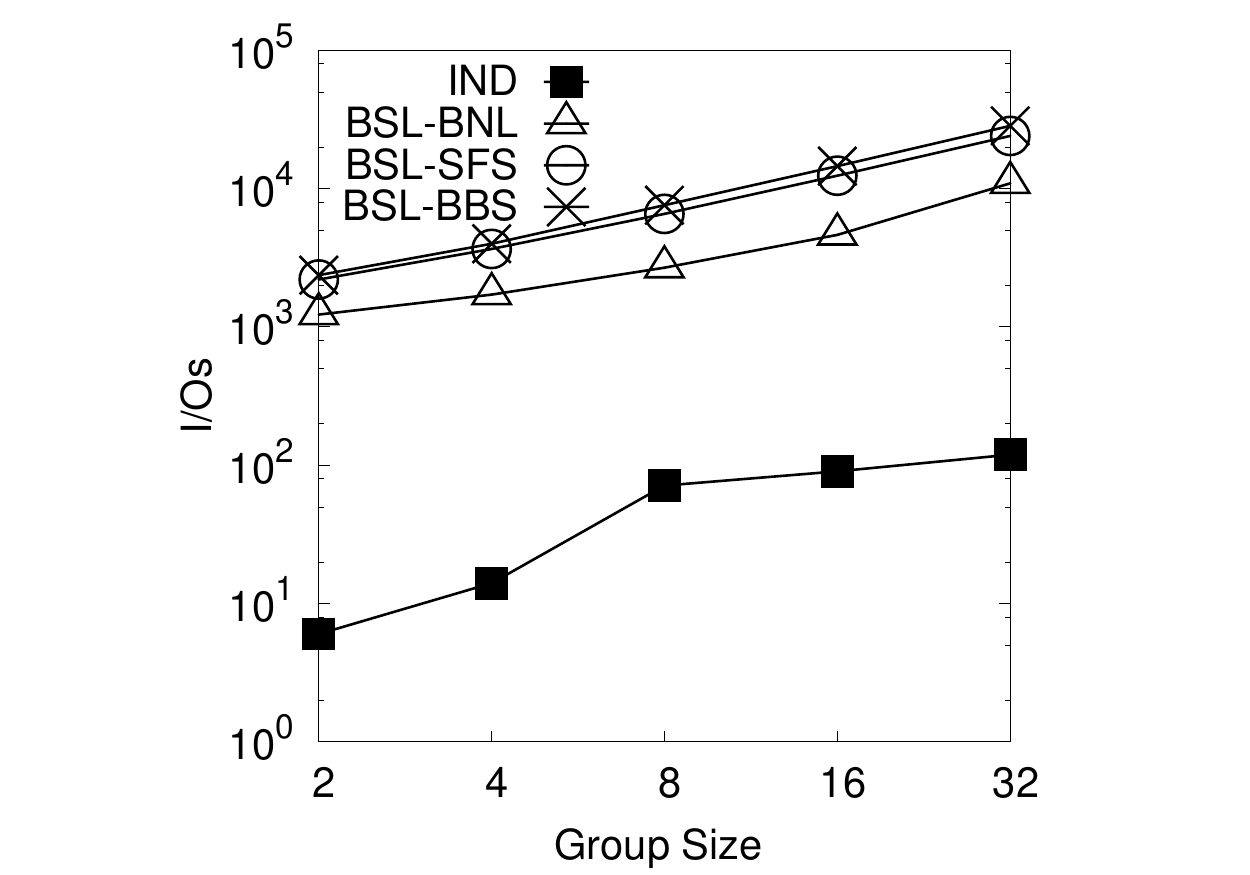}\label{fig:cars_r_io}}
\subfloat[Dom. Checks]{\hspace{-1.0 cm}\includegraphics[width=2.1in]{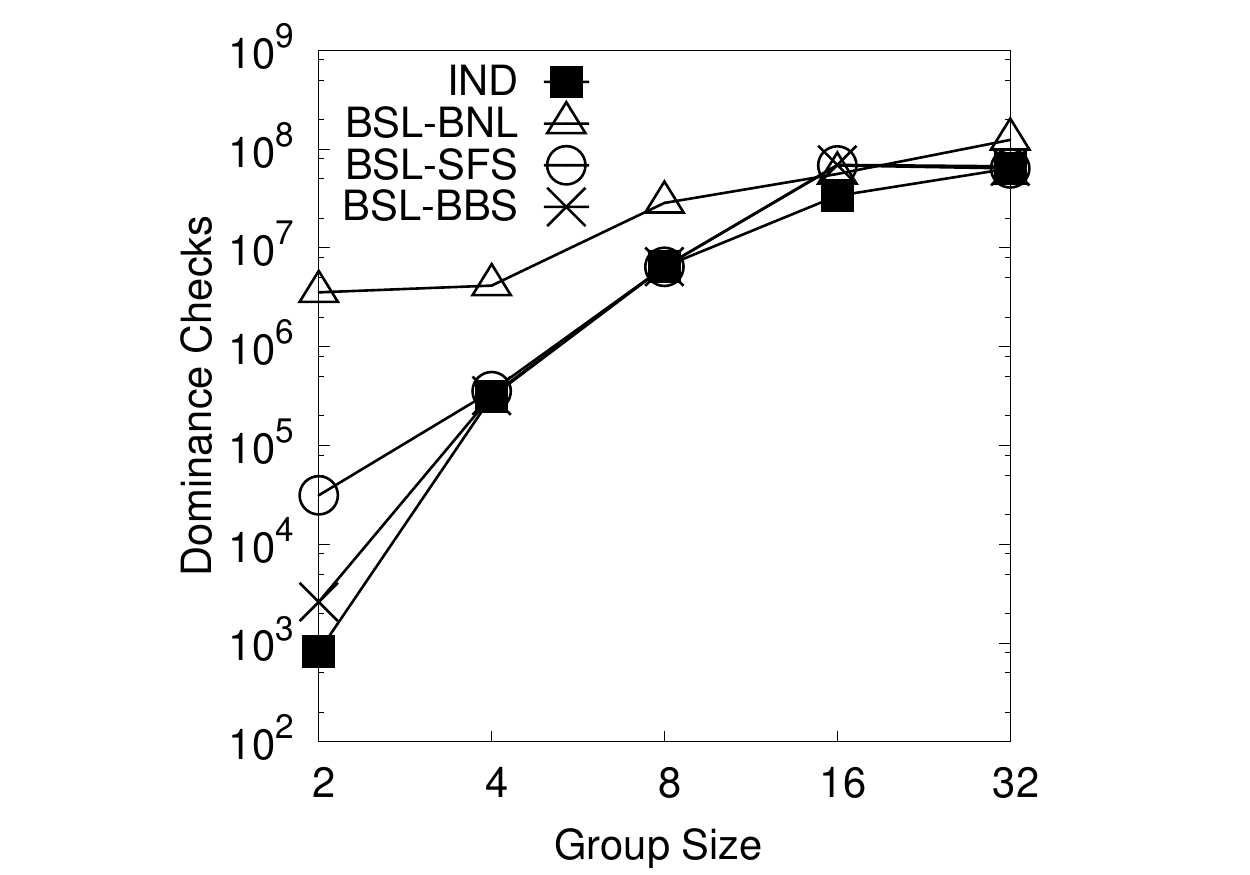}\label{fig:cars_r_dc}\hspace{-1.0 cm}}
\subfloat[Total Time]{\includegraphics[width=2.1in]{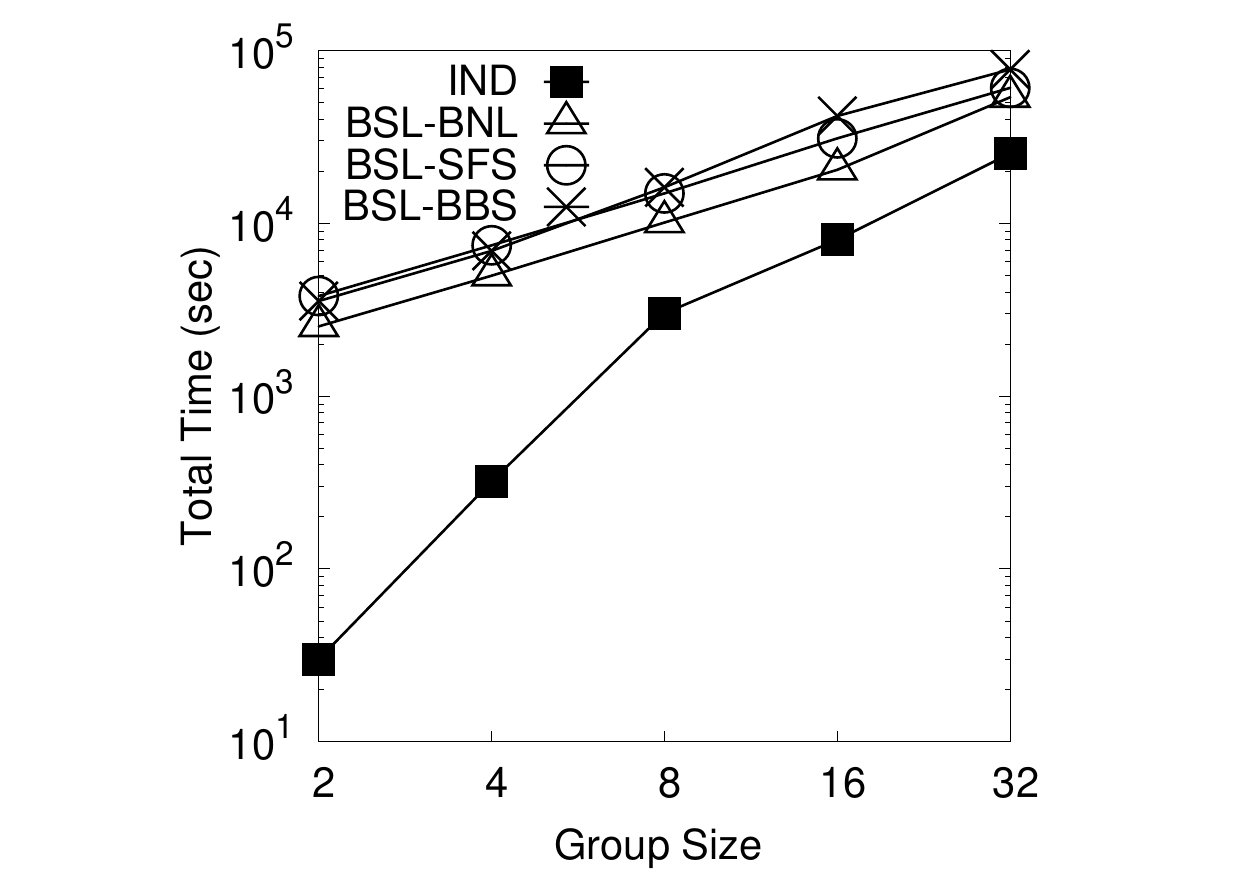}\label{fig:cars_r_time}}
\caption{GMCO algorithms, \textsf{Cars} (Real preferences): varying $|\U|$}
\label{fig:cars_r}
\vspace{-5pt}
\end{figure*}

\begin{figure*}[]
\hspace{-1.0 cm}
\subfloat[I/O Operations]{\includegraphics[width=2.1in]{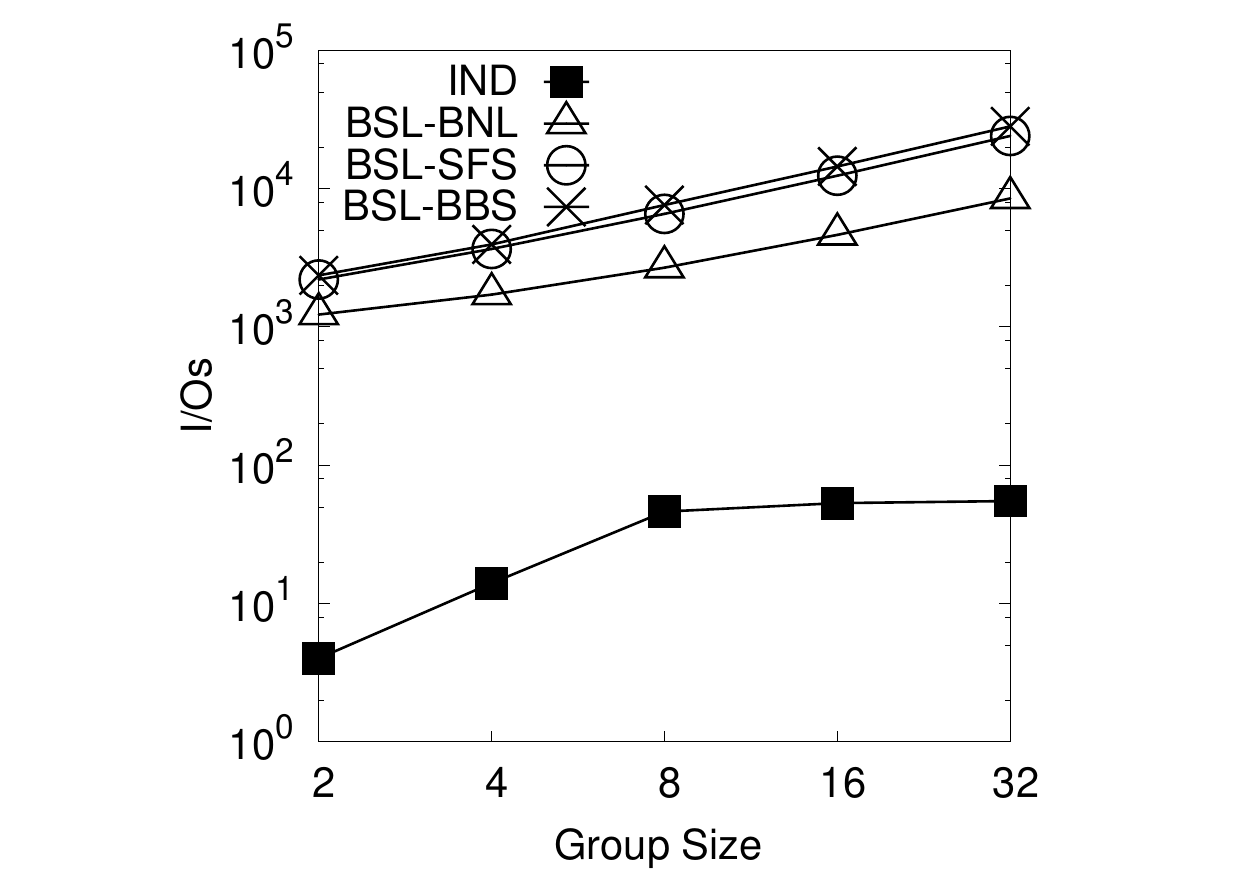}\label{fig:cars_s_io}}
\subfloat[Dom. Checks]{\hspace{-1.0 cm}\includegraphics[width=2.1in]{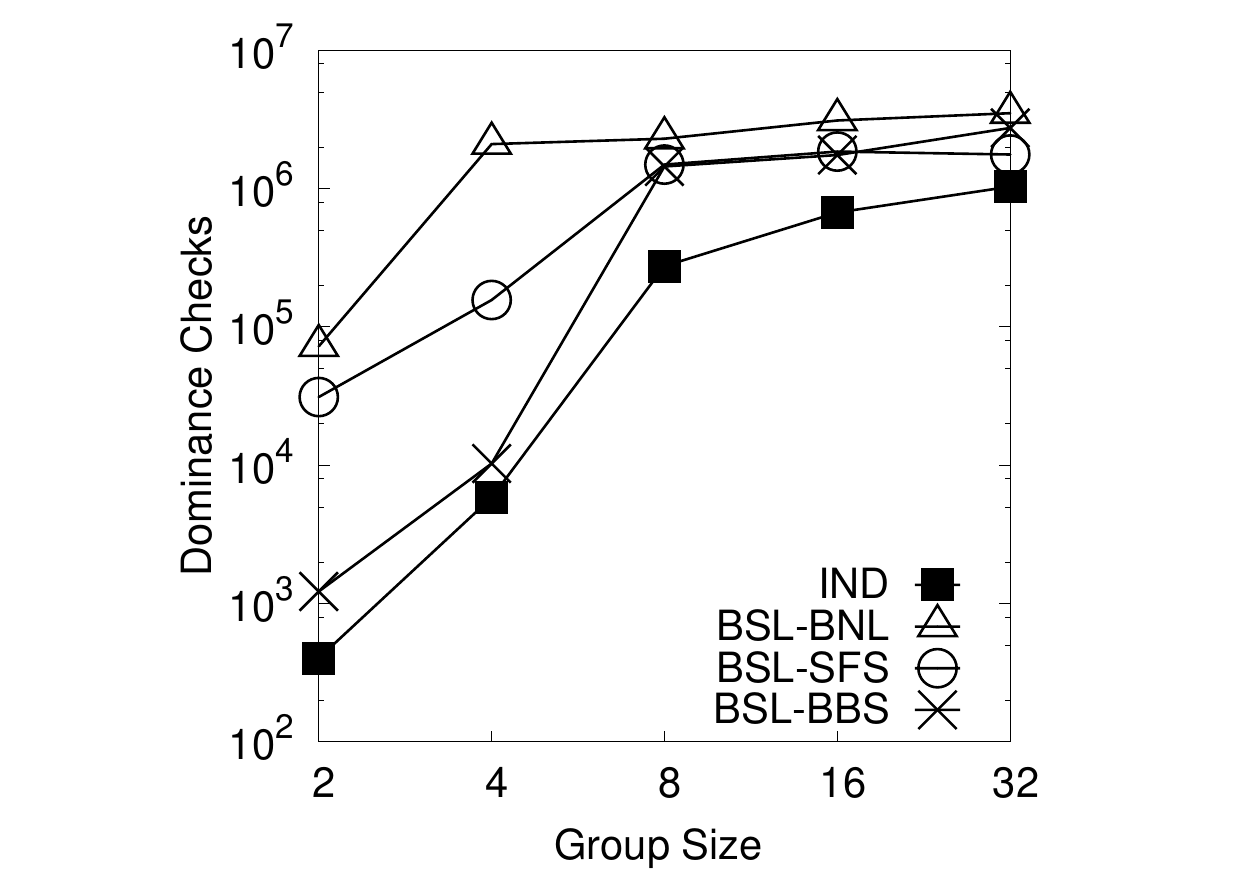}\label{fig:cars_s_dc}\hspace{-1.0 cm}}
\subfloat[Total Time]{\includegraphics[width=2.1in]{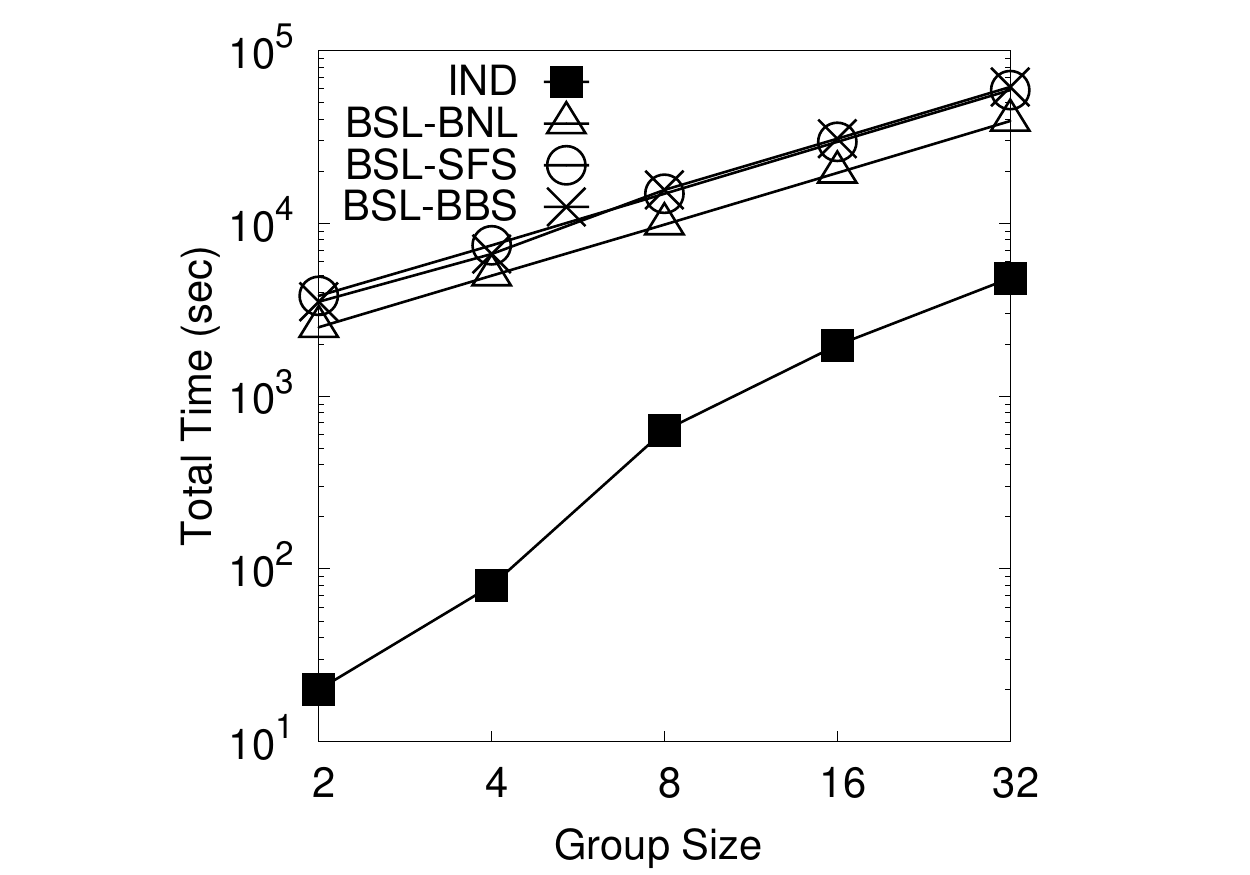}\label{fig:cars_s_time}}
\caption{GMCO algorithms, \textsf{Cars} (Synthetic preferences): varying $|\U|$}
\label{fig:cars_s}
\vspace{-5pt}
\end{figure*}

Figures~\ref{fig:rest_r}~\&~\ref{fig:rest_s} present the result for \textsf{RestaurantsF} dataset, 
for real and synthetic preferences, respectively. 
Similarly, Figures~\ref{fig:acm_r}~\&~\ref{fig:acm_s} present the result for \textsf{ACM} dataset, 
and Figures~\ref{fig:cars_r}~\&~\ref{fig:cars_s} for \textsf{Cars} dataset. 
As we can observe, the performance of the examined methods is almost similar for all datasets, real and synthetic. 
Also, similar performance is observed in real and synthetic user preferences. 
In most cases, IND outperforms the BSL methods by at least an order of magnitude in terms of I/Os and total time.
Additionally, IND performs less dominance checks than the BSL methods in almost all cases. 

Regarding BSL methods, BSL-BNL outperforms the others in terms of I/Os and total time; 
while BSL-SFS and BNL-BBS have the almost the same performance. 
Regarding the number of dominance checks, for less than 16 users 
BSL-BNL performs more dominance checks than other BSL methods; while for 
32 users, in many cases (Figures \ref{fig:rest_r_dc}, \ref{fig:rest_s_dc}, \ref{fig:acm_r_dc}) BSL-BNL performs the fewest dominance checks from BSL methods. 
Finally, for less than 8 users, BNL-BBS perform fewer dominance checks than other 
BSL methods.

\subsection{Efficiency of the {\large\textit{{p}}}-GMCO Algorithms}
\label{sec:e_pmcp}

In this section, we investigate the performance of the $p$-GMCO algorithms (\autoref{sec:pGMCO}). 
For the $p$-GMCO problem, we implement the respective extensions of all
algorithms (IND and BSL variants), distinguished by a $p$ prefix. 
As before, we measure the number of I/O operations, dominance checks and the total time.
In the following experiments, we use the three real datasets and vary the number of
users from 2 up to 1024, while $p=30\%$.
Also, we also vary the parameter $p$ from 10\% up to 50\%. 
However, the performance of all methods (in terms of I/Os and total time) 
remains unaffected by $p$; hence, the relevant figures are omitted.

\begin{figure*}[]
\hspace{-1.0 cm}
\subfloat[I/O Operations]{\includegraphics[width=2.1in]{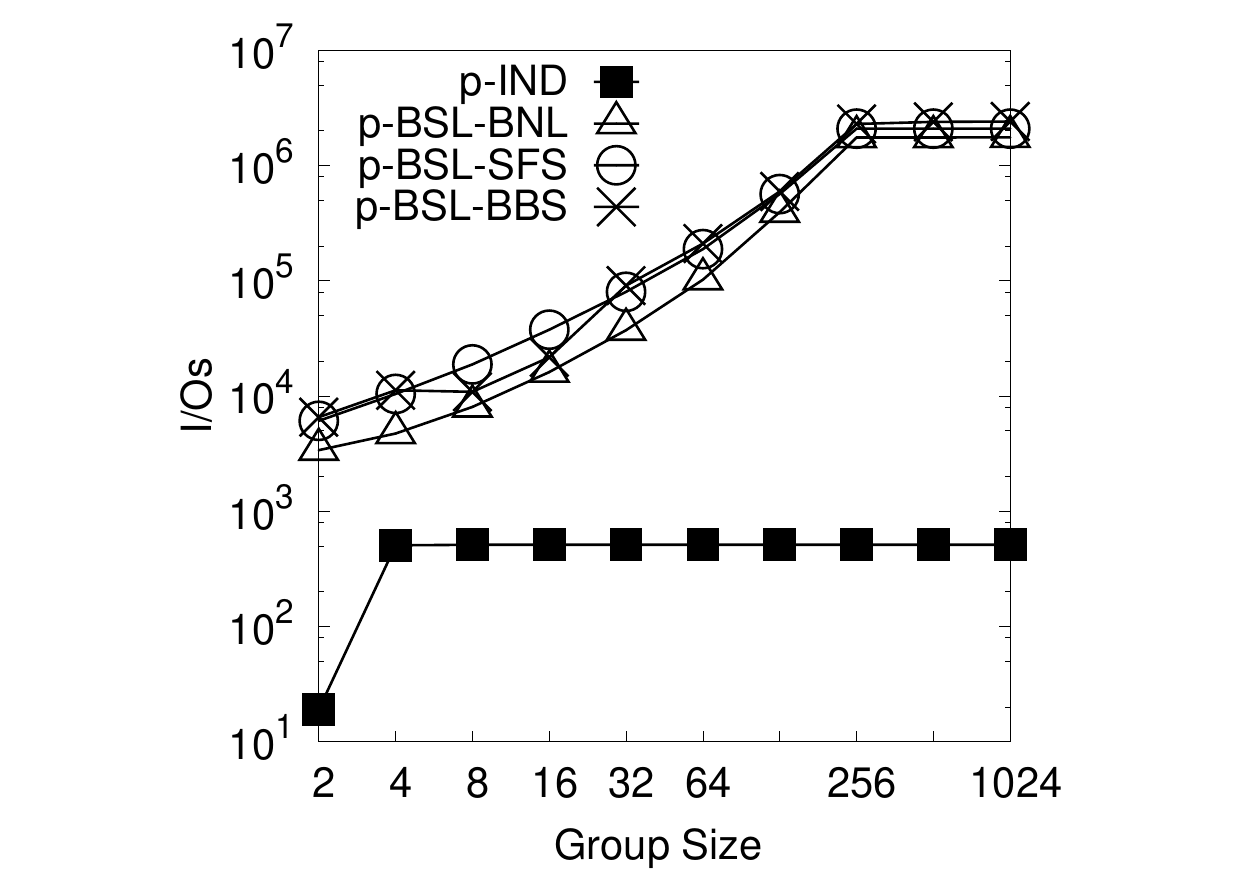}\label{fig:rest_rp_io}}
\subfloat[Dom. Checks]{\hspace{-1.0 cm}\includegraphics[width=2.1in]{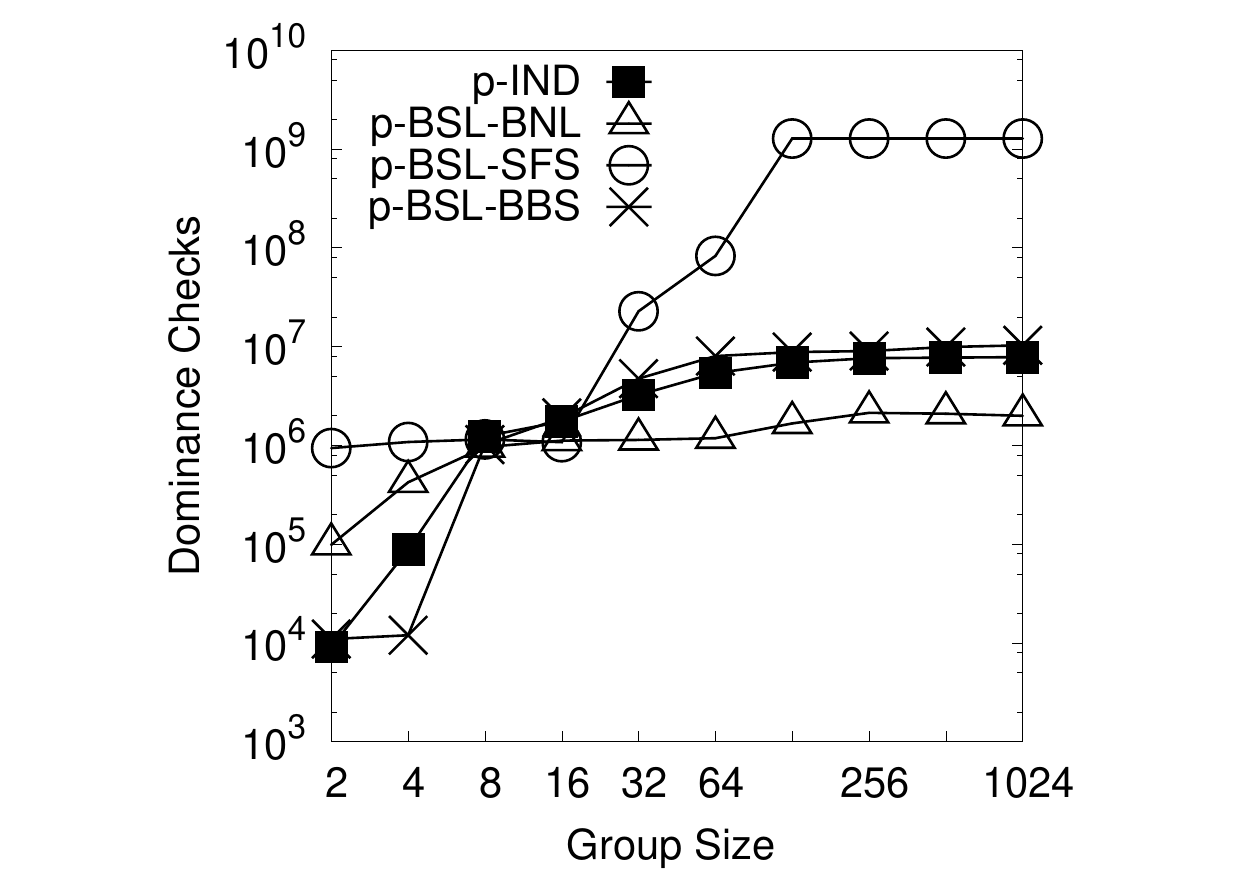}\label{fig:rest_rp_dc}\hspace{-1.0 cm}}
\subfloat[Total Time]{\includegraphics[width=2.1in]{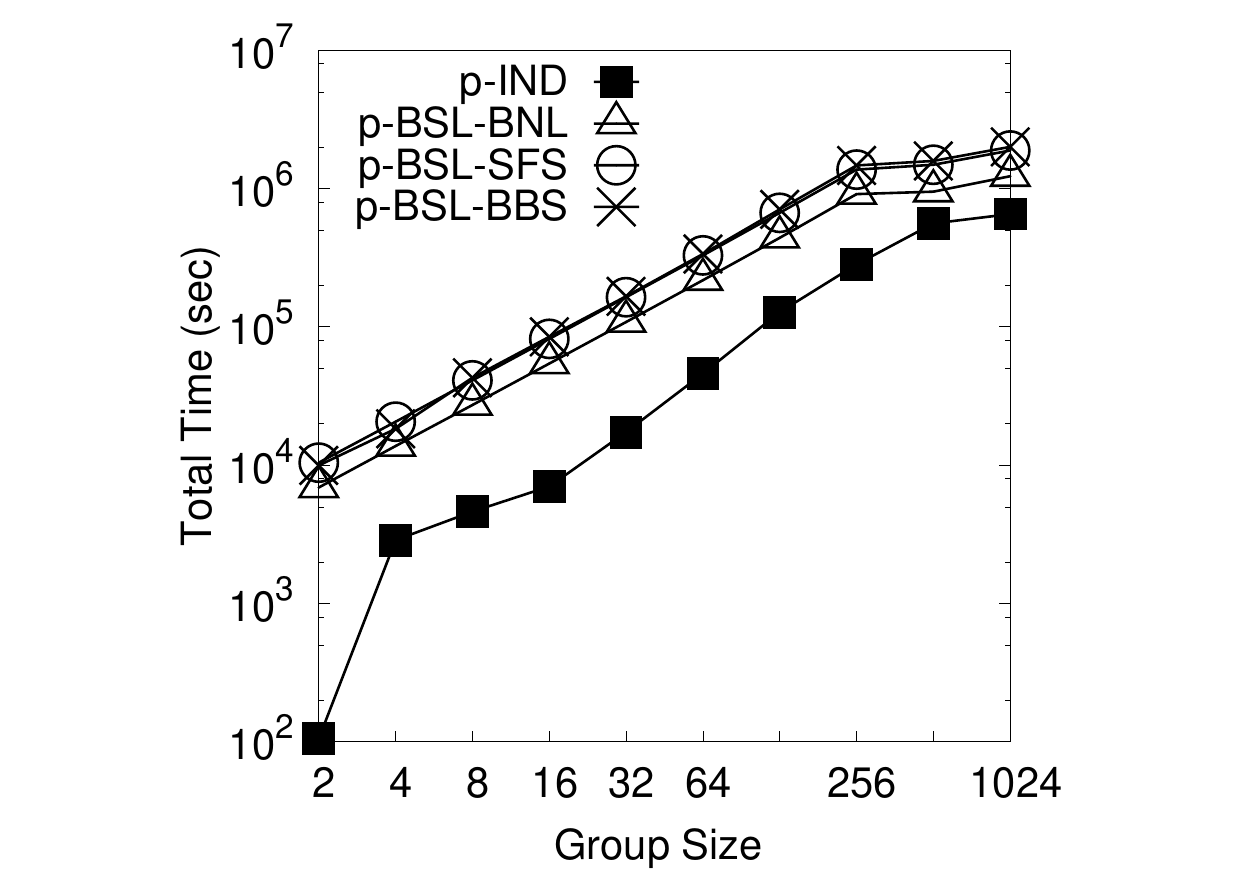}\label{fig:rest_rp_time}}
\caption{$p$-GMCO algorithms, \textsf{RestaurantsF} (Real preferences): varying $|\U|$}
\label{fig:rest_rp}
\vspace{-5pt}
\end{figure*}

\begin{figure*}[]
\hspace{-1.0 cm}
\subfloat[I/O Operations]{\includegraphics[width=2.1in]{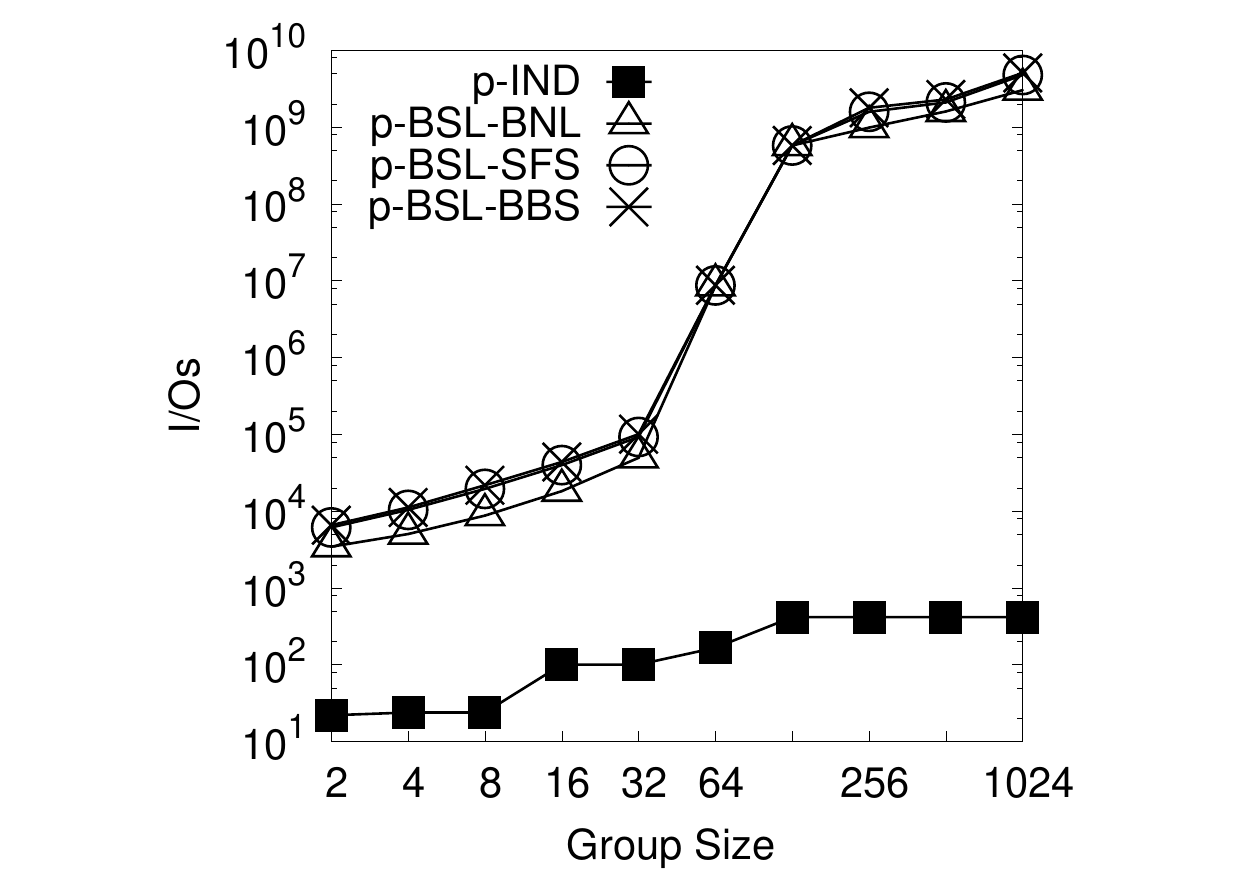}\label{fig:rest_sp_io}}
\subfloat[Dom. Checks]{\hspace{-1.0 cm}\includegraphics[width=2.1in]{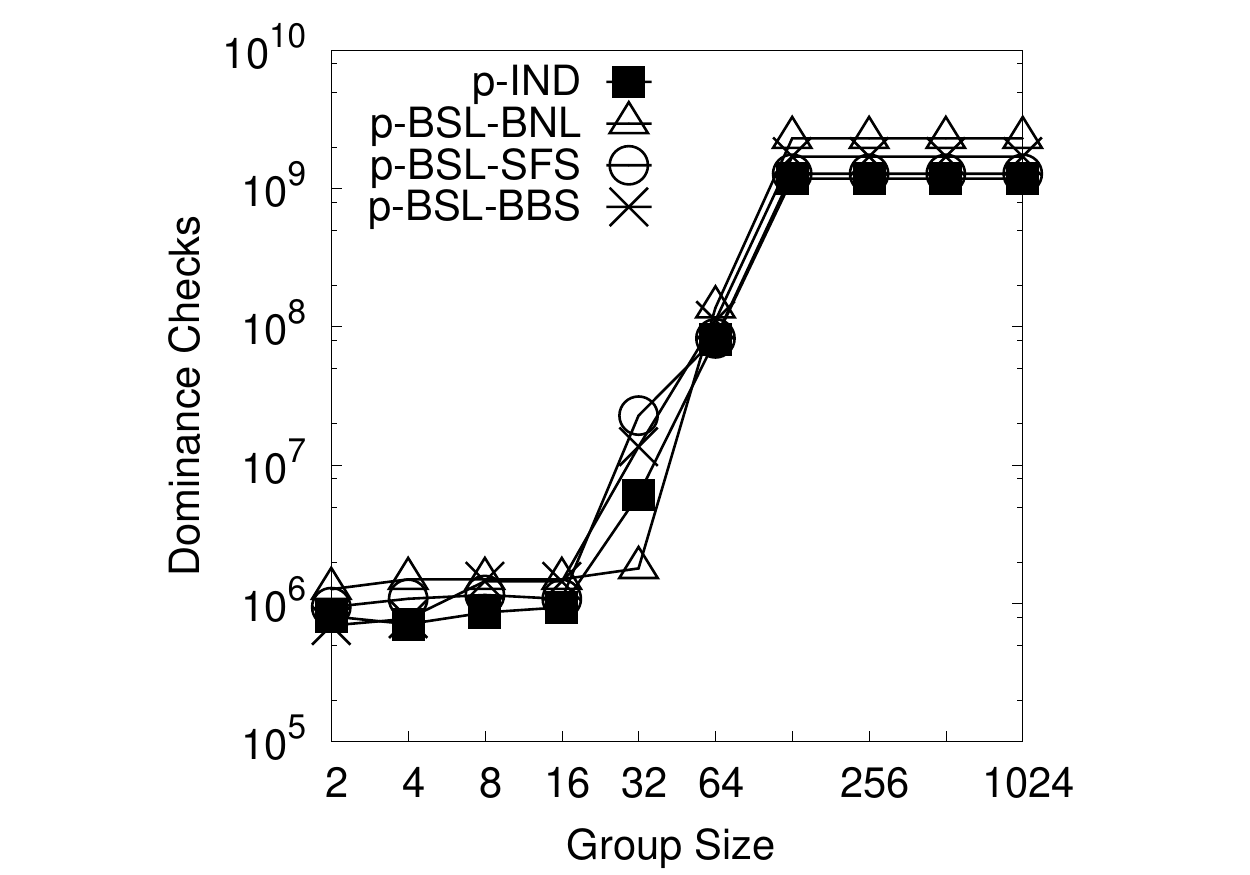}\label{fig:rest_sp_dc}\hspace{-1.0 cm}}
\subfloat[Total Time]{\includegraphics[width=2.1in]{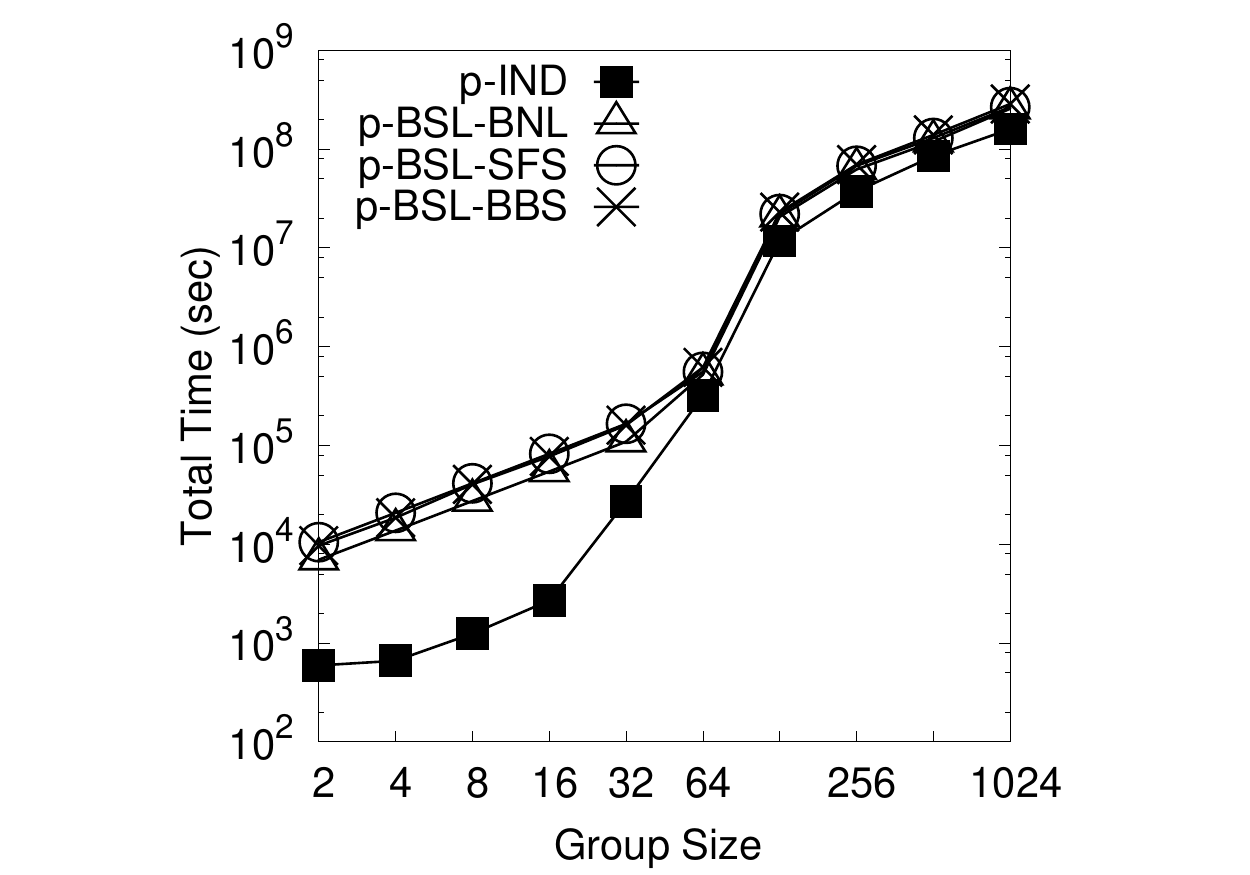}\label{fig:rest_sp_time}}
\caption{$p$-GMCO algorithms, \textsf{RestaurantsF} (Synthetic preferences): varying $|\U|$}
\label{fig:rest_sp}
\vspace{-5pt}
\end{figure*}

\begin{figure*}[]
\hspace{-1.0 cm}
\subfloat[I/O Operations]{\includegraphics[width=2.1in]{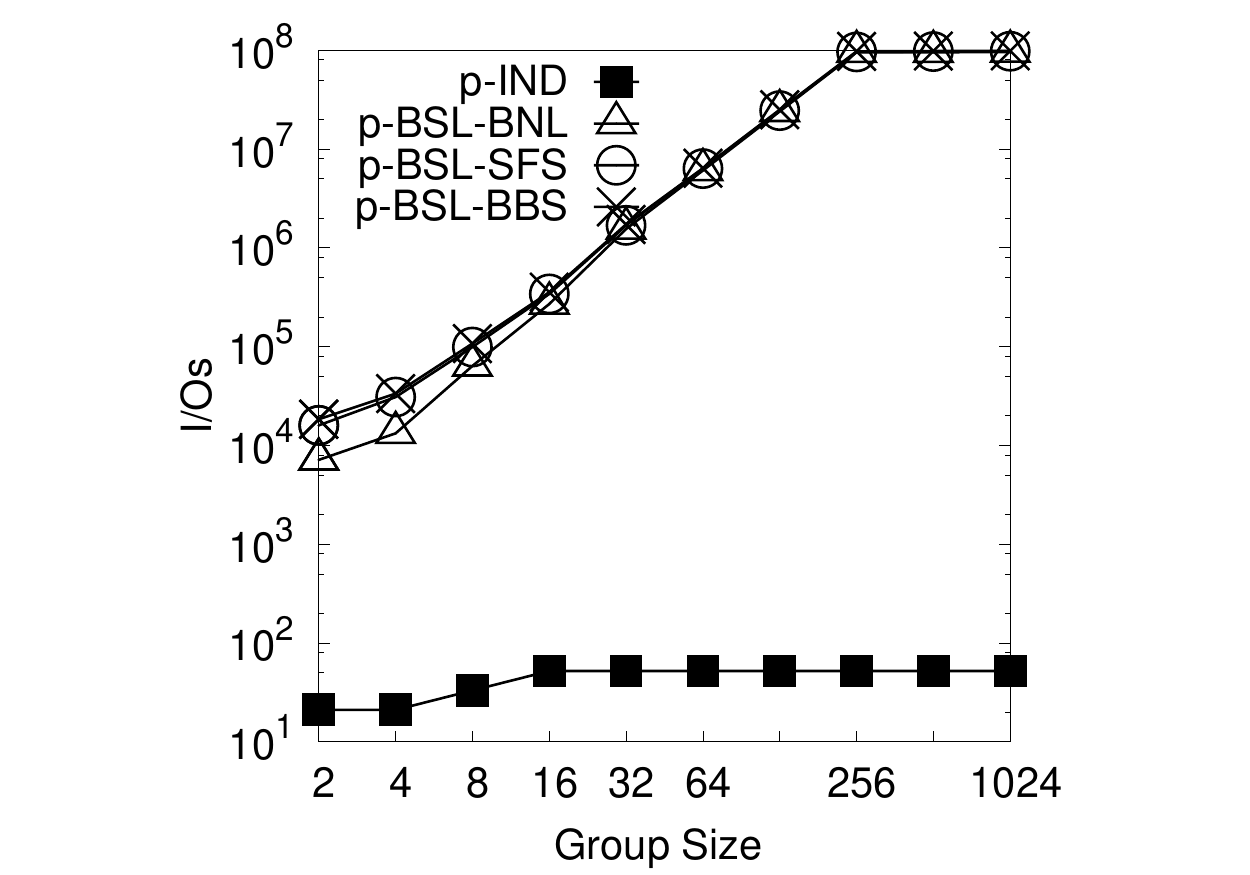}\label{fig:acm_rp_io}}
\subfloat[Dom. Checks]{\hspace{-1.0 cm}\includegraphics[width=2.1in]{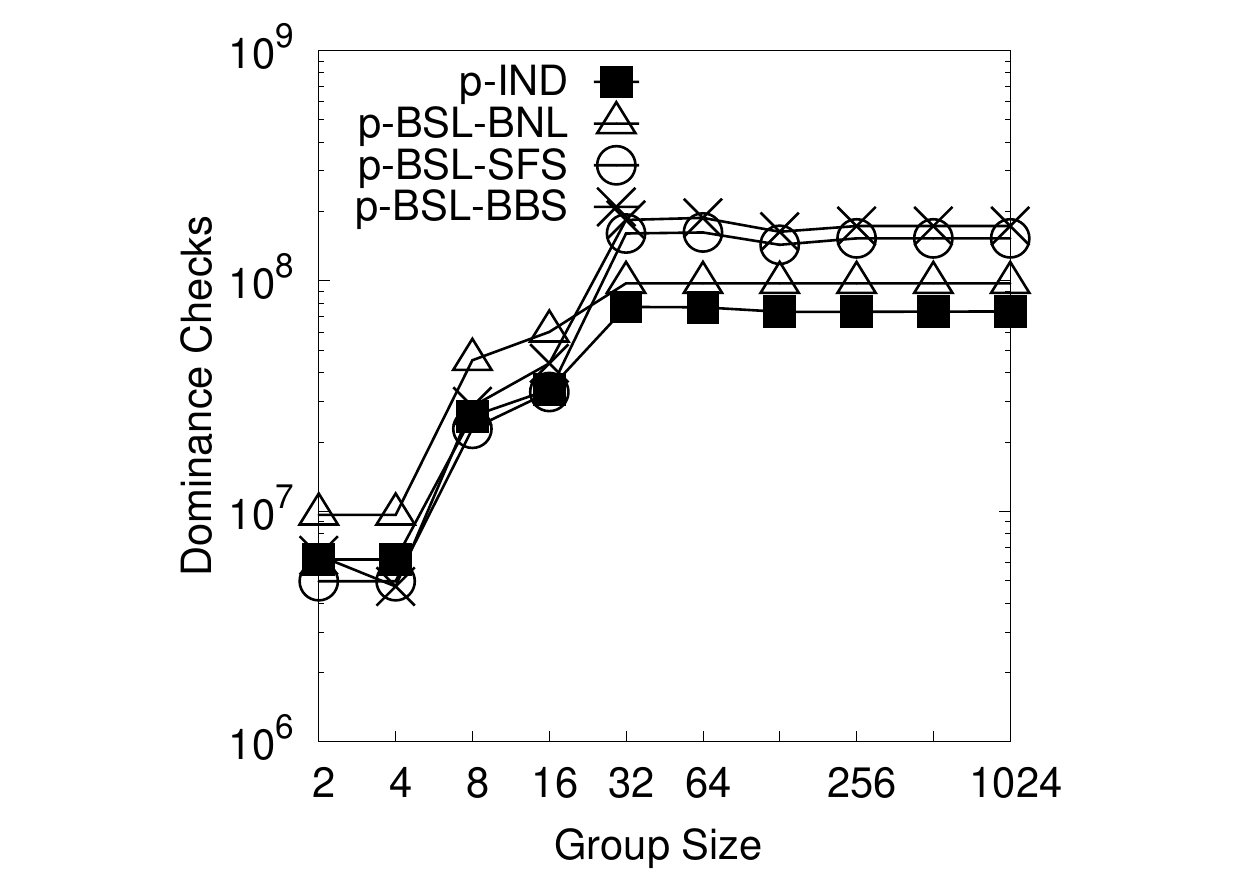}\label{fig:acm_rp_dc}\hspace{-1.0 cm}}
\subfloat[Total Time]{\includegraphics[width=2.1in]{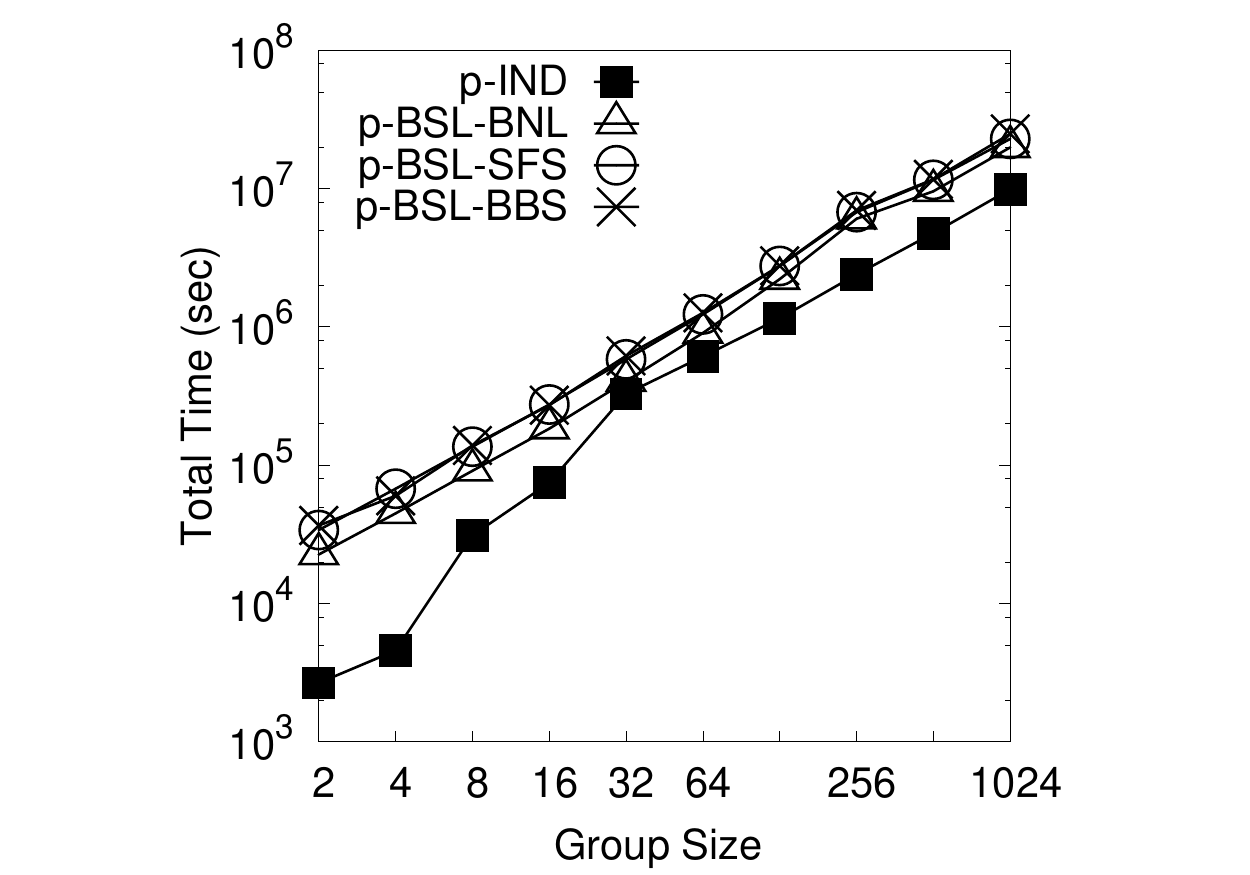}\label{fig:acm_rp_time}}
\caption{$p$-GMCO algorithms, \textsf{ACM} (Real preferences): varying $|\U|$}
\label{fig:acm_rp}
\end{figure*}

\begin{figure*}[]
\hspace{-1.0 cm}
\subfloat[I/O Operations]{\includegraphics[width=2.1in]{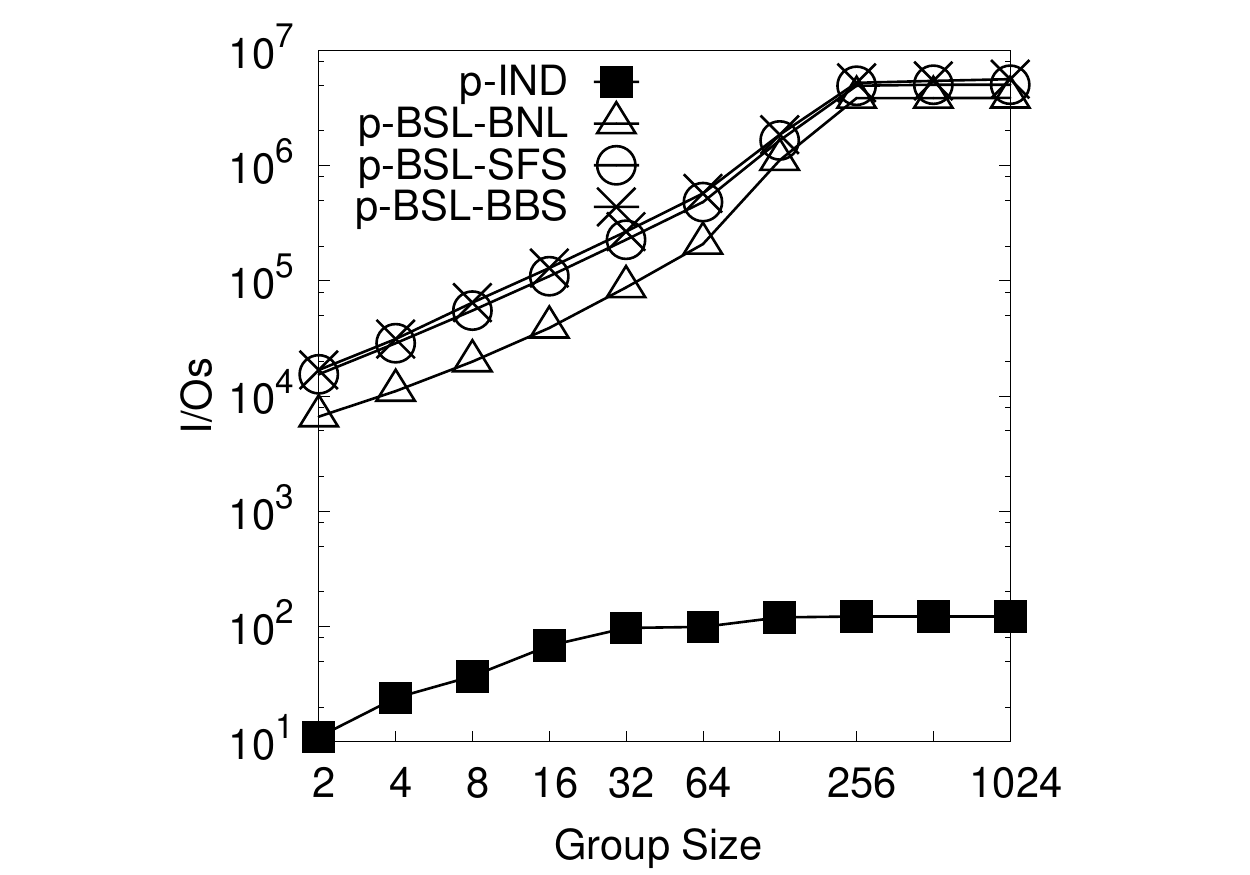}\label{fig:acm_sp_io}}
\subfloat[Dom. Checks]{\hspace{-1.0 cm}\includegraphics[width=2.1in]{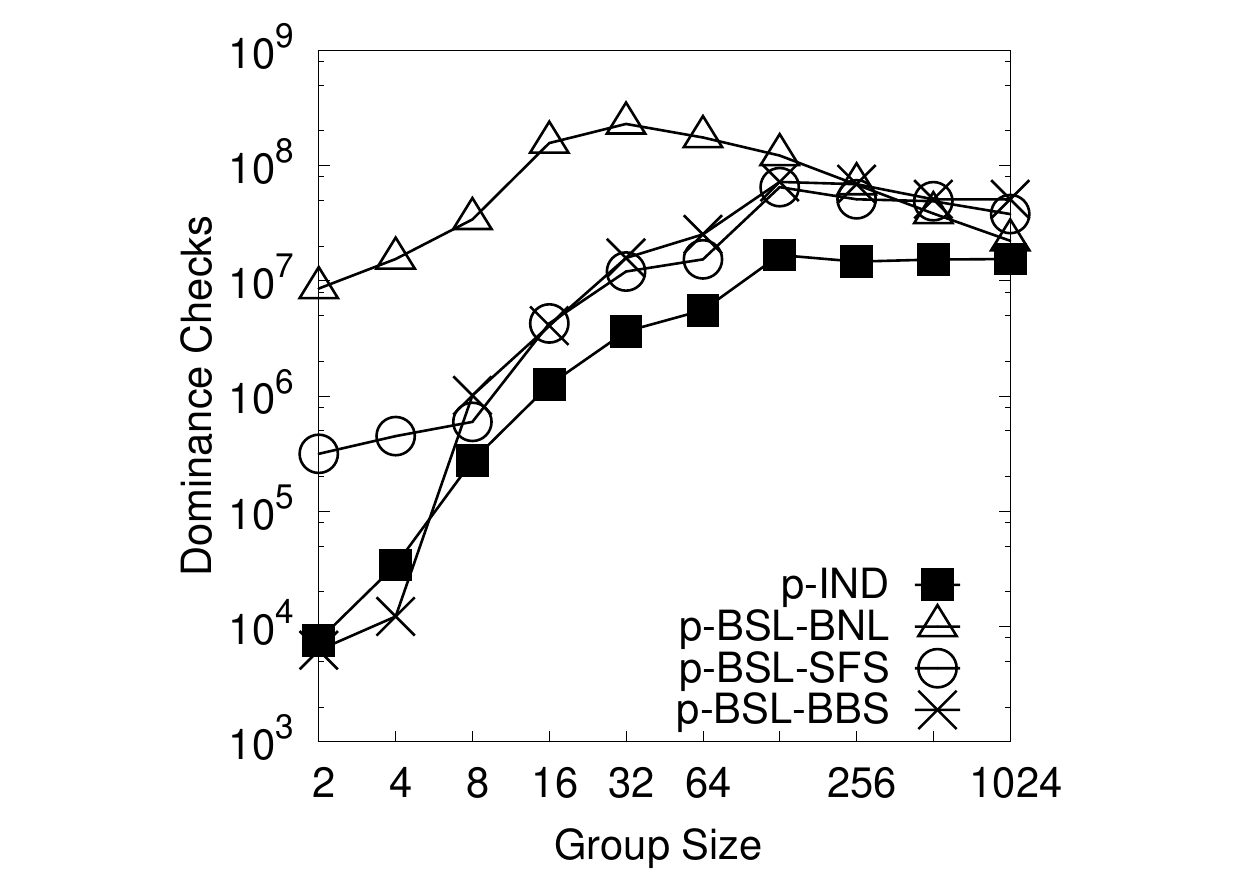}\label{fig:acm_sp_dc}\hspace{-1.0 cm}}
\subfloat[Total Time]{\includegraphics[width=2.1in]{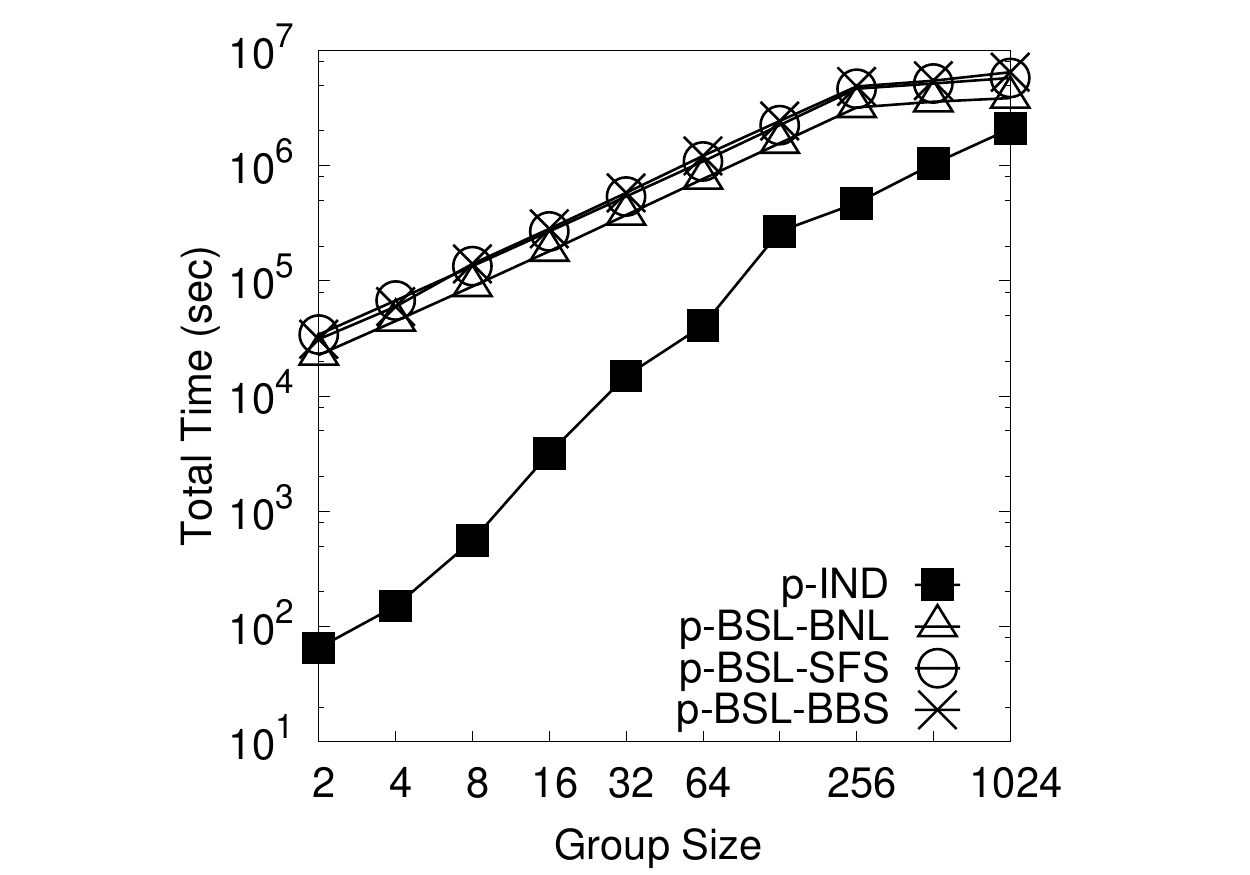}\label{fig:acm_sp_time}}
\caption{$p$-GMCO algorithms, \textsf{ACM} (Synthetic preferences): varying $|\U|$}
\label{fig:acm_sp}
\end{figure*}

\begin{figure*}[]
 \vspace{-0.5cm}
\hspace{-1.0 cm}
\subfloat[I/O Operations]{\includegraphics[width=2.1in]{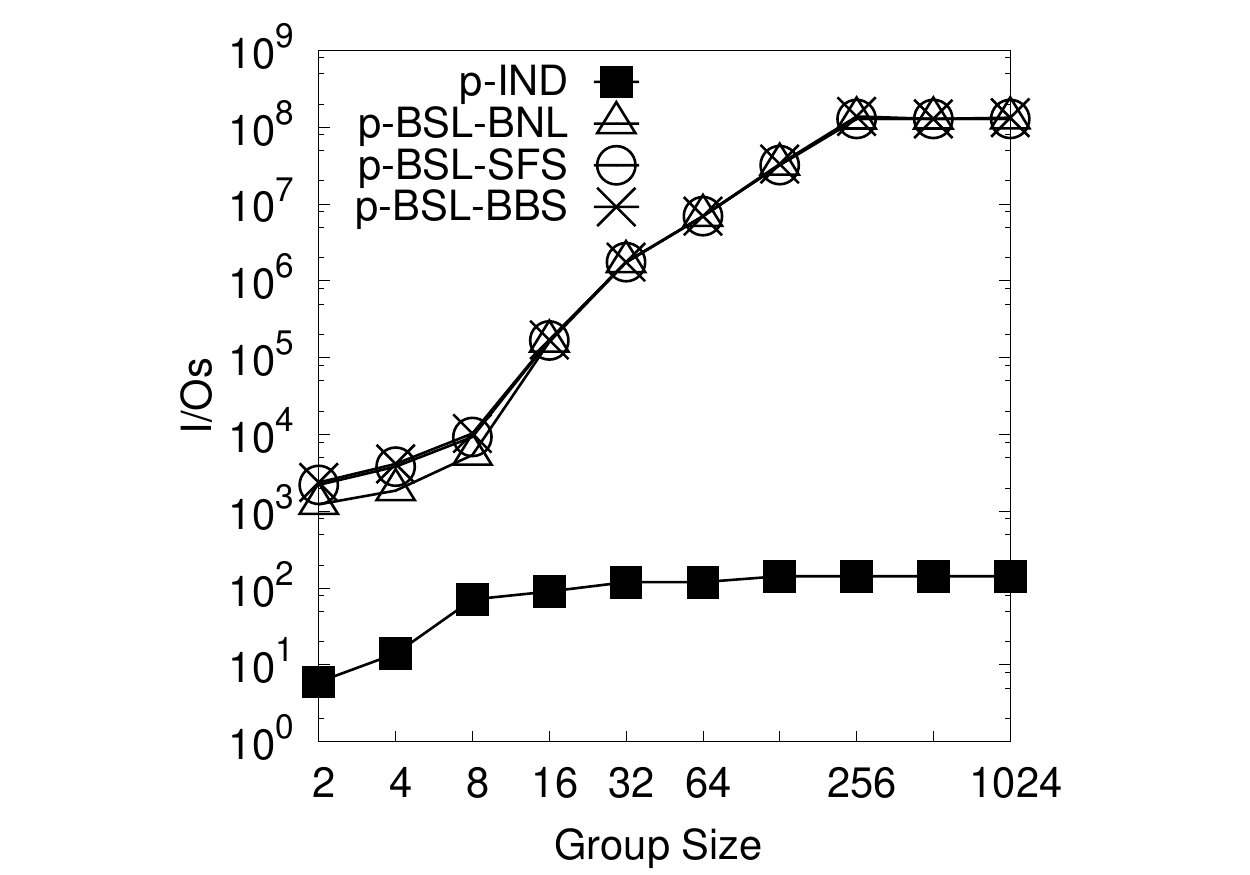}\label{fig:car_rp_io}}
\subfloat[Dom. Checks]{\hspace{-1.0 cm}\includegraphics[width=2.1in]{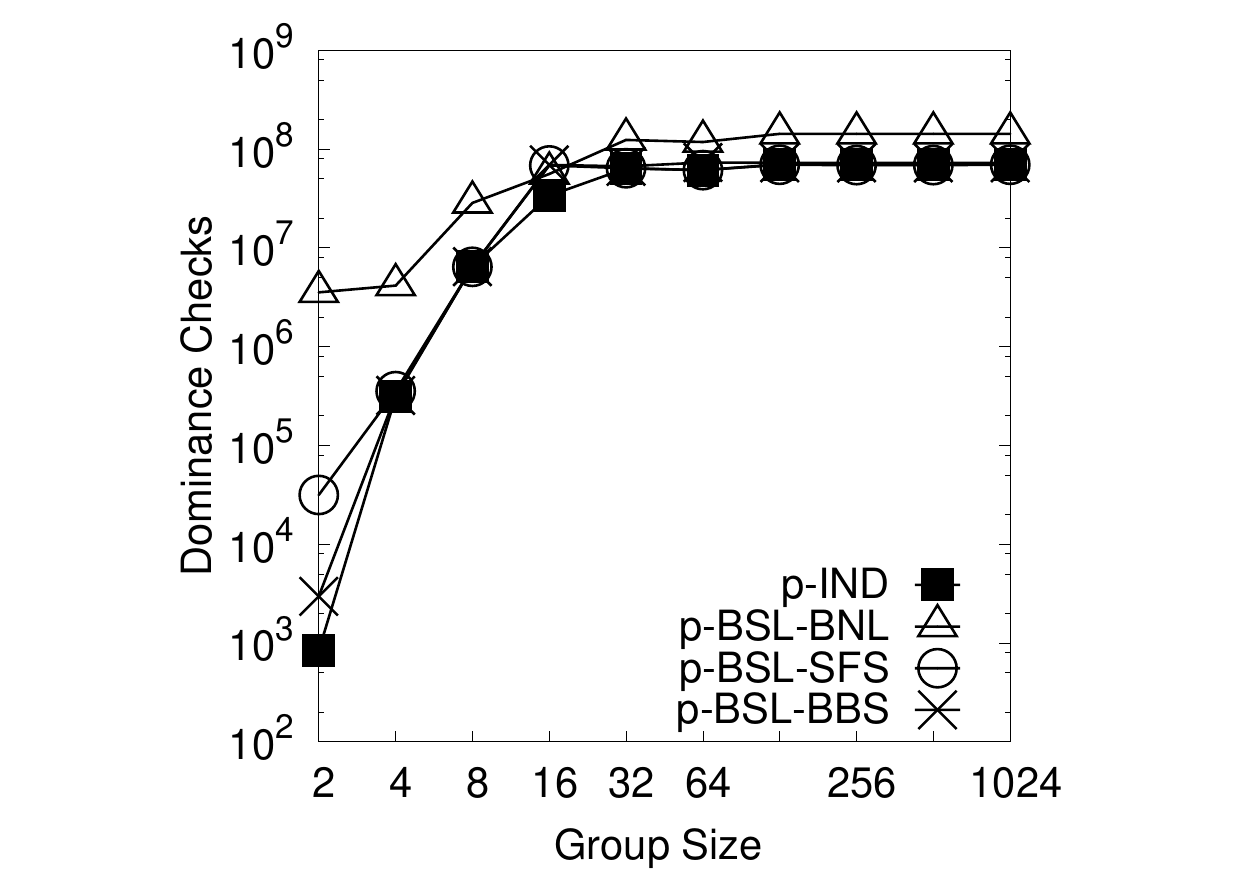}\label{fig:car_rp_dc}\hspace{-1.0 cm}}
\subfloat[Total Time]{\includegraphics[width=2.1in]{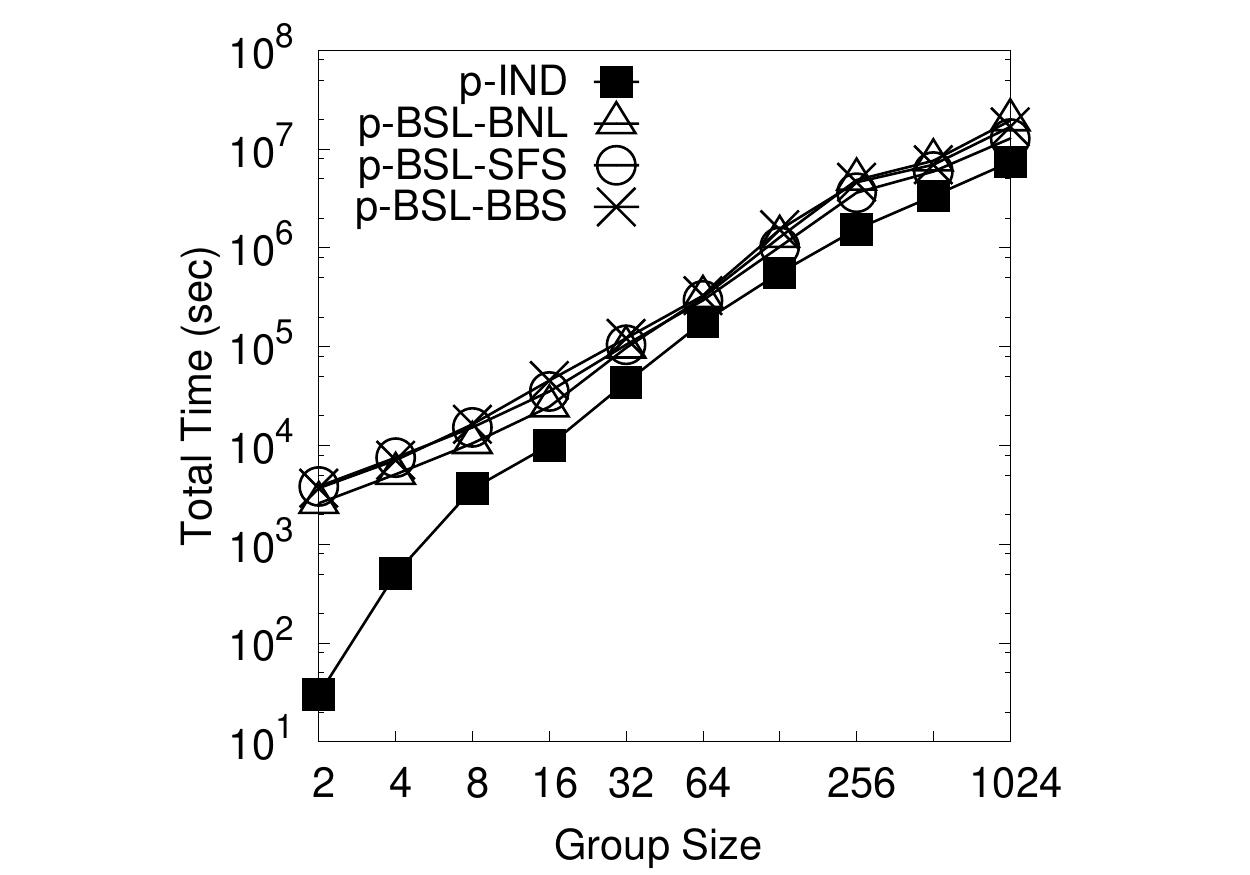}\label{fig:car_rp_time}}
\caption{$p$-GMCO algorithms, \textsf{Cars} (Real preferences): varying $|\U|$}
\label{fig:car_rp}
 \vspace{-0.5cm}
\end{figure*}

\begin{figure*}[]
\hspace{-1.0 cm}
\subfloat[I/O Operations]{\includegraphics[width=2.1in]{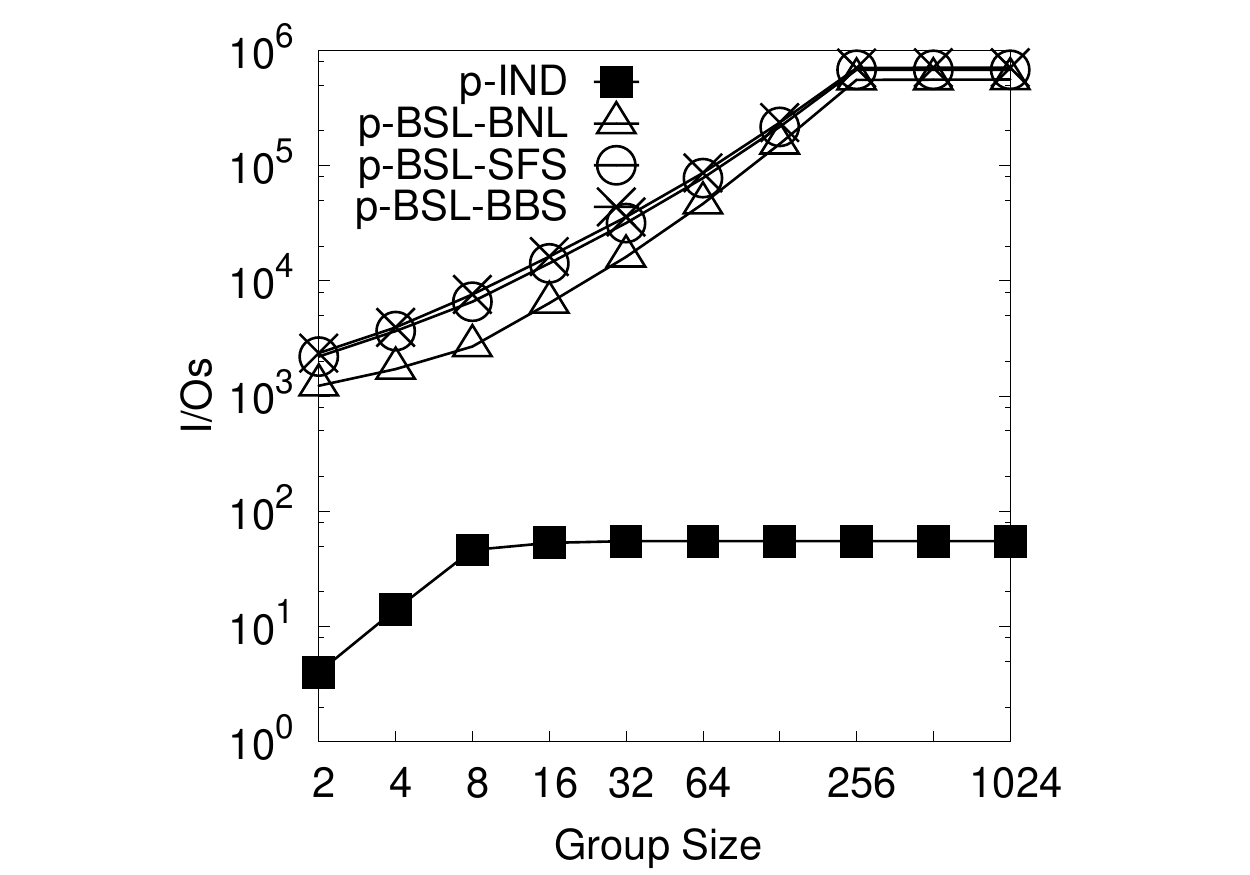}\label{fig:car_sp_io}}
\subfloat[Dom. Checks]{\hspace{-1.0 cm}\includegraphics[width=2.1in]{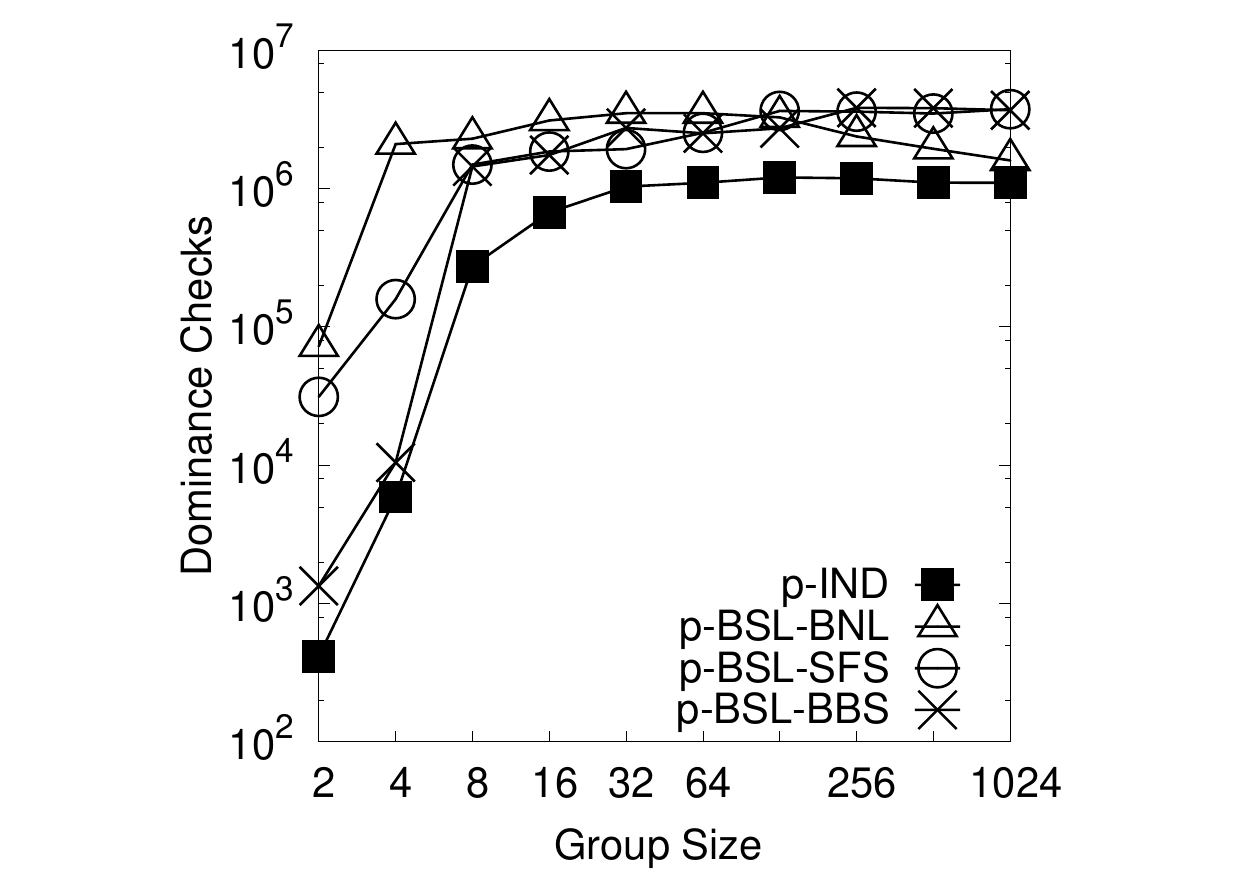}\label{fig:car_sp_dc}\hspace{-1.0 cm}}
\subfloat[Total Time]{\includegraphics[width=2.1in]{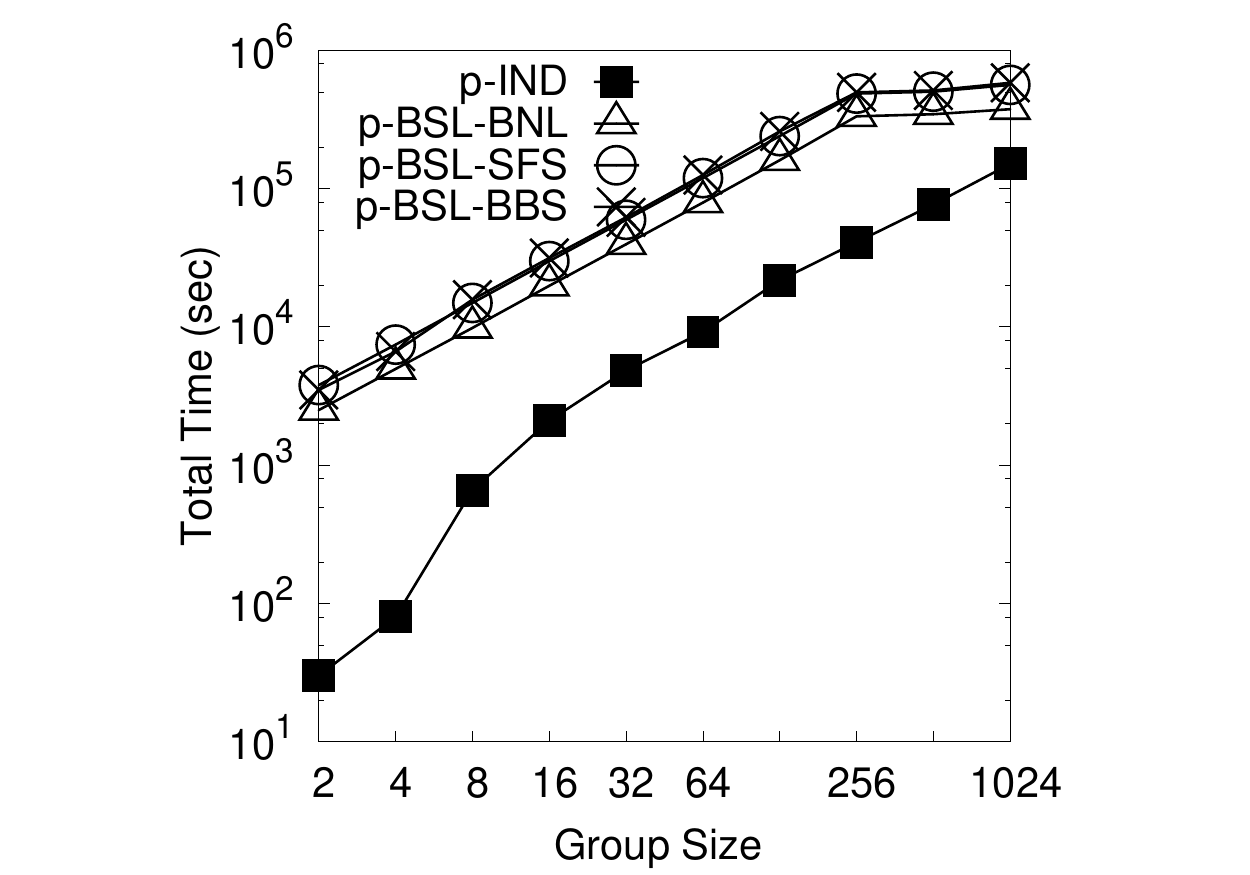}\label{fig:car_sp_time}}
\caption{$p$-GMCO algorithms, \textsf{Cars} (Synthetic preferences): varying $|\U|$}
\label{fig:car_sp}
\end{figure*}

Figures~\ref{fig:rest_rp}~\&~\ref{fig:rest_sp} present the result for \textsf{RestaurantsF} dataset, for real and synthetic preferences, respectively. 
Similarly, Figures~\ref{fig:acm_rp}~\&~\ref{fig:acm_sp} corresponds to the \textsf{ACM} dataset, and Figures~\ref{fig:car_rp}~\&~\ref{fig:car_sp} to \textsf{Cars}. 
 
As we can observe, IND outperforms the BSL methods in almost all cases.
Particularly, the number of I/O operations performed by IND is several order
of magnitude lower than the BSL variants. In addition, in almost all cases,
IND performs fewer dominance checks than the BSL methods. The number of I/Os
performed by IND remains stable for more than 16 users; while for BSL methods,
the I/O operations are constantly increased up to 256 users. Regarding
dominance check, the number of dominance checks increases with $|\U|$
following an almost similar trend for all methods.

Finally, regarding BSL methods, BSL-BNL outperforms the other BSL methods in terms of I/Os and total time; while BSL-SFS and BNL-BBS have almost the same performance. 
As far as dominance checks, in some cases (Figures~\ref{fig:rest_rp_dc} \& \ref{fig:acm_rp_dc}) BSL-BNL outperforms all BSL methods, 
while in other cases (Figures~\ref{fig:rest_sp_dc}, \ref{fig:acm_sp_dc}, \ref{fig:car_rp_dc}, \ref{fig:car_sp_dc}), BSL-BNL performs more dominance checks than the other BSL methods.

\eat{ 
\stitle{Parameters.} \autoref{tab:param_Rp} lists the parameters that we vary in Cars, and also
shows the range of values examined. As before, in each experiment we vary one parameter and set the
other to its default values.

\stitle{Varying the number of users.} \autoref{fig:p_cars_u} shows the effect of varying the
number of users from 8 up to 4096, while $p=30\%$. The required number of I/Os operations increases with $|\U|$ for both methods, as \autoref{fig:p_cars_u_io} shows; the rate of increase for all
$p$-BSL variants is much higher. \autoref{fig:p_cars_u_dc} shows that the number of dominance
check varies as we increase the number of users. Overall, \autoref{fig:p_cars_u_time} shows that
$p$-IND constantly outperforms all $p$-BSL variants by up to one order of magnitude.

\stitle{Varying parameter $p$.} 
We increase the parameter $p$ from 10\% up to 50\%. The performance
of all methods (in terms of I/Os and total time) remains unaffected by $p$, and thus we omit the
relevant figures in the interest of space.

\begin{table}[h!]
\centering
\vspace{-5pt}
\caption{Parameters (\textit{\textbf{p}}-GMCO, Cars)}
\vspace{2pt}
\label{tab:param_Rp}
\small
\begin{tabular}{ccc}
\toprule
\textbf{Symbol} & \textbf{Values} & \textbf{Default}\\ \midrule
$|\U|$ & $8, \ 16, \ 32, \ \dots, \ 4096$ & 256\\
$p$ & 10\%, 20\%, 30\%, 40\%, 50\% & 30\%\\
\bottomrule
\end{tabular}
\vspace{-5pt}
\end{table}
 }

\vspace{-4mm}

\subsection{Effectiveness of GRCO}
\label{sec:r_mcp}
In this section we study the effectiveness of the GRCO problem  (\autoref{sec:rank}). 
We compare our RANK-CM algorithm  (\autoref{sec:rank_method}) to nine popular aggregations strategies adopted by most group recommender systems \cite{CC12}.
Particularly, we  implement the following aggregation strategies:
\begin{itemize}
\item \textit{Additive}  (ADD): adds the individual matching degrees.
\item  \textit{Multiplicative} (MULT): multiplies the individual matching degrees.
\item  \textit{Least Misery} (MISERY):  considers the minimum of individual matching degrees.
\item  \textit{Most Pleasure} (PLEASURE): considers the maximum of individual matching degrees.
\item \textit{Average Without Misery} (AVG\_MISERY): takes the average matching degrees, excluding matching degrees below a  threshold;
\item \textit{Average Without Misery Threshold-free} (AVG\_MISERY+): is a strategy introduced here, similar to AVG\_MISERY, with the difference that the threshold is set to the minimum of individual matching degrees.
\item \textit{Copeland Rule} (COPELAND): counts the number of times an object has higher individual 
matching degrees than the rest of the  objects, minus the number of times the object has lower  individual matching degrees. 
\item \textit{Approval Voting} (APPROVAL): counts the number of  individual matching degrees with values greater than or equal to a threshold.
\item \textit{Borda Count} (BORDA): adds the scores computed per matching degree according to its rank in a user's preference list (the matching degree  with the lowest value gets a zero score, the next one point, and so on).
\end{itemize}

\eat{{
(1) \textit{Additive}  (ADD): adds the individual matching degrees.
(2) \textit{Multiplicative} (MULT): multiplies the individual matching degrees.
(3) \textit{Least Misery} (MISERY):  considers the minimum of individual matching degrees.
(4) \textit{Most Pleasure} (PLEASURE): considers the maximum of individual matching degrees.
(5) \textit{Average Without Misery} (AVG\_MISERY): takes the average matching degrees, excluding matching degrees below a  threshold;
(6) \textit{Average Without Misery Threshold-free} (AVG\_MISERY+): is a strategy introduced here, similar to AVG\_MISERY, with the difference that the threshold is set to the minimum of individual matching degrees.
(7) \textit{Copeland Rule} (COPELAND): counts the number of times an object has higher individual 
matching degrees than the rest of the  objects, minus the number of times the object has lower  individual matching degrees. 
(8) \textit{Approval Voting} (APPROVAL): counts the number of  individual matching degrees with values greater than or equal to a threshold.
(9) \textit{Borda Count} (BORDA): adds the scores computed per matching degree according to its rank in a user's preference list (the matching degree  with the lowest value gets a zero score, the next one point, and so on).}}
Note that, the threshold in AVG\_MISERY and APPROVAL strategies is set to 0.5.

To gauge the effectiveness of our ranking scheme, we use the
\textsf{RestaurantsR} dataset. We use the reviews from all users and
extract a ranked list of the most popular restaurants to serve as the ground
truth. Then, we compare the ranked lists returned by RANK-CM and the other aggregation strategies  to the ground truth, computing \textit{Precision} and the \textit{Generalized Spearman's Footrule} \cite{FKS03}, in several ranks and for different group sizes.
In order to construct group of users, for each group size, we randomly select users, composing 500 groups of the same size.
Hence, in each experiment the average measurements are presented.

\begin{figure*}[]
\hspace{-0 cm}
{\hspace{0cm} \includegraphics[width=4.5in]{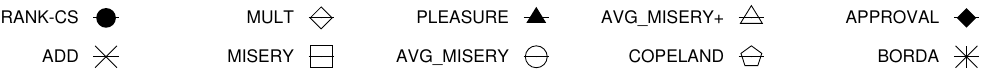}}\\
 \subfloat[Precision]{\hspace{-0 cm}\includegraphics[width=2.1in]{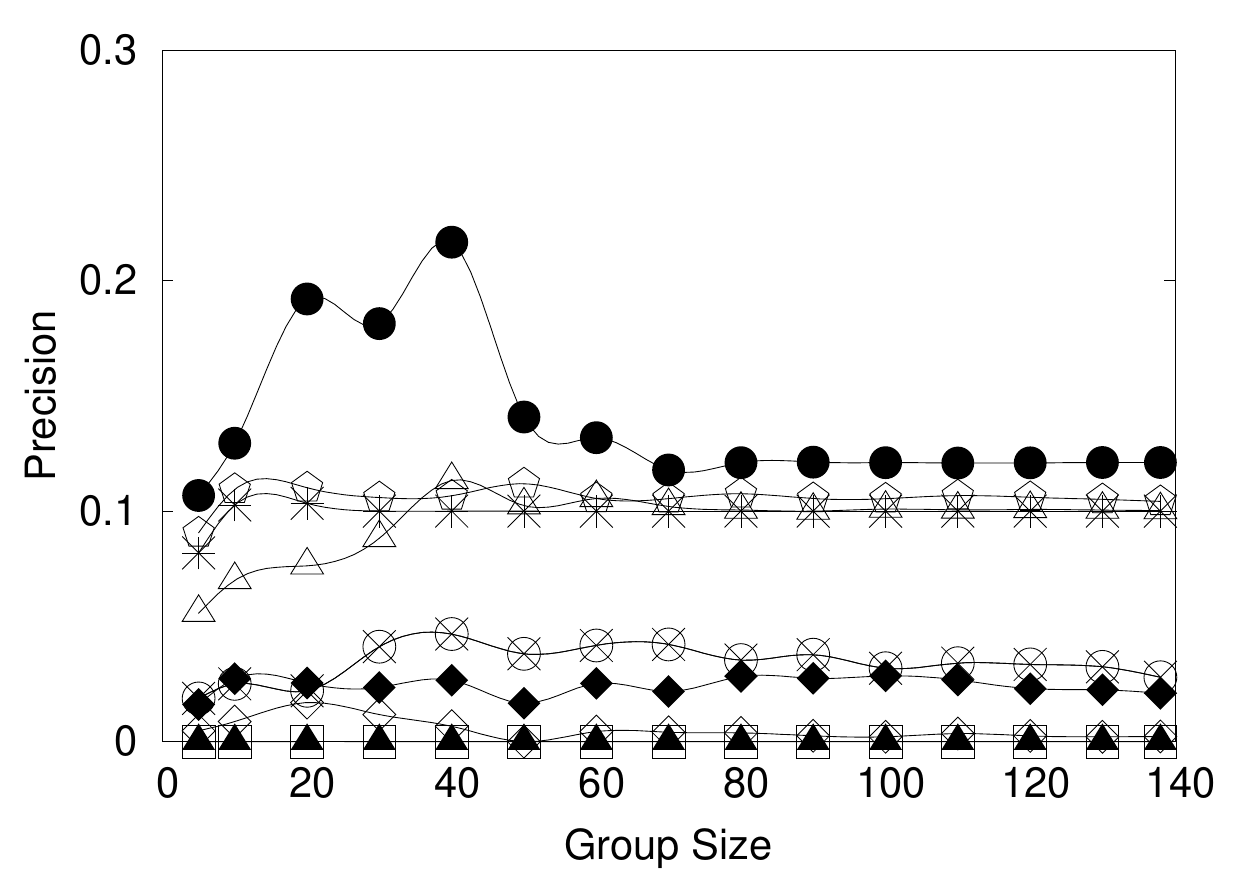}\label{fig:prec10}\hspace{1cm}}
\subfloat[Spearman's footrule ]{\includegraphics[width=2.1in]{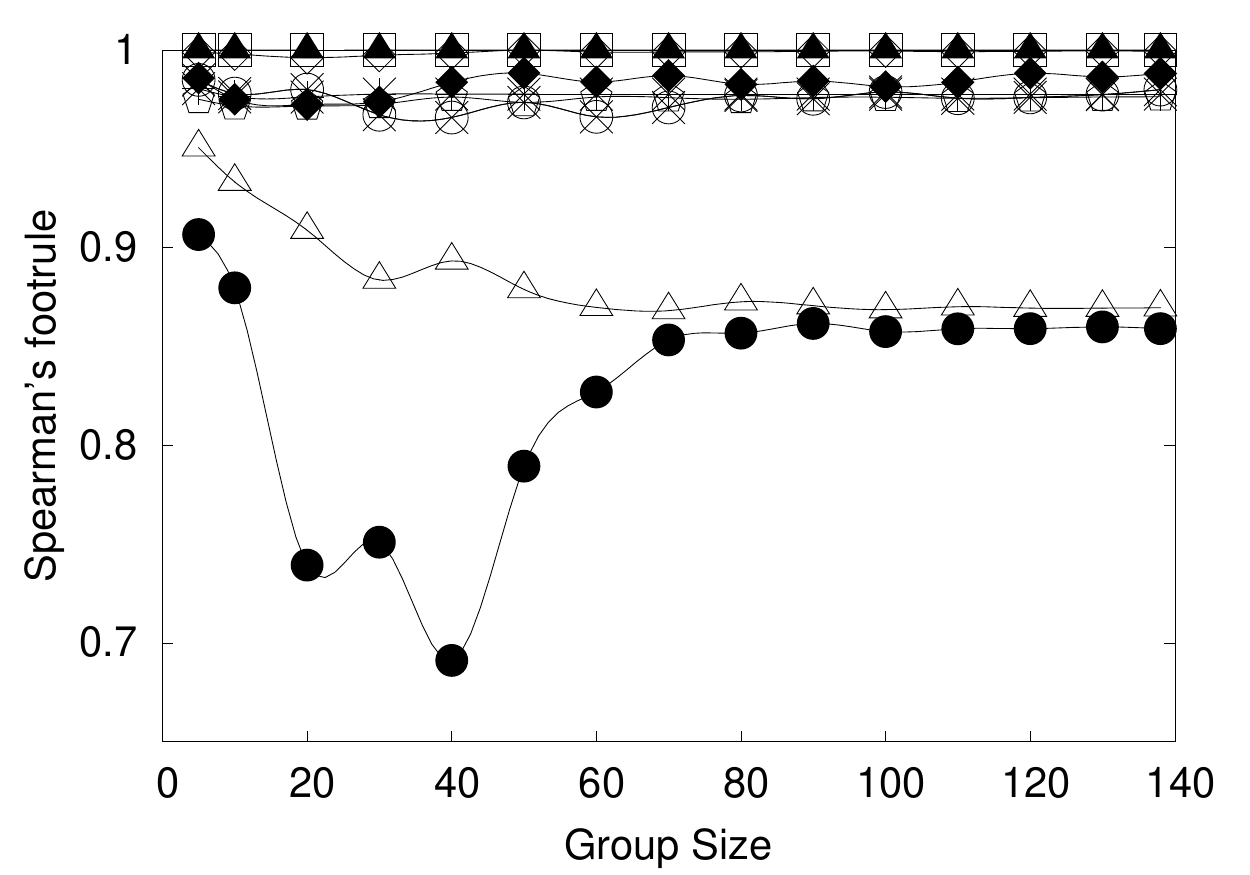}\label{fig:foot10}}
\caption{\textsf{RestaurantsR} (Rank 10): varying $|\U|$}
\label{fig:pf10_u}
\vspace{-5pt}
\end{figure*}

\begin{figure*}[]
\hspace{-0 cm}
 \subfloat[Precision]{\hspace{-0 cm}\includegraphics[width=2.1in]{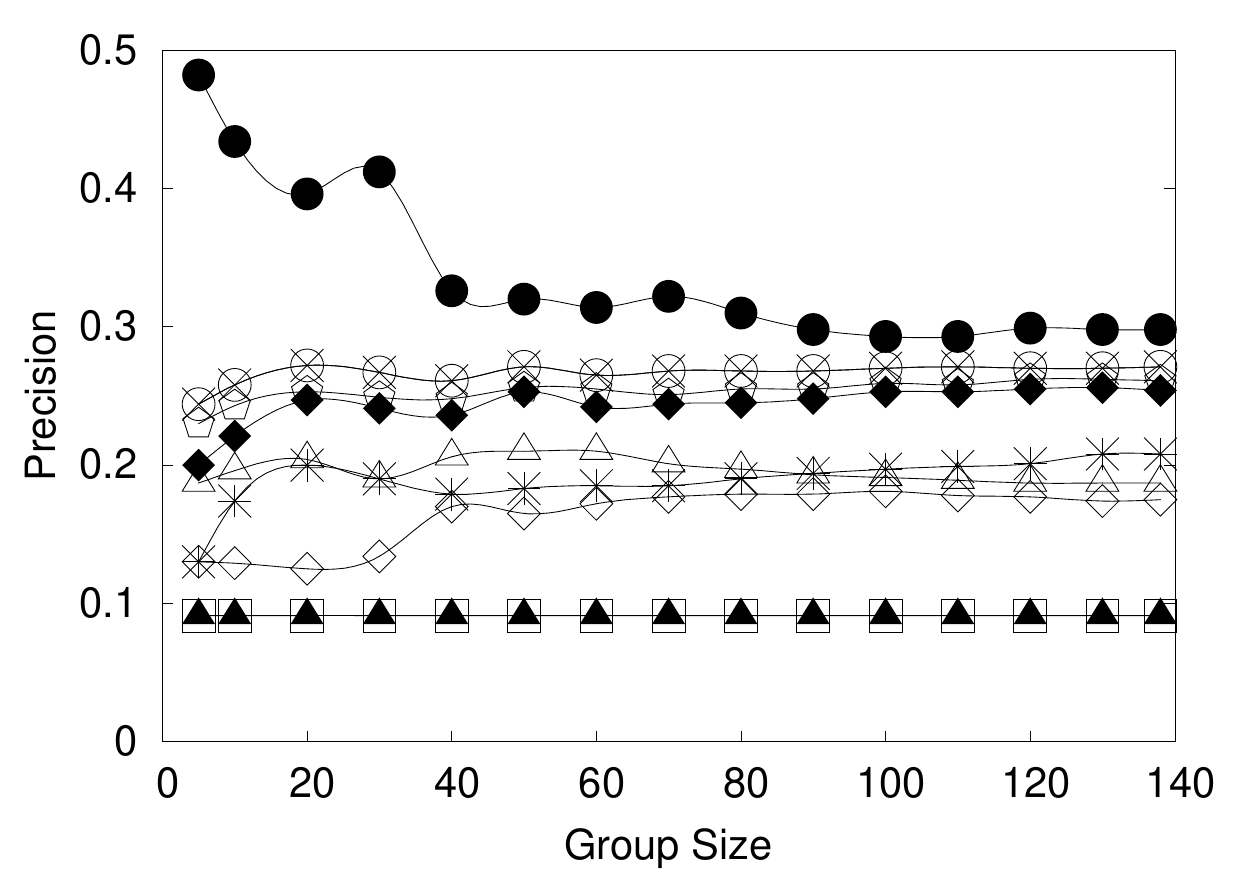}\label{fig:prec20}\hspace{1cm}}
\subfloat[Spearman's footrule]{\includegraphics[width=2.1in]{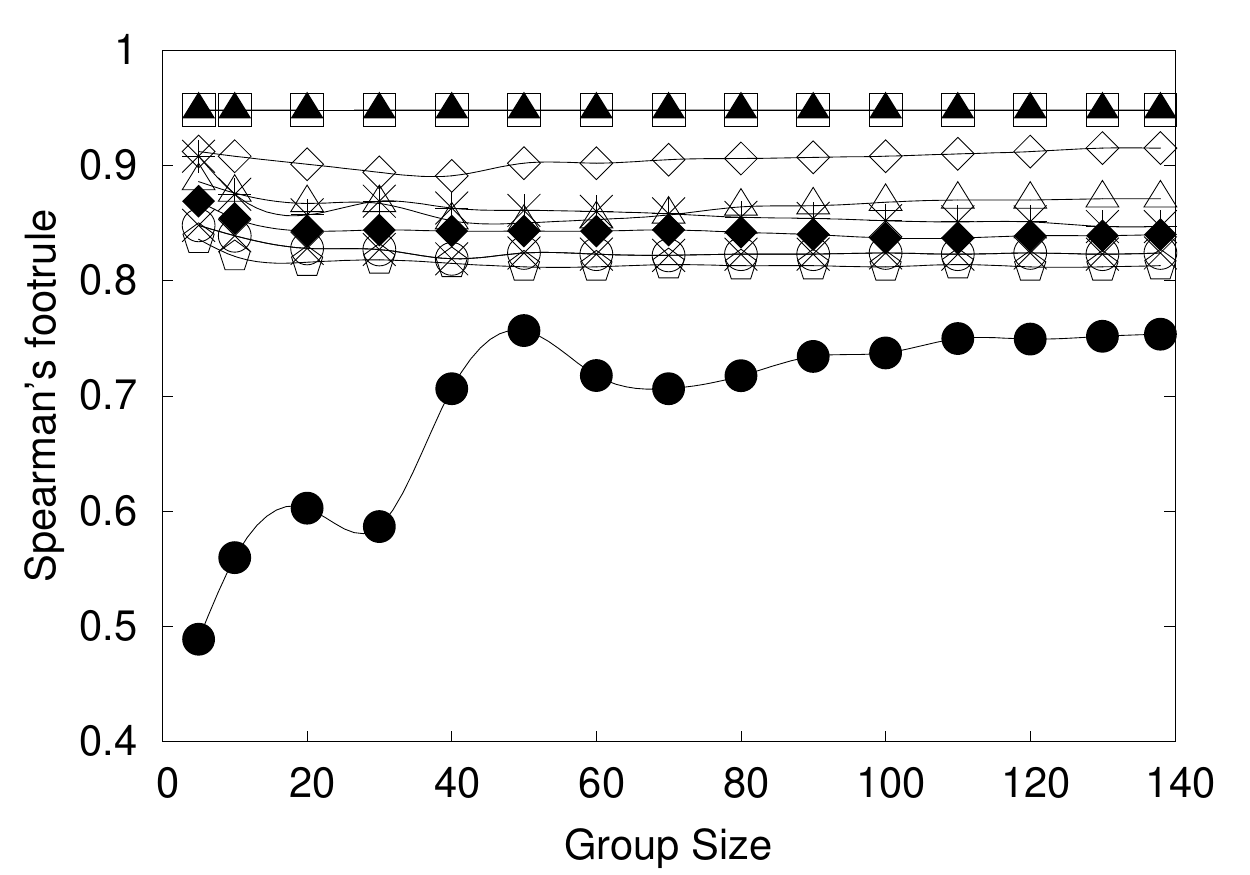}\label{fig:foot20}}
\caption{\textsf{RestaurantsR} (Rank 20): varying $|\U|$}
\label{fig:pf20_u}
\vspace{-5pt}
\end{figure*}

\stitle{Varying the group size.}
In the first experiment (Figures \ref{fig:pf10_u} \& \ref{fig:pf20_u}), 
we consider different group sizes, varying the number of users, from 5 to 138. 
We compute the precision and the Spearman's footrule for the ranked listed returned by all methods, 
compared to the ground truth list,  at rank 10 (\autoref{fig:pf10_u}) and rank 20 (\autoref{fig:pf20_u}).

In \autoref{fig:pf10_u}, we consider the first ten restaurants retrieved (i.e., at rank 10);
the precision for each method is defined as the number of common restaurants between
the ground truth list and the ranked  list returned by each method, divided by ten. 
For example, in \autoref{fig:prec10}, for the groups of 20 users, RANK-CM has precision around 0.2;
that is, among the first ten restaurants retrieved, RANK-CM retrieves in average two popular restaurants.
On the other hand, BORDA and COPELAND retrieve in average  one popular restaurant, and have precision around 0.1.

Regarding the results at rank 10, as we can observe from \autoref{fig:pf10_u},   
RANK-CM outperforms all other methods in both metrics. 
Note that, Spearman's footrule values range from 0 to 1, where
lower values indicate a better match to the ground truth (0 means that the two lists are identical). 
Regarding the other aggregation strategies, the best results are provided by COPELAND, BORDA and AVG\_MISERY+, while MISERY and PLEASURE performed the worst.

Similar results and observations hold at rank 20 (Figures~\ref{fig:pf20_u}), 
where RANK-CM outperforms all other methods, 
with COPELAND, ADD and AVG\_MISERY  being the best alternatives.

Overall, RANK-CM performs better in terms of precision 
and Spearman's footrule than the other strategies, in all cases. 
The COPELAND strategy seems to be the best alternative, 
while  MISERY and PLEASURE the worst.

\begin{figure*}[]
\hspace{-0 cm}
 \subfloat[Precision]{\hspace{-0 cm}\includegraphics[width=2.1in]{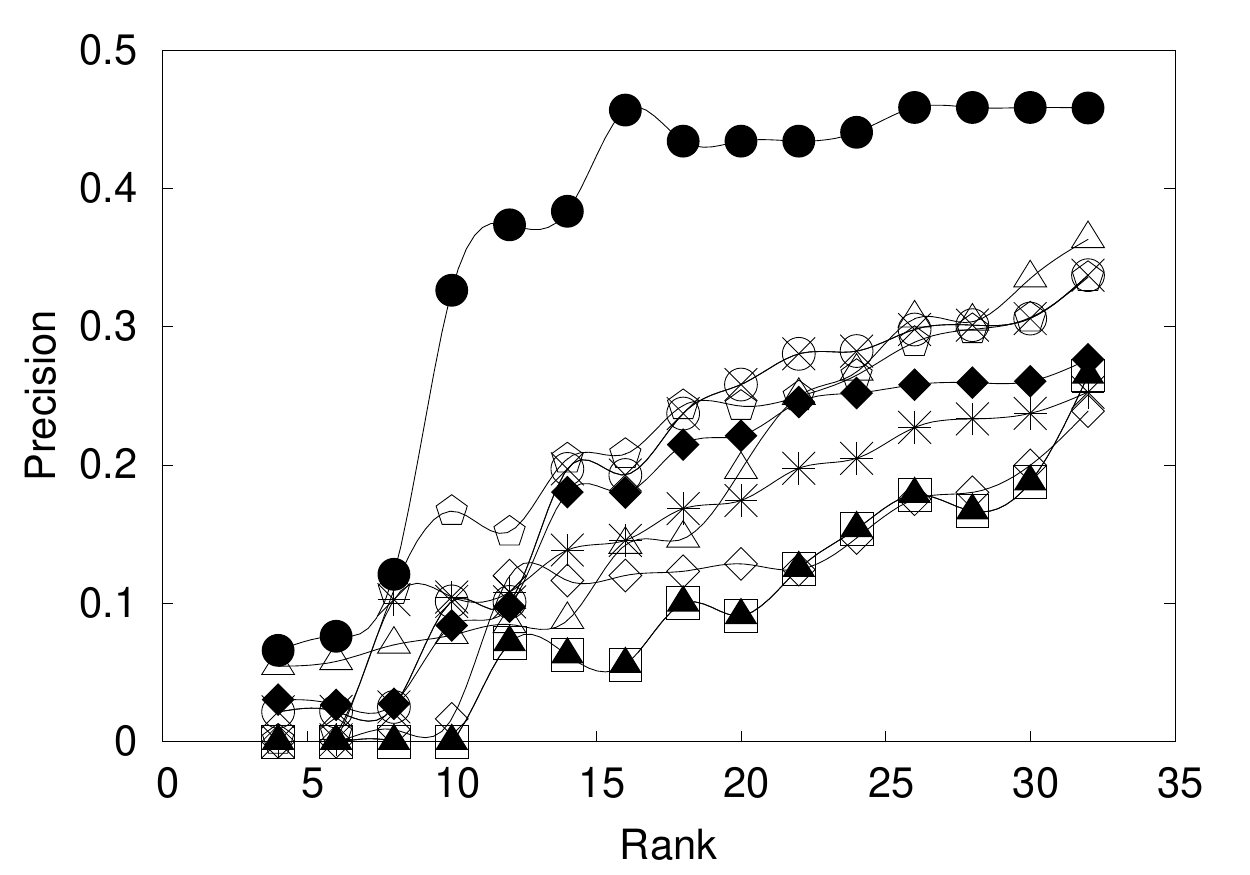}\label{fig:prec10u}\hspace{1cm}}
\subfloat[Spearman's footrule]{\includegraphics[width=2.1in]{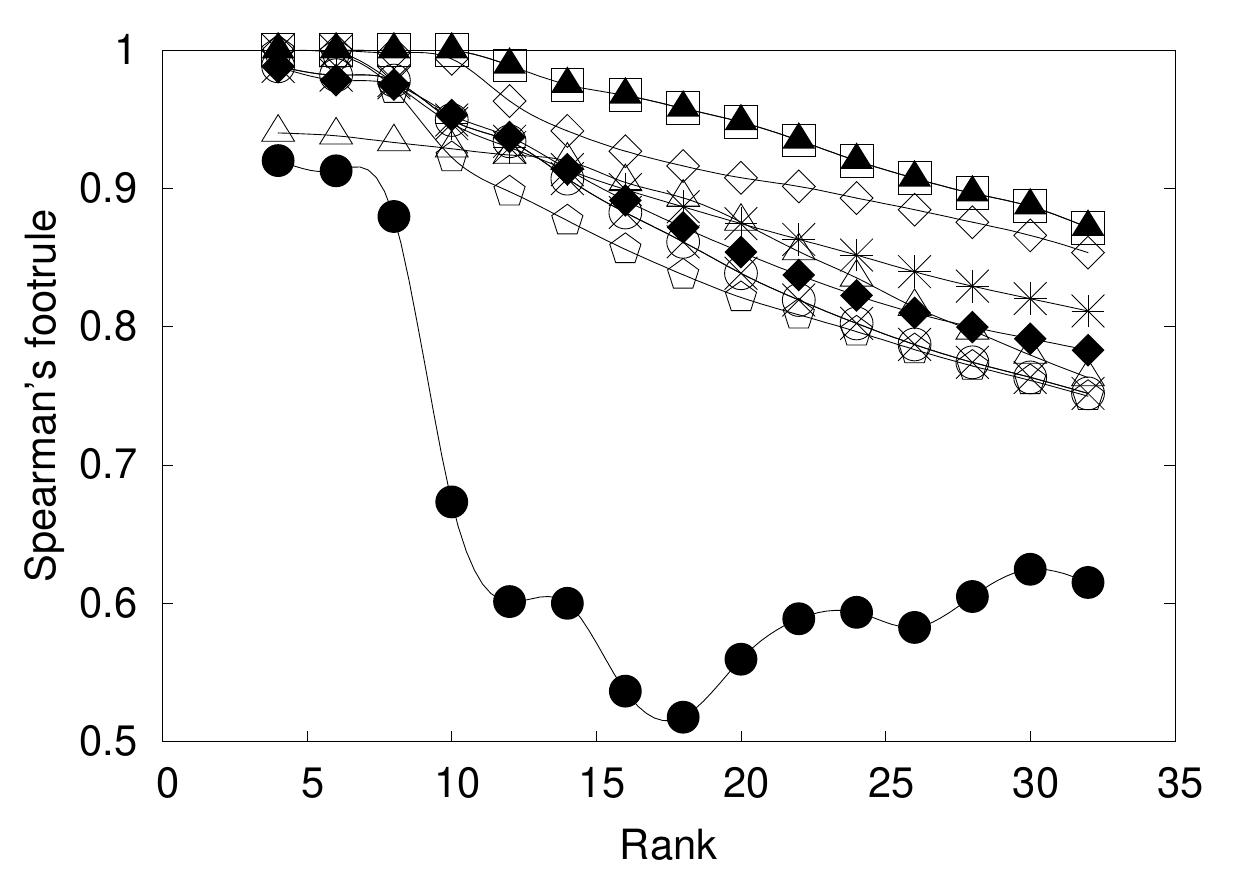}\label{fig:foot10u}}
\caption{\textsf{RestaurantsR} ($|\U| = 10$): varying rank}
\label{fig:pf10u}
\vspace{-10pt}
\end{figure*}

\begin{figure*}[]
\hspace{-0 cm}
 \subfloat[Precision]{\hspace{-0 cm}\includegraphics[width=2.1in]{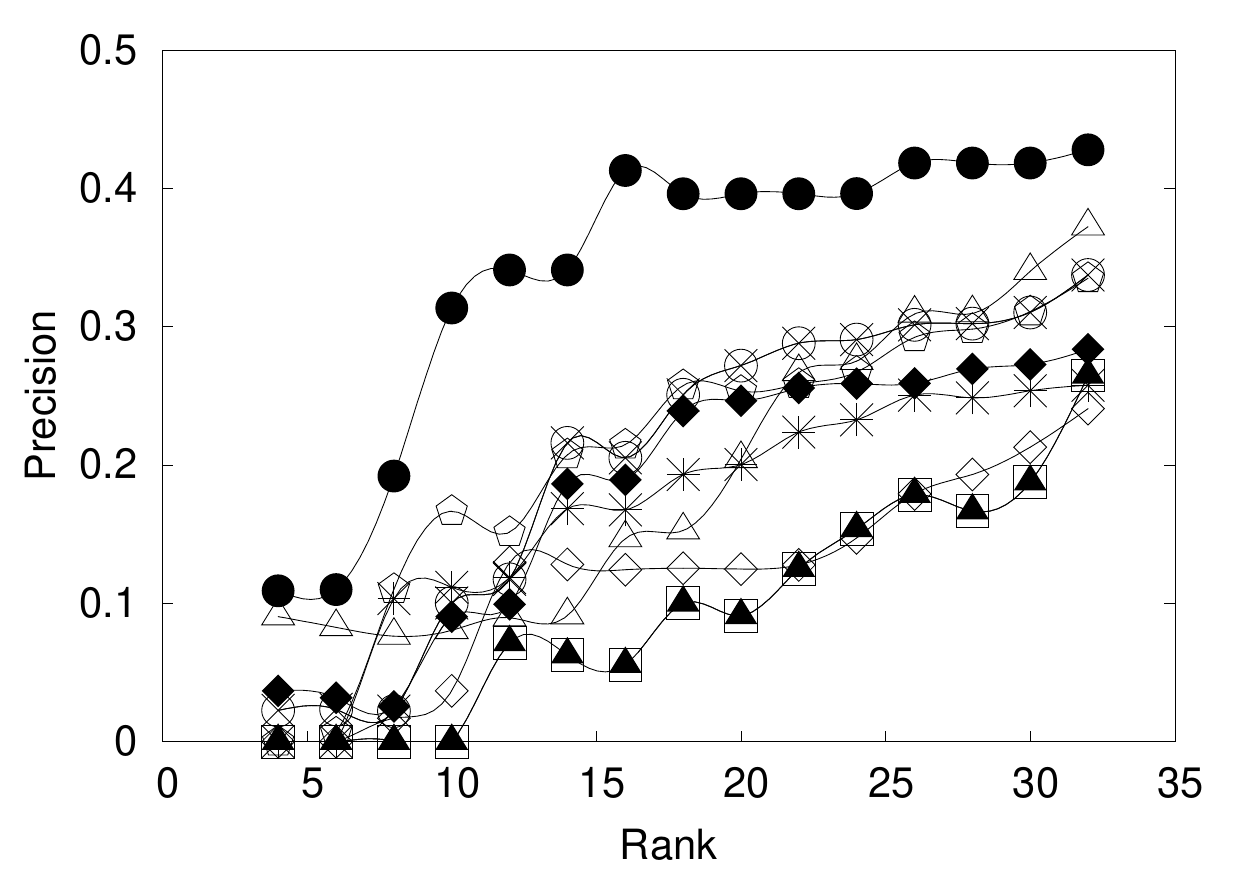}\label{fig:prec20u}\hspace{1cm}}
\subfloat[Spearman's footrule]{\includegraphics[width=2.1in]{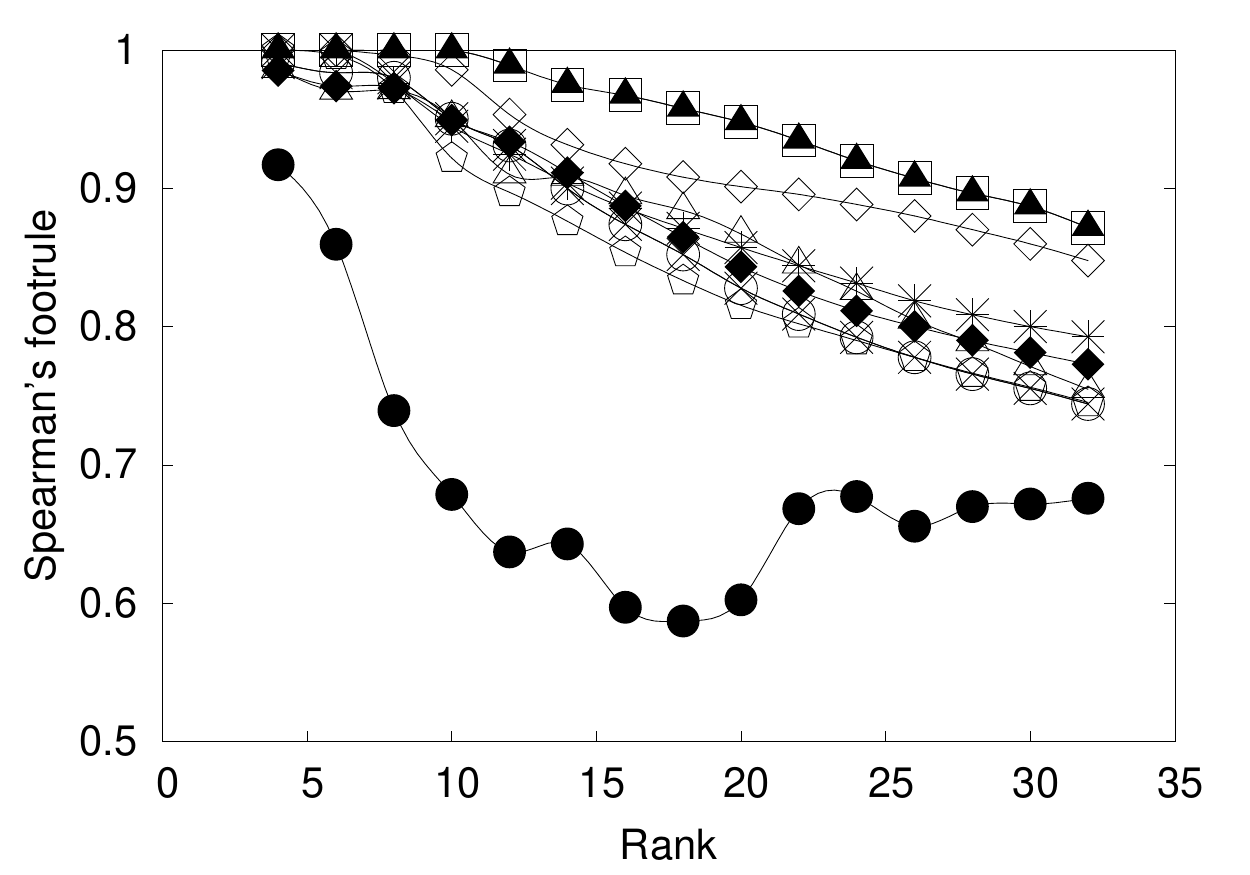}\label{fig:foot20u}}
\caption{\textsf{RestaurantsR} ($|\U| = 20$): varying rank}
\label{fig:pf20u}
\vspace{-10pt}
\end{figure*}

\begin{figure*}[]
\hspace{-0 cm}
 \subfloat[Precision]{\hspace{-0 cm}\includegraphics[width=2.1in]{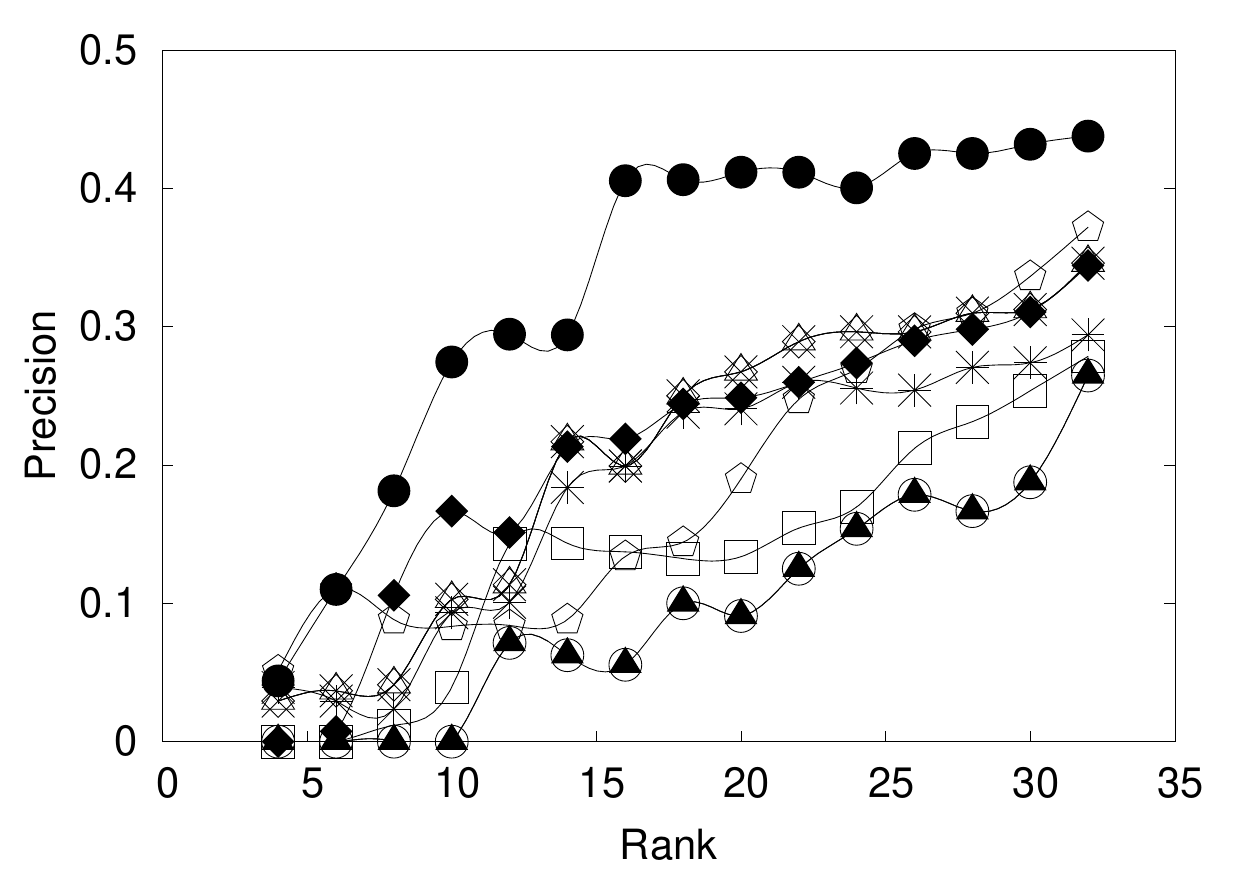}\label{fig:prec30u}\hspace{1cm}}
\subfloat[Spearman's footrule]{\includegraphics[width=2.1in]{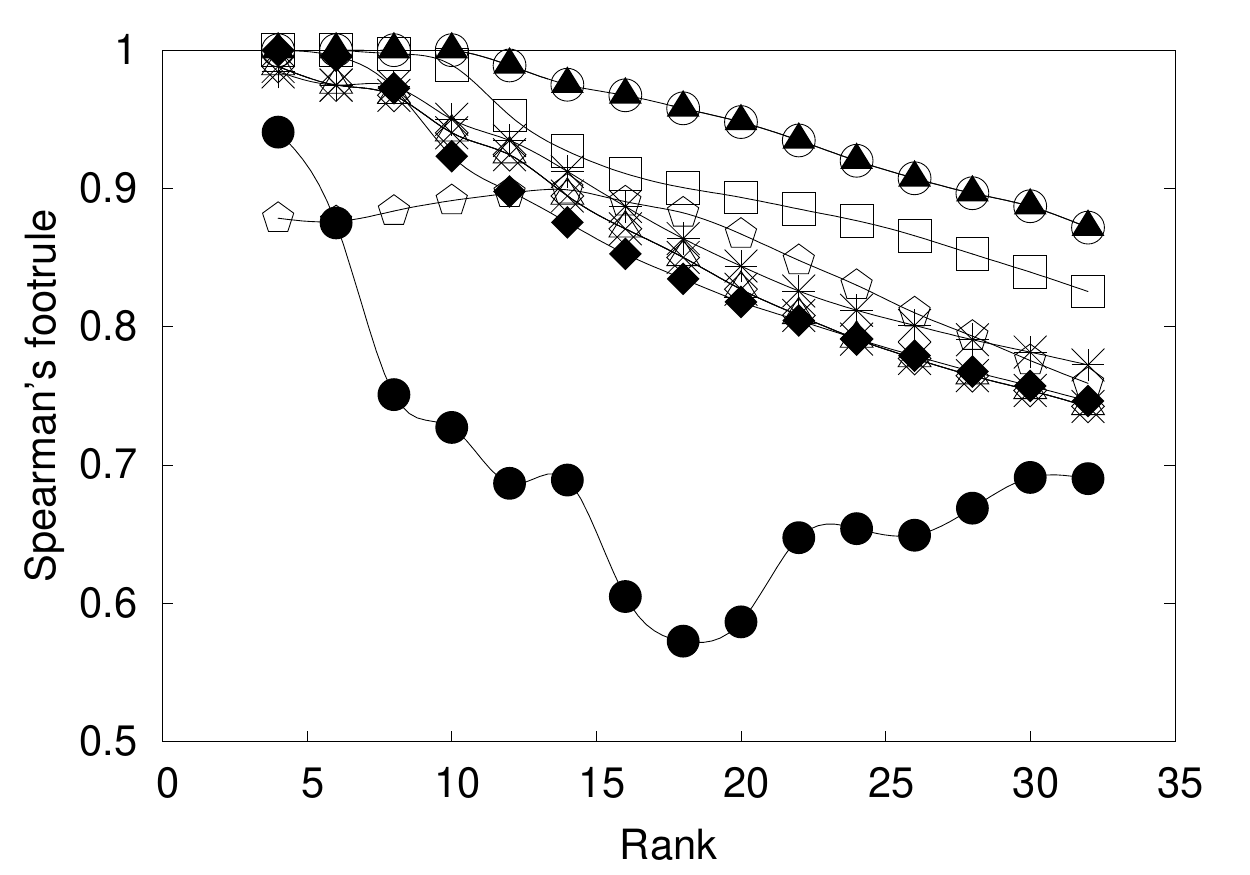}\label{fig:foot30u}}
\caption{\textsf{RestaurantsR} ($|\U| = 30$): varying rank}
\label{fig:pf30u}
\vspace{-10pt}
\end{figure*}

\stitle{Varying rank.}
In this experiment, we consider three different group sizes (i.e., 10, 20, 30)
and compute the precision and the Spearman's footrule from rank 4 to rank 32.
As we can observe from Figures \ref{fig:pf10u}, \ref{fig:pf20u} \& \ref{fig:pf30u}, 
the performance of all methods is almost similar for the examined group sizes. 
The  RANK-CM achieves better performance in terms of  precision and Spearman's footrule in almost all examined ranks, with the exceptions at ranks 4 and 6 for group sizes 10 and 30, where 
COPELAND achieves almost the same performance with RANK-CM. 
Regarding the other methods, the best performance is from COPELAND, ADD and  AVG\_MISERY+.

%

%% file: concl.tex
  
\section{Conclusions}
\label{sec:concl}
  
This work addressed objective ranking techniques for a group of preferences
over categorical attributes, where the goal is to rank objects based on what
is considered ideal by all users. In particular, we study three related
problems based on a double Pareto aggregation. The first is to return the set
of objects that are unanimously considered ideal by the entire group. In the
second problem, we relax the requirement for unanimity and only require a
percentage of users to agree. Then, in the third problem, we devise an
effective ranking scheme based on our double Pareto aggregation framework. The
proposed methods take advantage of a transformation of the categorical
attribute values in order to use a standard index structure. A detailed
experimental study verified the efficiency and effectiveness of our
techniques.

%% file: main.bbl
\begin{thebibliography}{10}

\bibitem{AT05}
G.~Adomavicius and A.~Tuzhilin.
\newblock {T}oward the {N}ext {G}eneration of {R}ecommender {S}ystems: {A}
  {S}urvey of the {S}tate-of-the-{A}rt and {P}ossible {E}xtensions.
\newblock {\em {IEEE} Transactions on Knowledge and Data Engineering ({TKDE})},
  17(6), 2005.

\bibitem{ABJ89}
R.~Agrawal, A.~Borgida, and H.~V. Jagadish.
\newblock {E}fficient {M}anagement of {T}ransitive {R}elationships in {L}arge
  {D}ata and {K}nowledge {B}ases.
\newblock In {\em Proc. of the {ACM} {SIGMOD} Intl. Conf. on Management of Data
  ({SIGMOD})}, 1989.

\bibitem{AW00}
R.~Agrawal and E.~L. Wimmers.
\newblock {A} {F}ramework for {E}xpressing and {C}ombining {P}references.
\newblock In {\em Proc. of the {ACM} {SIGMOD} Intl. Conf. on Management of Data
  ({SIGMOD})}, 2000.

\bibitem{MPJ07}
M.~D.~M. andJignesh M.~Patel and H.~V. Jagadish.
\newblock {E}fficient {S}kyline {C}omputation over {L}ow-{C}ardinality
  {D}omains.
\newblock In {\em Proc. of the Intl. Conf. on Very Large Databases ({VLDB})},
  2007.

\bibitem{ArdissonoGPST03}
L.~Ardissono, A.~Goy, G.~Petrone, M.~Segnan, and P.~Torasso.
\newblock {I}ntrigue: {P}ersonalized {R}ecommendation of {T}ourist
  {A}ttractions for {D}esktop and {H}and {H}eld {D}evices.
\newblock {\em Applied Artificial Intelligence}, 17(8-9), 2003.

\bibitem{A63}
K.~J. Arrow.
\newblock {\em {S}ocial {C}hoice and {I}ndividual {V}alues}.
\newblock Yale University Press, 2nd edition, 1963.

\bibitem{AM01}
J.~A. Aslam and M.~H. Montague.
\newblock {M}odels for {M}etasearch.
\newblock In {\em Proc. of the Intl. {ACM} {SIGIR} Conf. on Research and
  Development in Information Retrieval ({SIGIR})}, 2001.

\bibitem{BPD08}
E.-A. Baatarjav, S.~Phithakkitnukoon, and R.~Dantu.
\newblock {G}roup {R}ecommendation {S}ystem for {F}acebook.
\newblock In {\em OTM Workshops}, 2008.

\bibitem{BMR10}
L.~Baltrunas, T.~Makcinskas, and F.~Ricci.
\newblock {G}roup recommendations with rank aggregation and collaborative
  filtering.
\newblock In {\em ACM conference on Recommender systems, RecSys}, 2010.

\bibitem{gb02}
D.~G. Bar and O.~Glinansky.
\newblock {Family Stereotyping - A Model to Filter TV Programs for Multiple
  Viewers}.
\newblock In {\em Workshop on Personalization in Future TV}, 2002.

\bibitem{BCP08}
I.~Bartolini, P.~Ciaccia, and M.~Patella.
\newblock {E}fficient {S}ort-based {S}kyline {E}valuation.
\newblock {\em ACM Transactions on Database Systems ({TODS})}, 33(4), 2008.

\bibitem{BKSS90}
N.~Beckmann, H.-P. Kriegel, R.~Schneider, and B.~Seeger.
\newblock {T}he {R}*-{T}ree: {A}n {E}fficient and {R}obust {A}ccess {M}ethod
  for {P}oints and {R}ectangles.
\newblock In {\em Proc. of the {ACM} {SIGMOD} Intl. Conf. on Management of Data
  ({SIGMOD})}, 1990.

\bibitem{Bentley1990}
J.~L. Bentley, K.~L. Clarkson, and D.~B. Levine.
\newblock {F}ast {L}inear {E}xpected-{T}ime {A}lgorithms for {C}omputing
  {M}axima and {C}onvex {H}ulls.
\newblock In {\em Proc. of {ACM-SIAM} Symposium on Discrete Algorithms}, 1990.

\bibitem{BF10}
S.~Berkovsky and J.~Freyne.
\newblock {G}roup-based recipe recommendations: analysis of data aggregation
  strategies.
\newblock In {\em ACM conference on Recommender systems, RecSys}, 2010.

\bibitem{BBS14}
N.~Bikakis, K.~Benouaret, and D.~Sacharidis.
\newblock {R}econciling {M}ultiple {C}ategorical {P}references with {D}ouble
  {P}areto-based {A}ggregation.
\newblock In {\em Proc. of the Intl. Conf. on Database Systems for Advanced
  Applications ({DASFAA})}, 2014.

\bibitem{BikakisSS14}
N.~Bikakis, D.~Sacharidis, and T.~Sellis.
\newblock {A} {S}tudy on {E}xternal {M}emory {S}can-{B}ased {S}kyline
  {A}lgorithms.
\newblock In {\em Database and Expert Systems Applications - 25th International
  Conference {(DEXA)}}, 2014.

\bibitem{BOHG13}
J.~Bobadilla, F.~Ortega, A.~Hernando, and A.~Guti{\'e}rrez.
\newblock {R}ecommender systems survey.
\newblock {\em Knowl.-Based Syst.}, 46, 2013.

\bibitem{BC11}
L.~Boratto and S.~Carta.
\newblock {S}tate-of-the-{A}rt in {G}roup {R}ecommendation and {N}ew
  {A}pproaches for {A}utomatic {I}dentification of {G}roups.
\newblock In {\em Information Retrieval and Mining in Distributed
  Environments}. 2011.

\bibitem{BKS01}
S.~B{\"o}rzs{\"o}nyi, D.~Kossmann, and K.~Stocker.
\newblock {T}he {S}kyline {O}perator.
\newblock In {\em Proc. of the {IEEE} Intl. Conf. on Data Engineering
  ({ICDE})}, 2001.

\bibitem{CC12}
I.~Cantador and P.~Castells.
\newblock {G}roup {R}ecommender {S}ystems: {N}ew {P}erspectives in the {S}ocial
  {W}eb.
\newblock In {\em Recommender Systems for the Social Web}. 2012.

\bibitem{CET05}
C.~Y. Chan, P.-K. Eng, and K.-L. Tan.
\newblock {S}tratified {C}omputation of {S}kylines with {P}artially-{O}rdered
  {D}omains.
\newblock In {\em Proc. of the {ACM} {SIGMOD} Intl. Conf. on Management of Data
  ({SIGMOD})}, 2005.

\bibitem{CJT+06}
C.~Y. Chan, H.~V. Jagadish, K.-L. Tan, A.~K.~H. Tung, and Z.~Zhang.
\newblock {F}inding k-dominant {S}kylines in {H}igh {D}imensional {S}pace.
\newblock In {\em Proc. of the {ACM} {SIGMOD} Intl. Conf. on Management of Data
  ({SIGMOD})}, 2006.

\bibitem{CBC+00}
Y.-C. Chang, L.~D. Bergman, V.~Castelli, C.-S. Li, M.-L. Lo, and J.~R. Smith.
\newblock {T}he {O}nion {T}echnique: {I}ndexing for {L}inear {O}ptimization
  {Q}ueries.
\newblock In {\em Proc. of the {ACM} {SIGMOD} Intl. Conf. on Management of Data
  ({SIGMOD})}, 2000.

\bibitem{ChaoBF05}
D.~L. Chao, J.~Balthrop, and S.~Forrest.
\newblock {A}daptive radio: achieving consensus using negative preferences.
\newblock In {\em ACM Conference on Supporting Group Work}, 2005.

\bibitem{CL09}
L.~Chen and X.~Lian.
\newblock {E}fficient {P}rocessing of {M}etric {S}kyline {Q}ueries.
\newblock {\em {IEEE} Transactions on Knowledge and Data Engineering ({TKDE})},
  21(3), 2009.

\bibitem{C03}
J.~Chomicki.
\newblock {P}reference formulas in relational queries.
\newblock {\em ACM Transactions on Database Systems ({TODS})}, 28(4), 2003.

\bibitem{CGGL03}
J.~Chomicki, P.~Godfrey, J.~Gryz, and D.~Liang.
\newblock {S}kyline with {P}resorting.
\newblock In {\em Proc. of the {IEEE} Intl. Conf. on Data Engineering
  ({ICDE})}, 2003.

\bibitem{CBH02}
A.~Crossen, J.~Budzik, and K.~J. Hammond.
\newblock {F}lytrap: intelligent group music recommendation.
\newblock In {\em International Conference on Intelligent User Interfaces},
  2002.

\bibitem{DKNS01}
C.~Dwork, R.~Kumar, M.~Naor, and D.~Sivakumar.
\newblock {R}ank aggregation methods for the {W}eb.
\newblock In {\em Proc. of the Intl. World Wide Web Conf. ({WWW})}, 2001.

\bibitem{ElahiGRMB14}
M.~Elahi, M.~Ge, F.~Ricci, D.~Massimo, and S.~Berkovsky.
\newblock {I}nteractive {F}ood {R}ecommendation for {G}roups.
\newblock In {\em ACM Conference on Recommender Systems, RecSys}, 2014.

\bibitem{FKS03}
R.~Fagin, R.~Kumar, and D.~Sivakumar.
\newblock {C}omparing {T}op k {L}ists.
\newblock {\em SIAM J. Discrete Math.}, 17(1), 2003.

\bibitem{FV07}
M.~Farah and D.~Vanderpooten.
\newblock {A}n {O}utranking {A}pproach for {R}ank {A}ggregation in
  {I}nformation {R}etrieval.
\newblock In {\em Proc. of the Intl. {ACM} {SIGIR} Conf. on Research and
  Development in Information Retrieval ({SIGIR})}, 2007.

\bibitem{FS93}
E.~A. Fox and J.~A. Shaw.
\newblock {C}ombination of {M}ultiple {S}earches.
\newblock In {\em Proc. of the Text Retrieval Conf. ({TREC})}, 1993.

\bibitem{GSO11}
I.~Garcia, L.~Sebastia, and E.~Onaindia.
\newblock {O}n the design of individual and group recommender systems for
  tourism.
\newblock {\em Expert Syst. Appl.}, 38(6), 2011.

\bibitem{GXLBHMS10}
M.~Gartrell, X.~Xing, Q.~Lv, A.~Beach, R.~Han, S.~Mishra, and K.~Seada.
\newblock {E}nhancing group recommendation by incorporating social relationship
  interactions.
\newblock In {\em ACM international conference on Supporting group work,
  GROUP}, 2010.

\bibitem{GSG07}
P.~Godfrey, R.~Shipley, and J.~Gryz.
\newblock {A}lgorithms and analyses for maximal vector computation.
\newblock {\em The Intl. Journal on Very Large Data Bases ({VLDBJ})}, 16(1),
  2007.

\bibitem{HKP01}
V.~Hristidis, N.~Koudas, and Y.~Papakonstantinou.
\newblock {PREFER}: {A} {S}ystem for the {E}fficient {E}xecution of
  {M}ulti-parametric {R}anked {Q}ueries.
\newblock In {\em Proc. of the {ACM} {SIGMOD} Intl. Conf. on Management of Data
  ({SIGMOD})}, 2001.

\bibitem{IBS08}
I.~F. Ilyas, G.~Beskales, and M.~A. Soliman.
\newblock {A} survey of top-{\it k} query processing techniques in relational
  database systems.
\newblock {\em ACM Computing Surveys}, 40(4), 2008.

\bibitem{Jameson04}
A.~Jameson.
\newblock {M}ore than the sum of its members: challenges for group recommender
  systems.
\newblock In {\em Working conference on Advanced visual interfaces}, 2004.

\bibitem{JS07}
A.~Jameson and B.~Smyth.
\newblock {R}ecommendation to {G}roups.
\newblock In {\em The Adaptive Web}, 2007.

\bibitem{KannanIP14}
R.~Kannan, M.~Ishteva, and H.~Park.
\newblock {B}ounded matrix factorization for recommender system.
\newblock {\em Knowl. Inf. Syst.}, 39(3), 2014.

\bibitem{KW05}
J.~Kay and W.~Niu.
\newblock {A}dapting {I}nformation {D}elivery to {G}roups of {P}eople.
\newblock In {\em Workshop on New Technologies for Personalized Information
  Access}, 2005.

\bibitem{K02}
W.~Kie{\ss}ling.
\newblock {F}oundations of {P}references in {D}atabase {S}ystems.
\newblock In {\em Proc. of the Intl. Conf. on Very Large Databases ({VLDB})},
  2002.

\bibitem{KKYY10}
J.~K. Kim, H.~K. Kim, H.~Y. Oh, and Y.~U. Ryu.
\newblock {A} group recommendation system for online communities.
\newblock {\em International Journal of Information Management}, 30(3), 2010.

\bibitem{KRR02}
D.~Kossmann, F.~Ramsak, and S.~Rost.
\newblock {S}hooting {S}tars in the {S}ky: {A}n {O}nline {A}lgorithm for
  {S}kyline {Q}ueries.
\newblock In {\em Proc. of the Intl. Conf. on Very Large Databases ({VLDB})},
  2002.

\bibitem{KI04}
G.~Koutrika and Y.~E. Ioannidis.
\newblock {P}ersonalization of {Q}ueries in {D}atabase {S}ystems.
\newblock In {\em Proc. of the {IEEE} Intl. Conf. on Data Engineering
  ({ICDE})}, 2004.

\bibitem{Kung1975}
H.~T. Kung, F.~Luccio, and F.~P. Preparata.
\newblock {O}n {F}inding the {M}axima of a {S}et of {V}ectors.
\newblock {\em Journal of {ACM} ({JACM})}, 22(4), 1975.

\bibitem{LL87}
M.~Lacroix and P.~Lavency.
\newblock {P}references: {P}utting {M}ore {K}nowledge into {Q}ueries.
\newblock In {\em Proc. of the Intl. Conf. on Very Large Databases ({VLDB})},
  1987.

\bibitem{JS10}
J.~Lee and S.-w. Hwang.
\newblock {BS}ky{T}ree: scalable skyline computation using a balanced pivot
  selection.
\newblock In {\em Proc. of the Intl. Conf. on Extending Database Technology
  ({EDBT})}, 2010.

\bibitem{LYHSB12}
J.~Lee, G.~won You, S.~won Hwang, J.~Selke, and W.-T. Balke.
\newblock {I}nteractive skyline queries.
\newblock {\em Inf. Sci.}, 211, 2012.

\bibitem{LZLL07}
K.~C.~K. Lee, B.~Zheng, H.~Li, and W.-C. Lee.
\newblock {A}pproaching the {S}kyline in {Z} {O}rder.
\newblock In {\em Proc. of the Intl. Conf. on Very Large Databases ({VLDB})},
  2007.

\bibitem{LYZZ07}
X.~Lin, Y.~Yuan, Q.~Zhang, and Y.~Zhang.
\newblock {S}electing {S}tars: {T}he k {M}ost {R}epresentative {S}kyline
  {O}perator.
\newblock In {\em Proc. of the {IEEE} Intl. Conf. on Data Engineering
  ({ICDE})}, 2007.

\bibitem{LC10}
B.~Liu and C.-Y. Chan.
\newblock {ZINC}: {E}fficient {I}ndexing for {S}kyline {C}omputation.
\newblock {\em Proc. of the VLDB Endowment}, 4(3), 2010.

\bibitem{LB13}
C.~Lofi and W.-T. Balke.
\newblock {O}n {S}kyline {Q}ueries and {H}ow to {C}hoose from {P}areto {S}ets.
\newblock In {\em Advanced Query Processing (1)}. 2013.

\bibitem{LJZ11}
H.~Lu, C.~S. Jensen, and Z.~Zhang.
\newblock {F}lexible and {E}fficient {R}esolution of {S}kyline {Q}uery {S}ize
  {C}onstraints.
\newblock {\em {IEEE} Transactions on Knowledge and Data Engineering ({TKDE})},
  23(7), 2011.

\bibitem{JM04}
J.~Masthoff.
\newblock {G}roup {M}odeling: {S}electing a {S}equence of {T}elevision {I}tems
  to {S}uit a {G}roup of {V}iewers.
\newblock {\em User Model. User-Adapt. Interact.}, 14(1), 2004.

\bibitem{MAS11}
J.~Masthoff.
\newblock {G}roup {R}ecommender {S}ystems: {C}ombining {I}ndividual {M}odels.
\newblock In {\em Recommender Systems Handbook}. 2011.

\bibitem{McCarthy2002}
J.~F. McCarthy.
\newblock {P}ocket {R}estaurant {F}inder: {A} situated recommender systems for
  groups.
\newblock In {\em Workshop on Mobile Ad-Hoc Communication}, 2002.

\bibitem{McCarthyA98}
J.~F. McCarthy and T.~D. Anagnost.
\newblock {M}usic{FX}: {A}n {A}rbiter of {G}roup {P}references for {C}omputer
  {A}upported {C}ollaborative {W}orkouts.
\newblock In {\em ACM Conference on Computer Supported Cooperative Work}, 1998.

\bibitem{CMS07}
K.~McCarthy, L.~McGinty, and B.~Smyth.
\newblock {C}ase-{B}ased {G}roup {R}ecommendation: {C}ompromising for
  {S}uccess.
\newblock In {\em International Conference on Case-Based Reasoning, ICCBR},
  2007.

\bibitem{McCarthySCMSN06}
K.~McCarthy, M.~Salam{\'{o}}, L.~Coyle, L.~McGinty, B.~Smyth, and P.~Nixon.
\newblock {CATS: A} {S}ynchronous {A}pproach to {C}ollaborative {G}roup
  {R}ecommendation.
\newblock In {\em Florida Artificial Intelligence Research Society Conference},
  2006.

\bibitem{MA02}
M.~H. Montague and J.~A. Aslam.
\newblock {Condorcet Fusion for Improved Retrieval}.
\newblock In {\em Proc. of the Intl. Conf. on Information and Knowledge
  Management}, 2002.

\bibitem{NSNK12b}
E.~Ntoutsi, K.~Stefanidis, K.~N{\o}rv{\aa}g, and H.-P. Kriegel.
\newblock {F}ast {G}roup {R}ecommendations by {A}pplying {U}ser {C}lustering.
\newblock In {\em Proc. of the Intl. Conf. on Conceptual Modeling ({ER})},
  2012.

\bibitem{CCKR01}
M.~O'Connor, D.~Cosley, J.~A. Konstan, and J.~Riedl.
\newblock {P}oly{L}ens: {A} recommender system for groups of user.
\newblock In {\em European Conference on Computer Supported Cooperative Work,
  ECSCW}, 2001.

\bibitem{PDMK09}
D.~P, P.~M. Deshpande, D.~Majumdar, and R.~Krishnapuram.
\newblock {E}fficient skyline retrieval with arbitrary similarity measures.
\newblock In {\em Proc. of the Intl. Conf. on Extending Database Technology
  ({EDBT})}, 2009.

\bibitem{PTFS05}
D.~Papadias, Y.~Tao, G.~Fu, and B.~Seeger.
\newblock {P}rogressive skyline computation in database systems.
\newblock {\em ACM Transactions on Database Systems ({TODS})}, 30(1), 2005.

\bibitem{PPC08}
M.-H. Park, H.-S. Park, and S.-B. Cho.
\newblock {R}estaurant {R}ecommendation for {G}roup of {P}eople in {M}obile
  {E}nvironments {U}sing {P}robabilistic {M}ulti-criteria {D}ecision {M}aking.
\newblock In {\em Asia Pacific Conference on Computer Human Interaction}, 2008.

\bibitem{PiliponyteRK13}
A.~Piliponyte, F.~Ricci, and J.~Koschwitz.
\newblock {S}equential {M}usic {R}ecommendations for {G}roups by {B}alancing
  {U}ser {S}atisfaction.
\newblock In {\em User Modeling, Adaptation, and Personalization}, 2013.

\bibitem{PCCA05}
S.~Pizzutilo, B.~De~Carolis, G.~Cozzolongo, and F.~Ambruoso.
\newblock {G}roup {M}odeling in a {P}ublic {S}pace: {M}ethods, {T}echniques,
  {E}xperiences.
\newblock In {\em International Conference on Applied Informatics and
  Communications}, 2005.

\bibitem{R88}
W.~H. Riker.
\newblock {\em {L}iberalism {A}gainst {P}opulism}.
\newblock Waveland Press Inc, 1988.

\bibitem{RACDY10}
S.~B. Roy, S.~Amer-Yahia, A.~Chawla, G.~Das, and C.~Yu.
\newblock {S}pace efficiency in group recommendation.
\newblock {\em VLDB J.}, 19(6), 2010.

\bibitem{SPP09}
D.~Sacharidis, S.~Papadopoulos, and D.~Papadias.
\newblock {T}opologically {S}orted {S}kylines for {P}artially {O}rdered
  {D}omains.
\newblock In {\em Proc. of the {IEEE} Intl. Conf. on Data Engineering
  ({ICDE})}, 2009.

\bibitem{SLNX09}
A.~D. Sarma, A.~Lall, D.~Nanongkai, and J.~Xu.
\newblock {R}andomized {M}ulti-pass {S}treaming {S}kyline {A}lgorithms.
\newblock {\em Proc. of the VLDB Endowment}, 2(1), 2009.

\bibitem{SK13}
H.~Shang and M.~Kitsuregawa.
\newblock {S}kyline {O}perator on {A}nti-correlated {D}istributions.
\newblock {\em Proc. of the VLDB Endowment}, 6(9), 2013.

\bibitem{ST12}
C.~Sheng and Y.~Tao.
\newblock {W}orst-{C}ase {I/O}-{E}fficient {S}kyline {A}lgorithms.
\newblock {\em ACM Transactions on Database Systems ({TODS})}, 37(4), 2012.

\bibitem{SWT08}
D.~W. Sprague, F.~Wu, and M.~Tory.
\newblock {M}usic selection using the {P}arty{V}ote democratic jukebox.
\newblock In {\em Working Conference on Advanced Visual Interfaces}, 2008.

\bibitem{SKP11}
K.~Stefanidis, G.~Koutrika, and E.~Pitoura.
\newblock {A} survey on representation, composition and application of
  preferences in database systems.
\newblock {\em ACM Transactions on Database Systems ({TODS})}, 36(3), 2011.

\bibitem{TEO01}
K.-L. Tan, P.-K. Eng, and B.~C. Ooi.
\newblock {E}fficient {P}rogressive {S}kyline {C}omputation.
\newblock In {\em Proc. of the Intl. Conf. on Very Large Databases ({VLDB})},
  2001.

\bibitem{TDLP09}
Y.~Tao, L.~Ding, X.~Lin, and J.~Pei.
\newblock {D}istance-{B}ased {R}epresentative {S}kyline.
\newblock In {\em Proc. of the {IEEE} Intl. Conf. on Data Engineering
  ({ICDE})}, 2009.

\bibitem{Taylor05}
A.~D. Taylor.
\newblock {\em {S}ocial choice and the mathematics of manipulation}.
\newblock Cambridge University Press, 2005.

\bibitem{VKHA09}
E.~Vildjiounaite, V.~Kyll{\"o}nen, T.~Hannula, and P.~Alahuhta.
\newblock {U}nobtrusive dynamic modelling of {TV} programme preferences in a
  {F}innish household.
\newblock {\em Multimedia Syst.}, 15(3), 2009.

\bibitem{WFP+08}
R.~C.-W. Wong, A.~W.-C. Fu, J.~Pei, Y.~S. Ho, T.~Wong, and Y.~Liu.
\newblock {E}fficient skyline querying with variable user preferences on
  nominal attributes.
\newblock {\em Proc. of the VLDB Endowment}, 1(1), 2008.

\bibitem{YM07}
M.~L. Yiu and N.~Mamoulis.
\newblock {E}fficient {P}rocessing of {T}op-k {D}ominating {Q}ueries on
  {M}ulti-{D}imensional {D}ata.
\newblock In {\em Proc. of the Intl. Conf. on Very Large Databases ({VLDB})},
  2007.

\bibitem{YuHSD14}
H.~Yu, C.~Hsieh, S.~Si, and I.~S. Dhillon.
\newblock {P}arallel matrix factorization for recommender systems.
\newblock {\em Knowl. Inf. Syst.}, 41(3), 2014.

\bibitem{YZHG06}
Z.~Yu, X.~Zhou, Y.~Hao, and J.~Gu.
\newblock {TV} {P}rogram {R}ecommendation for {M}ultiple {V}iewers {B}ased on
  user {P}rofile {M}erging.
\newblock {\em User Model. User-Adapt. Interact.}, 16(1), 2006.

\bibitem{ZMC09}
S.~Zhang, N.~Mamoulis, and D.~W. Cheung.
\newblock {S}calable skyline computation using object-based space partitioning.
\newblock In {\em Proc. of the {ACM} {SIGMOD} Intl. Conf. on Management of Data
  ({SIGMOD})}, 2009.

\bibitem{ZMKC10}
S.~Zhang, N.~Mamoulis, B.~Kao, and D.~W.-L. Cheung.
\newblock {E}fficient {S}kyline {E}valuation over {P}artially {O}rdered
  {D}omains.
\newblock {\em Proc. of the VLDB Endowment}, 3(1), 2010.

\bibitem{ZXD05}
Y.~Zhiwen, Z.~Xingshe, and Z.~Daqing.
\newblock {A}n adaptive in-vehicle multimedia recommender for group users.
\newblock In {\em IEEE Vehicular Technology Conference}, 2005.

\end{thebibliography}
